\documentclass[10pt,a4paper,twoside,pdftex]{phdthesis}

\usepackage{palatino}
\usepackage{calligra}

\usepackage[utf8]{inputenc} 
\usepackage{fancyhdr}
\usepackage{amsmath}
\usepackage{amssymb}
\usepackage{amsthm}
\usepackage[english,dutch]{babel}
\usepackage[style=british]{csquotes}
\usepackage{graphicx}
\usepackage{color}
\usepackage{bbm}
\usepackage{pictexwd,dcpic}
\usepackage{hyphenat}

\usepackage[style=numeric-comp, 
bibstyle=bib/thesis, 
citestyle=numeric-comp, 
sorting=nyt,
backend=biber,
backref=false,
backrefstyle=two,
isbn=false,
doi=false,
maxnames=4,
block=nbpar,
firstinits=true
]{biblatex}
\usepackage{caption}
\usepackage[paperheight=24cm,paperwidth=17cm,vmarginratio=1:1,twoside,
vmargin=3cm,hmargin=1.50cm,bindingoffset=0.50cm]{geometry}

\usepackage{ifpdf}
\ifpdf
\RequirePackage[pdftex]{hyperref}
\else
\RequirePackage[hypertex]{hyperref}
\fi
\hypersetup{pdfpagemode={UseNone},draft=false}

\usepackage{chngcntr}
\counterwithout{footnote}{chapter}

\ifpdf\AtBeginDocument{%
  \hypersetup{
pdftitle={Piecewise Flat Gravity in 3+1 dimensions},
pdfauthor={Maarten van de Meent},
pdfkeywords={Piecewise flat gravity, General Relativity, Exact solutions, Gravitational waves, Quantum gravity, Cosmic strings}
}
}\fi

\graphicspath{{pics/}}

\newcommand{\set}[1]{\left\{#1\right\}}
\newcommand{\floor}[1]{\left\lfloor#1\right\rfloor}
\newcommand{\hh}[1]{\left(#1\right)}
\newcommand{\bb}[1]{\big[#1\big]}
\newcommand{\bhh}[1]{\big(#1\big)}

\newcommand{\abs}[1]{\lvert#1\rvert}
\newcommand{\norm}[1]{\lVert#1\rVert}
\newcommand{\map}[3]{#1: #2\rightarrow #3}
\newcommand{\RR}{\mathbbm{R}}
\newcommand{\CC}{\mathbbm{C}}

\newcommand{\ZZ}{\mathbbm{Z}}
\newcommand{\ii}{\mathbbm{i}}
\newcommand{\ee}{\mathbbm{e}}

\newcommand{\Id}{\mathbbm{I}}
\newcommand{\ISO}[1]{\textrm{ISO}(#1)}
\newcommand{\SL}[1]{\textrm{SL}(#1)}
\newcommand{\Sl}[1]{\mathfrak{sl}(#1)}
\newcommand{\So}[1]{\mathfrak{so}(#1)}
\newcommand{\GL}[1]{\textrm{GL}(#1)}
\newcommand{\PSL}[1]{\textrm{PSL}(#1)}
\newcommand{\SO}[1]{\textrm{SO}(#1)}
\newcommand{\md}{\mathrm{d}}

\newcommand{\cs}{\mathcal{M}}
\newcommand{\csp}{\bar{\mathcal{M}}}
\newcommand{\hf}{\tfrac{1}{2}}

\newcommand{\s}[1]{\mspace{#1mu}}

\newcommand{\lp}{\textrm{lam}}

\DeclareMathOperator{\tr}{Tr}
\DeclareMathOperator{\re}{Re}
\DeclareMathOperator{\im}{Im}
\DeclareMathOperator{\bigO}{O}
\DeclareMathOperator{\Ad}{Ad}

\DeclareMathOperator{\m}{-\!}

\DeclareMathOperator{\pathord}{\mathcal{P}}

\newtheorem{principle}{Principle}
\newtheorem{definition}{Definition}
\newtheorem{lemma}{Lemma}
\newcommand{\fullTeX}{}

\makeatletter
\renewcommand{\@makechapterhead}[1]{{%
   \vspace*{\@spacebeforechapterhead}%

{  
 \parindent 0pt  \Large\scshape
   \ifnum \c@secnumdepth >\m@ne
      \if@mainmatter
         \@chapapp\enspace\Huge\thechapter
      \fi
   \fi
}

   \interlinepenalty\@M
   \vspace*{\@spaceinchapterhead}%
   \hrule
   \vspace*{\@spaceinchapterhead}%

   {\hfill \huge \bfseries #1}%
   \mbox{}\par
   \mbox{}\par
   \mbox{}\par
   \mbox{}\par
   \mbox{}\par
   \mbox{}\par
}}

\renewcommand{\@makeschapterhead}[1]{{%
   \vspace*{\@spacebeforechapterhead}%
   {\parindent 0pt \Large\scshape
   \phantom{\@chapapp\Huge 1}}%
   \interlinepenalty\@M
   \vspace*{\@spaceinchapterhead}%
   \hrule
   \vspace*{\@spaceinchapterhead}%
   {\hfill \huge \bfseries #1}%
   \mbox{}\par
   \mbox{}\par
   \mbox{}\par
}}
\makeatother

\begin{document}

\selectlanguage{english}

\pagestyle{fancyplain}
\makeatletter
\renewcommand{\sectionmark}[1]{\markright{\thesection:\ \it #1}}
\renewcommand{\chaptermark}[1]{\markboth{ \textsc{\@chapapp}\ \thechapter:\ \it #1}{}}
\makeatother

\lhead[\fancyplain{}{\nouppercase{\leftmark}}]{}
\chead[]{}
\rhead[]{\fancyplain{}{\nouppercase{\rightmark}}}
\lfoot[\thepage]{} 
\cfoot[]{}
\rfoot[]{\thepage}

\pagenumbering{roman}
\cleardoublepage

\cleardoublepage

\thispagestyle{empty}
\begin{center}
\phantom{}
\vspace{2.0cm}

{\fontshape{sc}
\fontsize{32}{32}\selectfont
Piecewise flat gravity\\[9pt]
\fontsize{24}{24}\selectfont
in 3+1 dimensions}

\vspace{1.5cm}

\fontsize{14}{14}\selectfont Stuksgewijs vlakke zwaartekracht in 3+1 dimensies
\\[3pt]
{\fontshape{sl}\fontsize{11}{11}\selectfont (met een samenvatting in het Nederlands)}

\vfill

\Large Proefschrift

\vspace{1.2cm}

\parbox{0.8\textwidth}{
\normalsize
ter verkrijging van de graad van doctor aan de
Universiteit Utrecht op gezag van de rector magnificus, prof.\ dr.\
G.\ J.\ van der Zwaan, ingevolge het besluit van het college voor promoties
in het openbaar te verdedigen op maandag 19 december 2011 des middags te 4.15 uur}

\vspace{1.2cm}

\large door

\vspace{1.2cm}

{\LARGE\fontshape{sc}\selectfont Maarten van de Meent}

\vspace{0.5cm}

\large geboren op 29 juni 1982 te Delft

\end{center}

\newpage
\thispagestyle{empty}
\noindent \normalsize Promotor: Prof.\ dr.\ G.\ 't Hooft

\ \vfill

\cleardoublepage\markboth{Contents}{}
\setcounter{tocdepth}{1}
\tableofcontents

\cleardoublepage

\chapter*{Publications}\addcontentsline{toc}{chapter}{Publications}
This thesis is based on the following publications:
\begin{itemize}
\item\fullcite{meent:2010}.
\item\fullcite{meent:2010b}.
\item\fullcite{meent:2011}.
\end{itemize}

\noindent The original idea for the model studied in this thesis was published in:
\begin{itemize}
\item\fullcite{hooft:2008}.
\end{itemize}
Many of the ideas expressed therein are reproduced (and expanded upon) in chapter~\ref{ch:3+1gravity}.

\newcommand{\publ}{}
\cleardoublepage
\pagenumbering{arabic}


\setlength{\parskip}{0.5ex plus 0.5ex minus 0.2ex}
\setlength{\parindent}{0.5cm}
\cleardoublepage 
\ifx\fullTeX\undefined
\documentclass[11pt,a4paper]{article}
\usepackage[utf8]{inputenc} 
\usepackage[english]{babel}
\usepackage{graphicx}
\usepackage{hyperref}
\usepackage{color}
\usepackage{amsmath}
\usepackage{amsthm}
\usepackage{amssymb}
\usepackage{bbm}
\usepackage{pictexwd,dcpic}

\graphicspath{{../pics/}}

\title{Motivation: Gravity in 2+1 dimensions}
\author{Maarten van de Meent}
\date{\today} 
bla bla
\begin{document}
\maketitle
\else
\chapter{Motivation: Gravity in 2+1 dimensions}
\fi
Einsteinian gravity in 2+1 dimensions is much simpler than in 3+1 dimensions because it has no local gravitational degrees of freedom. This can be seen by considering the Riemann tensor. In any dimension the Riemann tensor satisfies the following (anti)-symmetric relations under exchange of its indices
\begin{equation}
\begin{aligned}
R_{\mu\nu\rho\sigma} &= -R_{\nu\mu\rho\sigma}\\
R_{\mu\nu\rho\sigma} &= -R_{\mu\nu\sigma\rho}\\
R_{\mu\nu\rho\sigma} &= R_{\rho\sigma\mu\nu}.
\end{aligned}
\end{equation}
Consequently we may write the Riemann tensor in $d$ dimensions as
\begin{equation}
 R_{\mu\nu\rho\sigma} = \omega^{A}_{\mu\nu}\omega^{B}_{\rho\sigma}Q_{AB},
\end{equation}
where the $\omega^{A}_{\mu\nu}$ with $A = 1,\ldots,\tfrac{1}{2}d(d-1)$ form a complete basis of 2-forms and $Q_{AB}$ is a symmetric 2-tensor. That is, we can write the Riemann tensor as a symmetric tensor on the bundle of 2-forms. For a 3-dimensional manifold $M$, Poincaré duality tells us that the bundle of 2-forms, $\Lambda^2(TM)$, is isomorphic to the tangent bundle and that a complete basis is given by the Levi-Civita tensors, $\omega^{A}_{\mu\nu} = \epsilon_{\mu\nu}^{\phantom{\mu\nu}A}$. The Ricci tensor is therefore given by
\begin{equation}
\begin{aligned}
 R_{\mu\nu} &= g^{\rho\sigma}R_{\mu\rho\nu\sigma}\\
&= g^{\rho\sigma}\omega^{A}_{\mu\nu}\omega^{B}_{\rho\sigma}Q_{AB}\\
&= g^{\rho\sigma}
	\epsilon_{\mu\rho}^{\phantom{\mu\rho}A}
	\epsilon_{\nu\sigma}^{\phantom{\nu\sigma}B}
	Q_{AB}\\
&= (g_{\mu\nu}g^{AB}-\delta_\mu^B\delta_\nu^A) Q_{AB}
&= Q^A_A g_{\mu\nu}-Q_{\mu\nu}.
\end{aligned}
\end{equation}
Consequently, we see that the Einstein tensor is equal to $G_{\mu\nu} = -Q_{\mu\nu}$. The Riemann tensor in 3 dimensions simply \emph{is} the Einstein tensor. Einstein's equation therefore completely fixes the curvature in terms of the energy--momentum tensor.

In particular, if the energy--momentum tensor vanishes so does the Riemann tensor. Vacuum solutions of the Einstein equations in 2+1 dimensions are flat and do not have any local structure. This means that there are no local gravitational degrees of freedom or gravitational waves.

If spacetime is simply connected, this implies that spacetime is diffeomorphic to (the covering space of an open subset of) Minkowski space. Non-local gravitational degrees of freedom do exist if the spacetime has a non-trivial topology. Witten \cite{Witten:1988hc} showed that, if spacetime has a topology of the form $\RR \times \Sigma_g$, where $\Sigma_g$ is a Riemann surface of genus $g$, the gravitational field equations can be solved exactly, and there is a $(12g-12)$-dimensional space of solutions.  The free parameters were identified as the holonomies of a flat $\ISO{2,1}$ connection around the non-contractable loops of the spacetime.

By writing gravity in 2+1 dimensions as a Chern-Simon theory of $\ISO{2,1}$,\footnote{A few years earlier Achucarro and Townsend \cite{Achucarro:1987vz} had already demonstrated this connection in the context of three dimensional supergravity.} Witten was able to provide a finite quantization of this system. Thereby he disproved earlier belief that gravity in 2+1 dimensions was non-renormalizible.

The relatively simple setting of pure gravity in 2+1 dimensions (i.e. without any kind of matter content) has proven to be a valuable testing ground for approaches to quantizing gravity. Many different approaches have been applied. When compared, the results do not always appear to be equivalent. (See \cite{Carlip:1998} for a review.)

The idea behind Witten's approach was to first use the classical constraints to reduce the infinite number of gravitational degrees of freedom to a finite number, and then quantize the resulting system.  This is the general idea behind `reduced phase space' quantizations. Since in 2+1 dimensions the gravitational phase space can be reduced to a finite dimension, quantization becomes a matter of traditional quantum mechanics making such approaches particularly powerful. 

Different approaches to describing 2+1 dimensional gravity and reducing its phase space have been employed. Some, like Witten, have used a `first order' description,\cite{Unruh:1993ft} while others have used `second order' ADM formalism.\cite{Martinec:1984fs, Hosoya:1989sy} Nelson and Regge have made an extensive study of the classical algebra of observables to use as the basis of a canonical quantization.\cite{Nelson:1989zd, Nelson:1990ba, Nelson:1991xj, Nelson:1991an, Nelson:1992tg, Nelson:1992dg, Nelson:1992pb, Nelson:1993md} A yet other approach is to use geometrical techniques to make a complete classification of Lorentzian manifolds with constant curvature.\cite{Mess:2007, Thurston:1997} These each lead to different approaches to quantizing the pure gravity theory,\cite{Martinec:1984fs, Deser:1989, Witten:1989sx, Hosoya:1989sy, Ashtekar:1989qd, Rovelli:1989za, Carlip:1990kp, Waelbroeck:1994iy} which do not necessarily seem to agree.\cite{Carlip:1993zi} 

One thing to learn from this is that understanding the space of classical solutions is important for the quantization. In particular, quantizing the pure gravity theory divorced from its coupling to the matter content may not lead to the same results as quantizing the full theory. It is therefore essential to understand the classical space of solutions of gravity coupled to matter.

\section{Point particles}
Arguably the simplest form of matter one could consider is a collection of point particles. The problem of describing moving point particles in (2+1)-dimensional gravity was first considered by Staruszkiewicz in 1963,\cite{Staruszkiewicz:1963zz}  providing a solution for the two body problem. The general $N$-body problem --- assuming a topologically trivial background --- was solved by Deser, Jackiw, and 't Hooft in 1983.\cite{Deser:1983tn}

In 2+1 dimensions the Schwarzschild metric describing the gravitational field of a point particle is
\begin{equation}\label{eq:SS2+1}
\md\sigma^2 = -A(r)\md t^2 + A(r)^{-1}\md r^2 + r^2 \md\phi^2.
\end{equation}
The vacuum Einstein equation implies that the function $A(r)$ is constant. If we set the constant value to $(1-\alpha/(2\pi))^2$, then we can rescale $t$ and $r$ by a factor $(1-\alpha/(2\pi))$ to obtain
\begin{equation}
\md\sigma^2 = -\md t^2 + \md r^2 +(1-\frac{\alpha}{2\pi})^2 r^2 \md\phi^2.
\end{equation}
This is the metric of a cone. The tip of the cone (at $r=0$) is not a traditional curvature singularity since the Riemann curvature (and hence any curvature invariant) vanishes in any neighbourhood of the point. Its singular nature becomes apparent if one calculates the holonomy of a loop around the origin. The holonomy of a loop that encloses the origin once does not vanish as the loop is contracted to a point --- as it should if the metric was regular at the origin. Instead it stays constant at a rotation of $\alpha$ degrees. The angle $\alpha$ is called the deficit angle of the conical singularity.

As shown by Regge \cite{Regge:1961} such a singularity should be associated with a finite amount of curvature, i.e. the curvature  is a Dirac delta peak. As a consequence, the Energy--momentum tensor will also contain a delta peak at this point. In units where $c=1$ and $G=\tfrac{1}{8\pi}$, we find that the conical singularity is associated with a stationary point particle of mass $\alpha$. This conclusion can be confirmed by smearing out the singularity to a smooth manifold and integrating the associated energy--momentum tensor.

\begin{figure}[tb]
\centering\includegraphics[width=\textwidth]{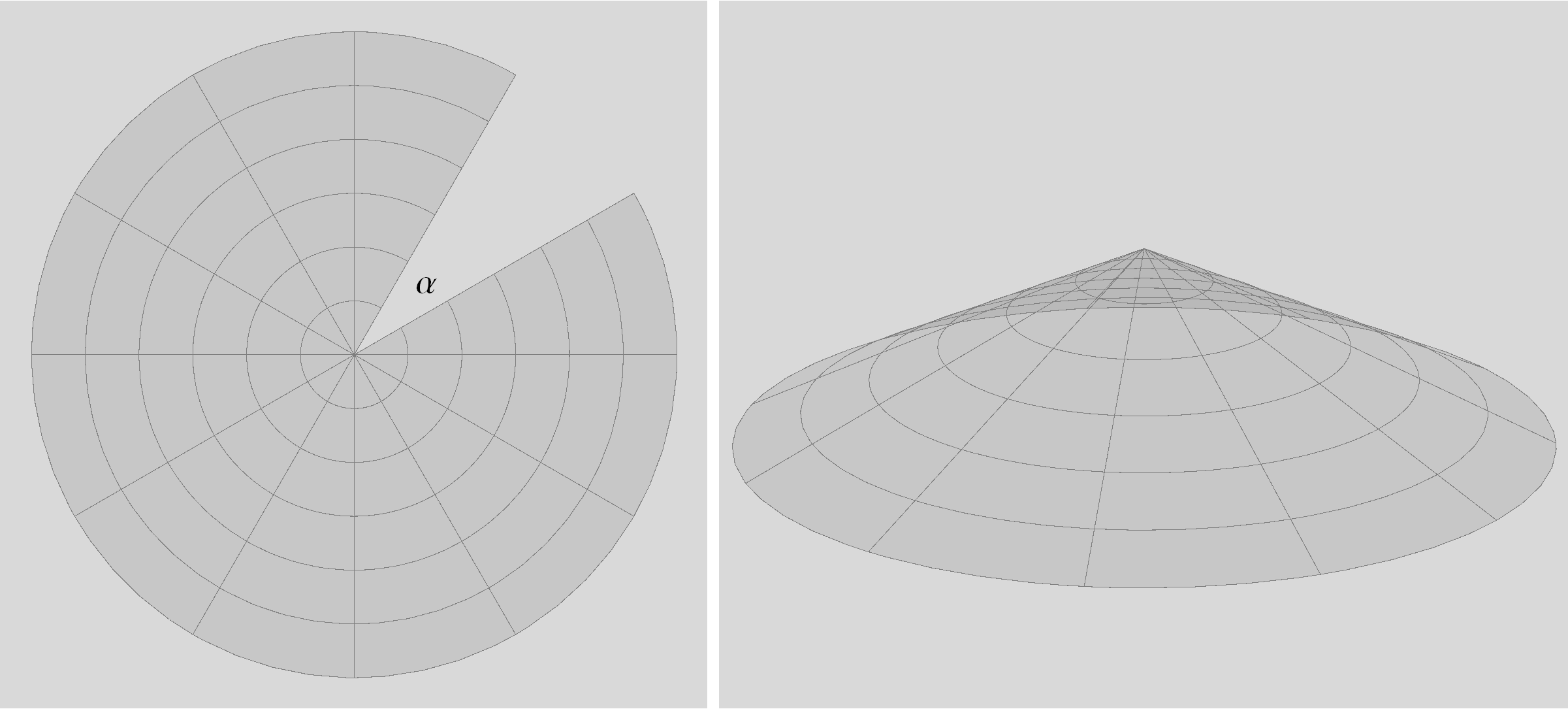}
\caption{Removing a wedge and identifying the opposing edges (on the left) creates a conical defect (on the right).}\label{fig:conicaldefect}
\end{figure}

A  cone can be constructed from the Euclidean plane by removing a wedge at the location of the particle and identifying the opposing edges of the wedge (see figure \ref{fig:conicaldefect}). The deficit angle is the angle of the removed wedge. This construction shows, in a geometrically explicit manner, that the space around the defect is locally flat.

\begin{figure}[tb]
\centering\includegraphics[width=\textwidth]{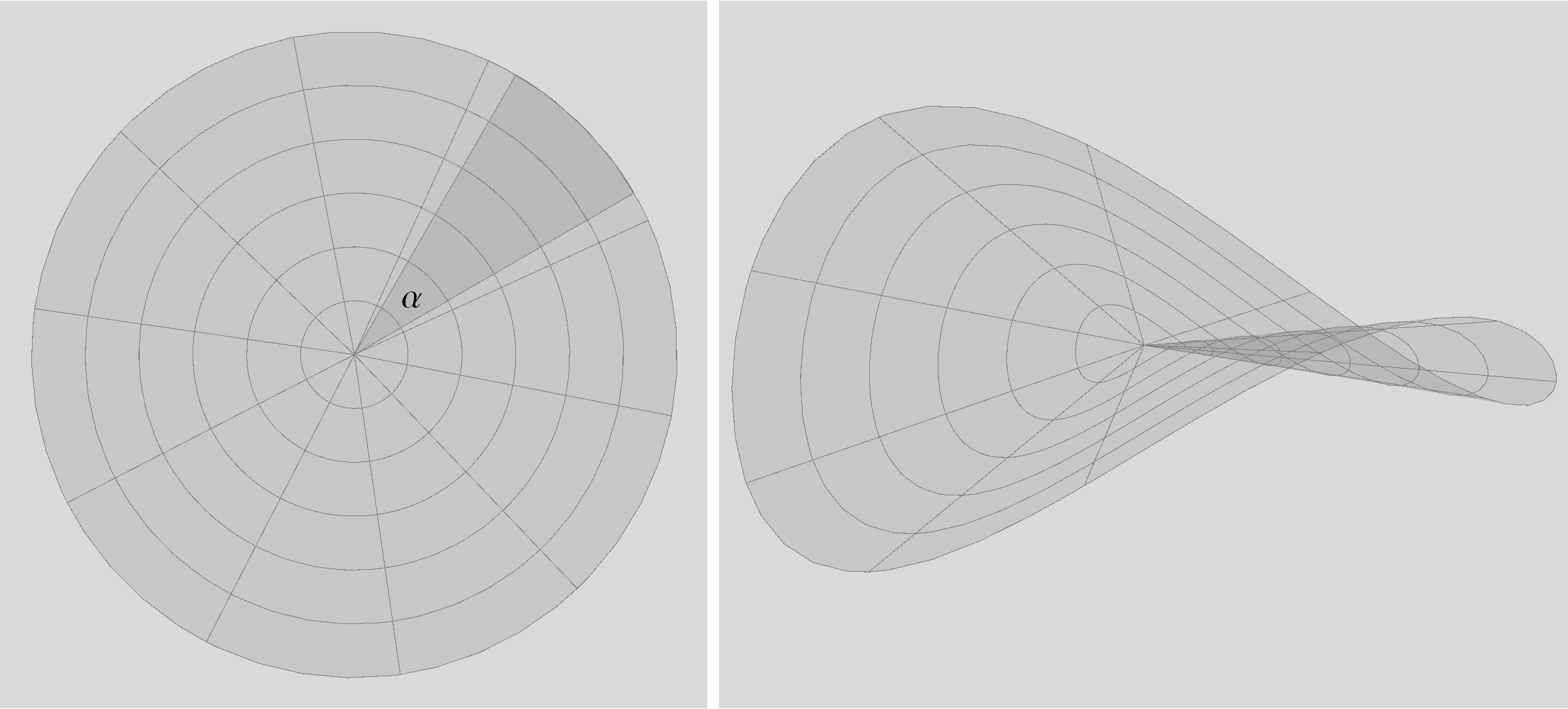}
\caption{Adding a wedge creates a conical defect with a surplus angle.}\label{fig:surplusangle}
\end{figure}

The conical defects constructed by removing a wedge are associated with positive curvature. Defects with negative curvature can be constructed from the Euclidean plane by --- instead of removing a wedge --- inserting a wedge along a line  (see figure \ref{fig:surplusangle}). This creates a defect with a surplus angle, which may be interpreted as a point particle with negative mass generating negative curvature.

By removing multiple wedges at different points one can create stationary configurations of multiple particles. If there are only positive mass particles, then the universe must be closed if the mass in any part exceeds $2\pi$.\cite{CohnVossen:1935} Application of the Gauss-Bonnet theorem then immediately implies that the total mass in the universe must be equal to $4\pi$.

Any long range interaction between particles must be accompanied by a field, which has non-zero energy--momentum, and therefore generates its own curvature. Therefore, if we assume that point particles are the only source of curvature, the particles can only move along  straight lines.

\begin{figure}[tb]
\centering\includegraphics[width=\textwidth]{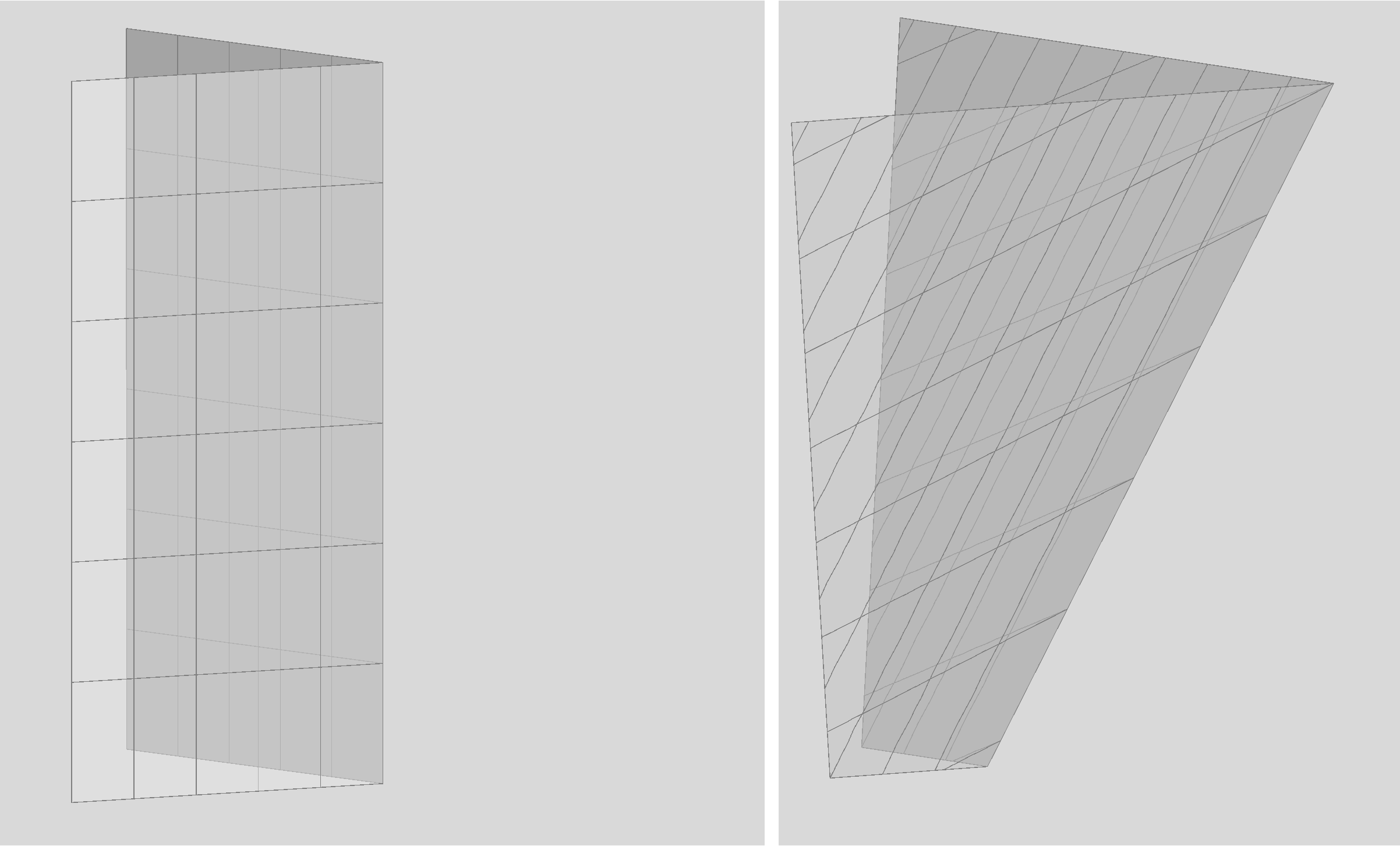}
\caption{A moving defect is obtained by boosting a stationary defect.}\label{fig:boosteddefect}
\end{figure}

The geometry generated by a moving particle can be obtained by Lorentz boosting the conical singularity of a stationary defect. If we start with a stationary particle at the origin, then the path of the particle through 3-dimensional spacetime after a boost $B$, is a straight timelike line that goes through the origin at $t=0$. The wedge of spacetime that was removed at the particle also gets tilted (see figure \ref{fig:boosteddefect}). Before the boost the map identifying the two sides of the wedge was a rotation $R$, the holonomy of a loop around the origin. The map $Q$ identifying the opposing sides of the wedge after the boost is a more general Lorentz transformation obtained by conjugation of $R$ by $B$, i.e. $Q=BRB^{-1}$. This is the holonomy of a loop around the moving particle.

The geometry corresponding to a particle located at another position than the origin at $t=0$ can be obtained from the previous case by applying an appropriate shift $S$. The map identifying opposing sides of the wedge no longer is a Lorentz transformation, but an element $P$ of the Poincaré group $\ISO{2,1}$ obtained by conjugation of $Q$ with $S$, i.e. $P= SQS^{-1}$. This is the so called Poincaré holonomy of a loop around the defect.\footnote{Because the spacetime around the defects is completely flat, it is possible to parallel transport not just the tangent space but the entire coordinate frame along a curve. Transporting the coordinate frame along a closed loop will in general result in a coordinate frame that differs from the one you started with by a Poincaré transformation.}

For a stationary particle, $R$ represented the mass of the particle.  After a boost $B$, the holonomy $Q$ encodes both the energy and the velocity of the moving particle. In fact, as shown by Matschull and Welling,\cite{Matschull:1997du} the holonomy of the defect may be interpreted as the covariant momentum of the particle. The Poincaré holonomy $P$ captures the complete data of the trajectory of the particle. In particular, the trajectory of the particle is given by the set of fixed points of the Poincaré holonomy. In this sense the Poincaré group may be viewed as the covariant phase space of a point particle in (2+1)-dimensional gravity.

\subsection{Timelike loops}
In equation \eqref{eq:SS2+1} we considered a Schwarzschild-like solution in 2+1 dimensions; the most general static and rotationally symmetric solution one can write down. More generally, one can relax the condition of the metric being static to being stationary and allow a Kerr-like solution (see \cite{Clement:1983nk} for a derivation)
\begin{equation}
\md\sigma^2 = -\md t^2 +2\omega \md t\md\phi + \md r^2 +\bb{(1-\alpha/(2\pi))^2 r^2 -\omega^2}\md\phi^2.
\end{equation}
This metric corresponds to a spinning point particle. The corresponding geometry can be obtained from Minkowski space by removing a wedge of spacetime like in the static case, but identifying the opposing sides with a shift in time. The Poincaré holonomy of such a defect therefore includes a timelike shift --- even in the frame in which it is at rest.

This shift opens up the possibility of creating timelike loops. For example, in the above metric the curve given by the constant values $r=r_0 < \omega/(1-\tfrac{\alpha}{2\pi})$ and $t=t_0$ is closed and timelike. Since closed timelike curves are irreconcilable with causality, this strongly implies that such sources should not be physically allowed.

This, however, is not enough to bar the occurrence of closed timelike curves. As pointed out by Deser, Jackiw, and 't Hooft in 1983,\cite{Deser:1983tn} the geometry of a pair of particles moving past each other resembles that of a single pointlike source with the same angular momentum. In 1991, Gott  \cite{Gott:1990zr} showed that if one constructs a pair of particles that move past each other with a sufficient velocity, then it is possible to find closed timelike curves around the pair of particles.  Deser, Jackiw, and 't Hooft \cite{Deser:1991ye} quickly pointed out that such a solution would have unacceptable boundary conditions at infinity.

In principle, this left open the option that a Gott pair could form dynamically from the decay of slow moving particles. Carroll, Farhi, Guth, and Olum \cite{Carroll:1991nr, Carroll:1994hz} showed that this would be impossible in an open universe, since no part of an open universe could contain enough energy to create a Gott pair. In the case of a closed universe 't Hooft \cite{hooft:1992} showed by considering a Cauchy surface tesselated by polygons that although Gott pairs could form, the universe would collapse in a big crunch before a closed timelike curve could form.

\subsection{Quantization}

The exact solubility of a system of gravitating point particles in 2+1 dimensions invites a reduced phase space approach to quantizing the system. That is, one first classically solves the equations of motion for the gravitational field to obtain a restricted phase space for the point particles, and then quantizes this reduced system. This obviously requires a thorough understanding of the classical phase space.

The polygon tessellation by Cauchy surfaces introduced by 't Hooft \cite{hooft:1992} allowed him to completely formulate the dynamics of the system in terms of the edge lengths and their rapidities. The quantization of this system was studied in.\cite{hooft:1993nj} Because the Hamiltonian of the system is given by a periodic angle, it was observed that time was discretized after quantization.  Due to Lorentz invariance one would expect space to be discretized as well, but this was not the case because the momenta (the rapidities of the edges) are not periodic. In a different parametrization using the coordinates of the vertices of the polygons rather than the edges length as coordinates, 't Hooft \cite{hooft:1996uc} found that the conjugate momenta lay on a sphere, and consequently found that in the canonical quantization spacetime was given by a lattice. 

Using different methods  Matschull and Welling \cite{Matschull:1997du} showed that the covariant momentum space of a gravitating point particle in 2+1 dimensions should be taken to be the spin-1/2 representation of the Lorentz group rather than the $S^1\times S^2$ topology proposed by 't Hooft.\footnote{In fact, it is more inline with 't Hooft's covariant momenta used in \cite{hooft:1993nj} which used one angle and two hyperbolic angles.} Besides the discretization of time, Matschull and Welling remarked that the curvature of momentum space leads to non-commuting operators for the coordinates in the quantum theory --- non-commutative geometry seems to be generated naturally. 

Waelbroeck and Zapata \cite{Waelbroeck:1996sg} have argued that the conclusion whether time is discretized in the quantization of the polygon model depends on the details of the quantization procedure used. In particular,  care needs to be taken in the implementation of gauge fixings. Further study of the phase space of the polygon model by Kadar \cite{Kadar:2003ie,Kadar:2004im} and Eldering \cite{Eldering:2006kk} have revealed further subtleties which complicate the quantization of the model.

Meanwhile other approaches to quantum gravity in 2+1 dimensions have yielded conflicting results with respect to the presence of any fundamental discreteness. A loop quantum gravity based analysis \cite{Freidel:2002hx} indicates that the length of spacelike intervals should be continuous, whereas the length of timelike intervals should be discrete. However, a more recent quantization using Dirac variables  associated to physical lengths and time intervals has found no indication of any discreteness.\cite{Budd:2009kf} 

\section{Towards 3+1 dimensions}
The broad idea of the work described in this thesis is to generalize the 2+1 dimensional model of gravitating point particles to 3+1 dimensions. The naive approach would be to study a system of gravitating point particles interacting according to the rules of general relativity in 3+1 dimensions. Of course, such an approach would not share the nice properties of the system of gravitating point particles in 2+1 dimensions. In fact, in 3+1 dimensions one cannot even restrict one's attention to point particles because their gravitational interaction would inevitably lead to the production of gravitational waves.

Instead we take a more unconventional approach and attempt to construct a model that preserves some of the essential features of the model of gravitating point particles in 2+1 dimensions. In particular, we want to preserve the local finiteness of the model, which is key for its quantization. In essence this means that we want to preserve the property of gravity in 2+1 dimensions that the Einstein tensor completely determines the geometry. In particular, we want that regions of spacetime that are completely devoid of matter (i.e. where the energy--momentum tensor vanishes) are (Riemann) flat.

We find that this is possible, if we study a model of propagating straight cosmic strings propagating in 3+1 dimensions according to the rules of general relativity. 

Although our motivation for studying this model is the possibility of formulating a theory of quantum gravity in 3+1 dimensions, we completely focus on the classical aspects of the model. As the history of quantization of 2+1 gravity shows, complete and correct knowledge of the classical phase space of a model is essential for its (reduced phase space) quantization. Understanding the classical behaviour of the model is already quite a challenge.

The general plan of this thesis is as follows. Chapter \ref{ch:3+1gravity} establishes the various foundational issues of the model studied, and describes a number of ways in which configurations of straight cosmic strings may be parametrized. In chapter \ref{ch:collisions} we then study what happens in the model when two straight cosmic strings collide. It is found that the model may not be consistent for certain extreme collisions. Ignoring these issues we continue to study the continuum limit of the model using a linearized approximation in chapter \ref{ch:cont}, finding restrictions on the types of matter that may be approximated by the model. In chapter \ref{ch:gravwave} we then show that the model reproduces gravitational waves as an emergent phenomenon in continuum limit. Finally, in chapter \ref{ch:openproblems} we will discuss the open problems with the model and how these relate to a possible quantization.

\section{Conventions}
Throughout this thesis, we shall use the following conventions unless explicitly noted otherwise.
\begin{itemize}
\item Spacetime metrics in 3+1 dimensions have a signature $(- + + +)$.
\item All quantities are expressed in natural units such that $c =\hbar=8\pi G=1$.
\item Lightcone coordinates $u$ and $v$ are defined as
\begin{equation}
u = \frac{z+t}{\sqrt{2}},\quad\text{and}\quad v = \frac{z-t}{\sqrt{2}}.
\end{equation}
In particular the Minkowski metric in lightcone coordinates is
\begin{equation}
 \md s^2 = 2\md u\md v +\md x^2+\md y^2.
\end{equation}
\item The Fourier transform of a function $f(x)$ is defined as
\begin{equation}
\bar{f}(k) = \int_{-\infty}^\infty \md x f(x)\exp(-2\pi\ii k x).
\end{equation}
Consequently, the inverse Fourier transform is 
\begin{equation}
f(x) = \int_{-\infty}^\infty \md k \bar{f}(k)\exp(2\pi\ii k x).
\end{equation}
\end{itemize}
\nocite{hooft:1988yr,hooft:1992iw,hooft:1993cl,hooft:1993gz,hooft:2009}

\ifx\fullTeX\undefined

\bibliographystyle{../bib/utcaps}
\bibliography{../bib/thesis}
\end{document}
\fi

\cleardoublepage 
\ifx\fullTeX\undefined
\documentclass[11pt,a4paper]{article}
\usepackage[utf8]{inputenc} 
\usepackage[english]{babel}
\usepackage{graphicx}
\usepackage{hyperref}
\usepackage{color}
\usepackage{amsmath}
\usepackage{amsthm}
\usepackage{amssymb}
\usepackage{bbm}
\usepackage{pictexwd,dcpic}

\graphicspath{{../pics/}}

\title{Piecewise flat gravity in 3+1 dimensions}
\author{Maarten van de Meent}
\date{\today} 

\begin{document}
\maketitle
\else
\chapter{Foundational matters}\label{ch:3+1gravity}
\fi
Unlike in 2+1 dimensions, the Riemann tensor in 3+1 dimensions cannot be completely expressed in terms of the Einstein tensor. Consequently, the Einstein equation does not completely fix the geometry in terms of the matter content. As a result general relativity in 3+1 dimensions has geometrical local degrees of freedom, which are dynamical. These degrees of freedom make the (3+1)-dimensional theory much richer than the (2+1)-dimensional one. For example, they allow the appearance of gravitational waves and long range gravitational fields.

However, these local degrees of freedom are also notoriously problematic when trying to quantize general relativity. There are many ways to phrase the problems that occur. A simple one is that from a field theory point of view these degrees of freedom form a spin-2 gauge field, the graviton. The coupling constant of this field, Newton's constant, has negative mass dimension. Consequently, a naive power counting tells us that the resulting Feynman diagrams in perturbative quantum field theory will have non-renormalizable UV divergences. This can be confirmed by explicitly doing the loop integrations up to two loops.\cite{Hooft:1974bx,Goroff:1985th}

Many possible ways around this problem have been investigated over the years. For example, in supergravity \cite{Freedman:1976xh} the hope is that the addition of new fields together with supersymmetry would help cancel the divergent terms.\footnote{This hope was somewhat dampened by the discovery that the cancellations found do not generically persist beyond loop order.\cite{Deser:1977nt,Kallosh:1980fi} More recently however, it was found that in the specific case of maximal ($N=8$) supergravity the cancellations persist to much higher orders, leading to the suggestion that it might in fact be finite.\cite{Stelle:2009zz, Dixon:2010gz,Bern:2011qn}} In other approaches, like string theory, general relativity and its metric only appear as an effective field theory of some more fundamental quantum theory that is renormalizable. Another possibility --- expressed in theories like loop quantum gravity \cite{Ashtekar:1986yd,Rovelli:2010bf} --- is that the renormalization problems disappear if the theory is quantized using a different set of variables. Yet others blame the use of perturbative methods and hope that the issues with perturbative quantum gravity may be circumvented by using a non-perturbative approach --- for example as used in causal dynamical triangulations.\cite{Ambjorn:2006jf}

Our take in this matter  is that the local gravitational/geometrical degrees of freedom are themselves the root of the problem. The relative success of the quantization of (2+1)-dimensional gravity relies on the fact that a system having a finite number of particles will only have a finite number of degrees of freedom. Empty space is essentially featureless. We would like to recreate this situation in a theory of gravity in 3+1 dimensions.

Our approach to reach this will be somewhat heavy handed. We simply impose as a new principle that empty space should be featureless.
\begin{principle}
\emph{No local structure}.
The Riemann tensor in empty space\footnote{We define ``empty space''  as a region of spacetime where the energy--momentum tensor vanishes.} vanishes.
\end{principle}
At first sight, this is a substantial departure from conventional general relativity. How nevertheless general relativity is recovered in the continuum limit is the subject of chapters \ref{ch:cont} and \ref{ch:gravwave} of this thesis. In addition, we require that the Einstein equation continues to hold in all of spacetime ---  including regions that are not empty. Combined with  principle 1 this requirement puts very stringent limitations on the ``matter'' contents of the theory. For example, one cannot add point particles, because the Einstein equation would then imply that the surrounding empty space should have the Schwarzschild metric, which does not have a vanishing Riemann curvature.

One can however add conical curvature defects. Conical defects have codimension 2, so in 3+1 dimensions they are 2-dimensional planes. These are  the only degrees of freedom that we will allow in the model. We want to interpret these 2-dimensional planes as the world sheet of a 1-dimensional straight line propagating through space. A timelike conical defect  thereby corresponds to a line defect travelling at a subluminal speed, while a spacelike conical defect corresponds to a line defect travelling at a superluminal speed. In the latter case, this means that there exists a Lorentz frame in which the spacelike plane lies in a single time slice. It would therefore represent a defect instantaneously appearing and disappearing in an extended region. This is unacceptable behaviour for a physical excitation from the point of view of local causality. This leads to the second pillar of our model:

\begin{principle}
\emph{Local Causality}. The only allowed degrees of freedom are non-spacelike conical curvature defects.
\end{principle}
The only allowed defects are therefore either (1+1)-dimensional or lightlike surfaces. We can view these as 1-dimensional defect lines propagating through space. Physically, they correspond to straight cosmic strings moving at a constant velocity.

Although our model is motivated as a step towards a renormalizable theory of quantum gravity, we are not actually going to discuss any quantization of the model. Instead we are taking the lesson from (2+1)-dimensional gravity to heart, that thorough knowledge of the classical phase space is required for a proper quantization. The rest of this thesis is devoted to developing the classical description of a system of moving line defects, and discussing the subtleties that this involves.

In this chapter we discuss the foundational aspects of the model, and the necessary formalism needed to describe the possible configurations. In section \ref{sec:statdefect} we first discuss the geometry of a single stationary defect. We continue with the discussion of moving defects in section \ref{sec:movingdefects}. Section \ref{sec:paramdefect} then discusses  different ways of parametrizing a single defect. In particular we discuss the use of the holonomy of a loop around a defect to identify the defect. We discuss some different types of defects in section \ref{sec:generaldefects}, and elaborate on the possibility of massless lightlike defects. In section \ref{sec:multipledefects} we discuss geometries with multiple defects. This opens the possibility of junctions of defects, which are discussed in the following  section (\ref{sec:junctions}). Instead of focussing on the curvature defects, the considered geometries can also be described with a focus on the local flatness by describing them as piecewise flat manifolds, this is discussed in section \ref{sec:PWFmanifolds}. The 
\hyperref[sec:otherapproaches]{final section} of this chapter discusses the relation between the gravity model discussed in this thesis and other piecewise flat approaches to gravity.

\begin{figure}[tb]
\centering\includegraphics[width=10cm]{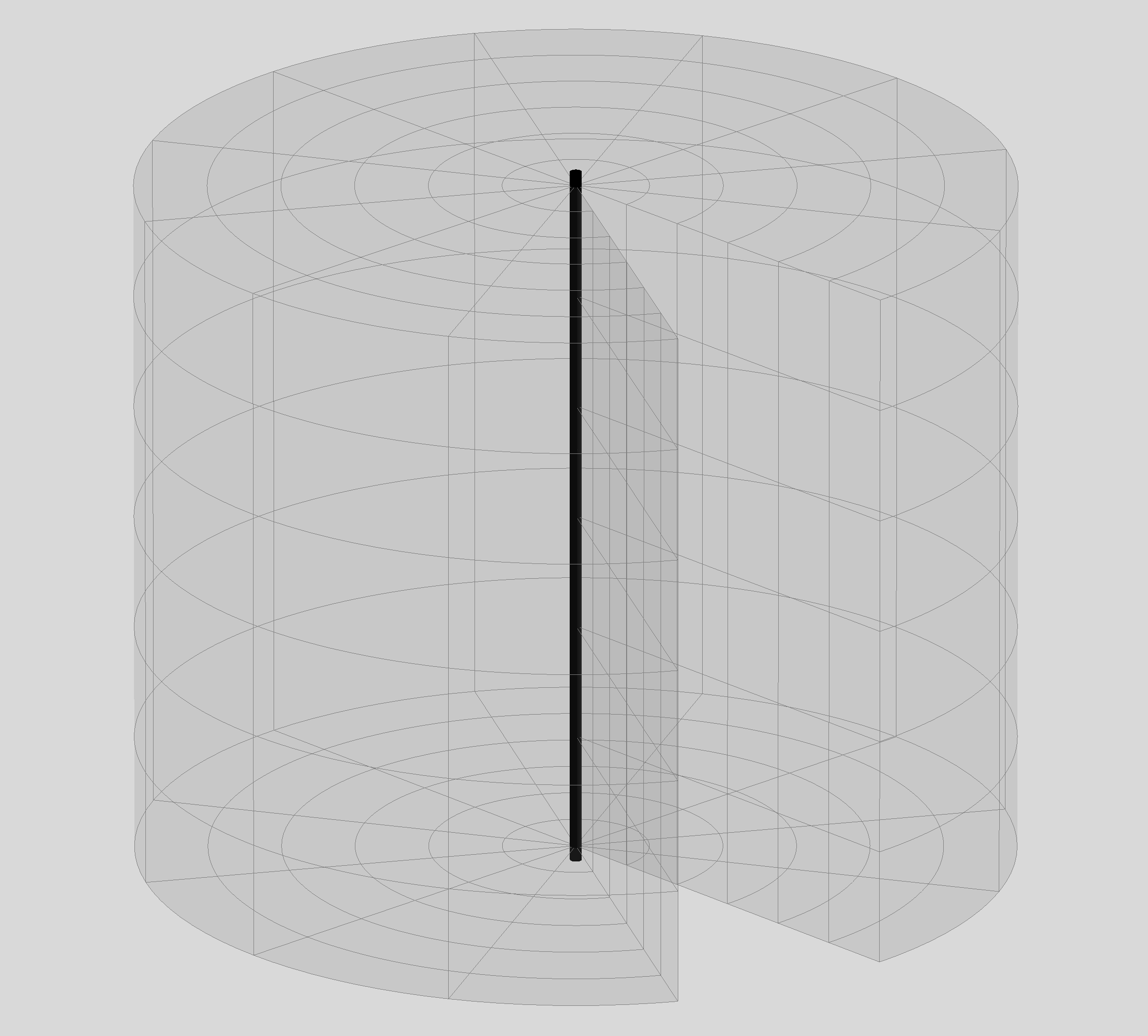}
\caption{A conical singularity in 3 spatial dimensions can be constructed by removing a 3-dimensional wedge.}\label{fig:3+1statdefect}
\end{figure}

\section{Stationary defects}\label{sec:statdefect}
We first describe a single stationary conical defect in 3+1 dimensions. Since the defect is stationary, all time slices are identical, and we just need to consider a conical defect in 3-dimensional Euclidean space. We can obtain the geometry of such a defect as follows (see figure \ref{fig:3+1statdefect}). Like a conical defect in 2 dimensions, we start from a 3-dimensional Euclidean space, $\RR^3$. From this space we remove a single wedge, and identify the opposing sides --- in this case planes.

In flat polar coordinates $(r,\theta,z)$ the effect of removing a wedge is that at the location where the wedge was removed the azimuthal coordinate $\theta$ will make  a jump, say from $\beta$ to $\beta+\alpha$, where $\alpha$  is the angular size of the removed wedge, $\alpha$; the deficit angle of the conical defect.  The metric on the remaining region (including the time direction) is still the common Minkowski metric,
\begin{equation}
\md s^2 = -\md t^2 +\md r^2 +r^2\md \theta^2 +\md z^2.
\end{equation}
However, because the azimuthal angle $\theta$ makes a jump, its period is reduced to $2\pi -\alpha$. Alternatively, we can rescale the azimuthal coordinate $\theta$ by a factor $1-\tfrac{\alpha}{2\pi}$ such that its period remains $2\pi$. As a result the metric becomes,
\begin{equation}\label{eq:staticdefectmetric}
\md s^2 = -\md t^2 +\md r^2 +(1-\frac{\alpha}{2\pi})^2 r^2\md \theta^2 +\md z^2.
\end{equation}
Note that this metric does not record the value of $\beta$. This angle, which gives the direction of the missing wedge, plays no role in the resulting geometry. In fact, one could have removed multiple wedges along the same axis, and obtained the same result as long as the total angle of the removed wedges  was the same.

Since this metric is locally isometric to the Minkowski metric, its curvature vanishes everywhere, except possibly at the coordinate singularity at $r=0$. That the curvature does not vanish at $r=0$ can be inferred from calculating the holonomy of a loop around the $r=0$ axis. Consider a path $\gamma^\mu(\lambda) = (0,\tfrac{2\pi}{2\pi-\alpha},\lambda,0)$. The holonomy $Q^\mu_{\phantom{\mu}\nu}\in \SO{3,1}$ of this path is
\begin{equation}\label{eq:holstatic}
\begin{aligned}
Q^\mu_{\phantom{\mu}\nu}&=\exp\hh{-\int_\gamma \Gamma^{\mu}_{\sigma\nu}\gamma^\sigma(\lambda) \md\lambda}\\
&=\begin{pmatrix} 1 & 0 & 0 & 0 \\
0 & \cos\alpha & \sin\alpha & 0\\
0 & -\sin\alpha & \cos\alpha & 0\\
0 & 0 & 0 & 1 
\end{pmatrix},
\end{aligned}
\end{equation}
where $\Gamma^{\mu}_{\sigma\nu}$ is the Levi-Civita connection of the metric \eqref{eq:staticdefectmetric}. Not coincidently, this rotation about the $z$-axis is the map that identifies the opposing sides of the wedge.\footnote{More correctly, it is the pull-back of that map to the tangent space, but since locally we are in Minkowski space these two can be identified.} Since this loop has a non-trivial holonomy, there must be curvature somewhere inside the loop. Since the space outside the $r=0$ axis is completely flat, this curvature must be located on the axis.

To calculate the curvature associated to the axis it is convenient to change to a set of coordinates where the $r$ coordinate is rescaled such that the azimuthal part of the metric takes the canonical form $r^2 \md \theta^2$,
\begin{equation}
\md s^2 = -\md t^2 +\bhh{\frac{2\pi}{2\pi-\alpha}}^2\md r^2 + r^2\md \theta^2 +\md z^2.
\end{equation}
In these coordinates, we can ``smear out'' the singularity at $r=0$ by turning $\alpha$ into a function of $r$,
\begin{equation}
\alpha_\epsilon(r) \equiv \begin{cases}
 \alpha_0 \frac{r}{\epsilon}& \text{for $r<\epsilon$}\\
 \alpha_0&  \text{otherwise.}\end{cases}
\end{equation}
For $r>\epsilon$ this is the same metric as before, but as $r$ approaches zero the metric continuously approaches the Minkowski metric. The holonomy of a loop of radius $r$ is a rotation about the $z$-axis of $\alpha_\epsilon(r)$. The holonomy thus becomes trivial at the $r=0$ axis, and the space is regular.

For this non-singular metric we can calculate the Einstein tensor. For $r>\epsilon$, the metric is unchanged, and the Einstein tensor vanishes. For $r<\epsilon$ the Einstein tensor is,
\begin{equation}
G_{\mu\nu} = \begin{pmatrix}
\frac{\alpha_0}{2\pi\epsilon}(\frac{1}{r}- \frac{\alpha_0}{2\pi\epsilon})& 0 & 0 & 0\\
0 & 0 & 0 & 0\\
0 & 0 & 0 & 0\\
0 & 0 & 0 & -\frac{\alpha_0}{2\pi\epsilon}(\frac{1}{r}- \frac{\alpha_0}{2\pi\epsilon})
\end{pmatrix}.
\end{equation}
The total curvature of the smeared out singularity is therefore,
\begin{equation}\label{eq:Tstationarystring}
\int_0^{\epsilon}\int_0^{2\pi} G_{\mu\nu} \frac{2\pi r}{2\pi-\alpha_0 \frac{r}{\epsilon}}\md\theta\md r= \begin{pmatrix}
\alpha_0 & 0 & 0 & 0\\
0 & 0 & 0 & 0\\
0 & 0 & 0 & 0\\
0 & 0 & 0 & -\alpha_0 
\end{pmatrix}.
\end{equation}
This is independent of the smearing parameter $\epsilon$. Consequently, we can associate a delta peaked curvature to the conical singularity. If we set $8\pi G =1$, then the Einstein equation identifies \eqref{eq:Tstationarystring} as the energy--momentum tensor of the conical defect. This is the well-known energy--momentum tensor of a stationary straight cosmic string (see for example \cite{Brandenberger:2007ae}): it has  a linear mass/energy density equal to its deficit angle $\alpha$, and an equal tension. 

Note that in this discussion there is no need for $\alpha$ to be a positive number. Geometrically, a negative value of $\alpha$ corresponds to opening the space along some hypersurface and inserting a wedge of angle $\abs\alpha$, which creates a defect with a surplus angle. Normally, cosmic strings with a negative mass would be highly unstable, since negative mass implies a positive pressure along the string. Any deviation from straight would cause the string to  buckle immediately. In our model, however, the principle that the space surrounding the defect is locally flat stabilizes negative mass defects because it forces the defects to be always straight; the destabilizing fluctuations simply cannot occur within the model.

Still, one might object to the model including objects with negative mass, since these are in obvious violation of the various energy conditions that one generally expects realistic classical matter to satisfy. However, in the model that we are describing, the line defects are not just describing matter excitations, they also represent the gravitational excitations of the model. If this model is to have any hope of representing vacuum spacetimes at large scales, where on average the Einstein tensor vanishes, but the Weyl tensor has a non-zero average, then it needs to have both defects that add positively and negatively to the Einstein tensor. In chapter \ref{ch:gravwave} we will see that having both types of defects is, in fact, essential in reproducing gravitational waves.

\section{Moving defects}\label{sec:movingdefects}
Since the spacetime around a stationary defect is locally Minkowski, we can obtain the geometry of a moving defect by applying a Lorentz boost to the geometry of a stationary defect. A (3+1)-dimensional wedge can be parametrized by four (unit) 4-vectors: two 4-vectors $v_0$ and $d_0$ that span the leading edge of the wedge and two 4-vectors $e_1$ and $e_2$ perpendicular to $v_0$ and $d_0$ that give the directions of the sides of the wedge. For the wedge removed from a stationary defect, $v_0= (1,0,0,0)$ and $d_0$, $e_1$, and $e_2$ are chosen to lie in the $t=0$ hyperplane.

\begin{figure}[tbp]
\centering\includegraphics[width=7cm]{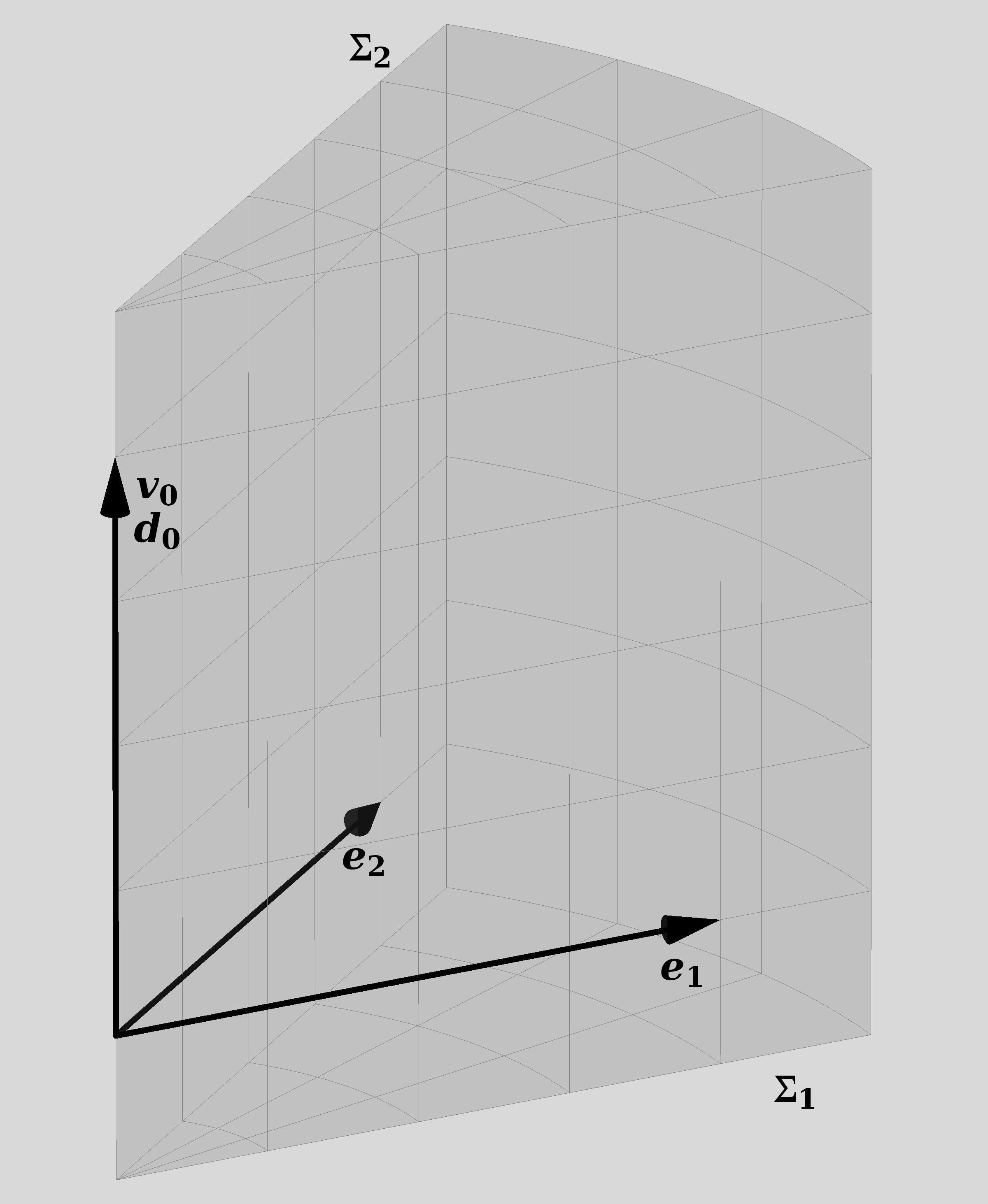}
\caption{A (3+1)-dimensional wedge can be parametrized by four (unit) 4-vectors: two 4-vectors $v_0$ and $d_0$ that span the leading edge of the wedge and two 4-vectors $e_1$ and $e_2$ perpendicular to $v_0$ and $d_0$ that give the directions of the sides of the wedge.}\label{fig:wedgeparam}
\end{figure}

The sides of the wedge are 3-dimensional half-hyperplanes $\Sigma_i$ spanned by $v_0$, $d_0$, and $e_i$,
\begin{equation}
\begin{aligned}
\Sigma_1 &= \set{\sigma_1(\tau,\lambda,\mu)= \tau v_0 +\lambda d_0 +\mu e_1 | \tau,\lambda\in\RR, \mu\in\RR_{\geq 0}}\\
\Sigma_2 &= \set{\sigma_2(\tau,\lambda,\mu)= \tau v_0 +\lambda d_0 +\mu e_2 | \tau,\lambda\in\RR, \mu\in\RR_{\geq 0}}.
\end{aligned}
\end{equation}
The conical curvature defect is created by identifying the sides $\Sigma_1$ and $\Sigma_2$ through the identification
\begin{equation}\label{eq:wedgeidentify}
\sigma_1(\tau,\lambda,\mu) \sim\sigma_2(\tau,\lambda,\mu)
\end{equation}
for all $\tau$, $\lambda$, and $\mu$.
We obtain the construction for a moving defect by applying a Lorentz boost $B$ to this whole set-up. That is, we create a new wedge parametrized by the vectors
\begin{equation}
\begin{aligned}
v'_0 &= B\cdot v_0\\
d'_0 &= B\cdot d_0\\
e'_1 &= B\cdot e_1\\
e'_2 &= B\cdot e_2,
\end{aligned}
\end{equation}
and identify the edges,
\begin{equation}
\begin{aligned}
\Sigma'_1 &= \set{\sigma'_1(\tau,\lambda,\mu)= \tau v'_0 +\lambda d'_0 +\mu e'_1 | \tau,\lambda\in\RR, \mu\in\RR_{\geq 0}}\\
\Sigma'_2 &= \set{\sigma'_2(\tau,\lambda,\mu)= \tau v'_0 +\lambda d'_0 +\mu e'_2 | \tau,\lambda\in\RR, \mu\in\RR_{\geq 0}},
\end{aligned}
\end{equation}
through the identification
\begin{equation}\label{eq:wedgeidentify2}
\sigma'_1(\tau,\lambda,\mu) \sim \sigma'_2(\tau,\lambda,\mu)
\end{equation}
for all $\tau$, $\lambda$, and $\mu$.

The first thing we note is that, if the boost $B$ is in the $d_0$ direction, then $e'_i = B\cdot e_i =e_i$ and this procedure is simply a reparametrization of the hyperplanes $\Sigma_1$ and $\Sigma_2$. This has no effect on the identification \eqref{eq:wedgeidentify}. We thus see that conical defects are invariant under boosts in the direction of the defect line.

This allows us to restrict our attention to boosts in the $e_1e_2$-plane. Let $B_\eta$ be the boost with velocity $\eta\in\mathrm{span}(e_1,e_2)$. Then  $B_\eta$ leaves $d_0$ invariant, while 
\begin{equation}
\begin{alignedat}{2}
v'_0 &= B_\eta\cdot v_0 &&= \gamma (v_0 + \eta),\\
e'_i &= B_\eta\cdot e_i &&= {e_i}_{\perp\eta} + \gamma {e_i}_{\parallel\eta} + \gamma(e_i\cdot\eta)v_0,
\end{alignedat}
\end{equation}
where $\gamma$ is the Lorentz factor of the boost, $(1-\abs{\eta})^{-1/2}$ and ${e_i}_{\perp\eta}$ and ${e_i}_{\parallel\eta}$ denote the parts of $e_i$ that are respectively perpendicular and parallel to $\eta$. In general, the identification \eqref{eq:wedgeidentify2} will identify points  that do not lie in the same $t' = constant$ plane. More specifically, the difference in $t'$ is given by
\begin{equation}
\begin{aligned}
\Delta t'(\tau,\lambda,\mu) 
&= \bhh{\sigma'_2(\tau,\lambda,\mu)-\sigma'_1(\tau,\lambda,\mu)}\cdot v_0\\
&= \mu(e'_2-e'_1)\cdot v_0\\
&= \mu\gamma(e_1\cdot\eta-e_2\cdot\eta).
\end{aligned}
\end{equation}
Therefore, the identification \eqref{eq:wedgeidentify2} will only identify points on the same time slice if $e_1\cdot\eta=e_2\cdot\eta$ or equivalently if $\eta \propto e_1+e_2$.  That is, if $\eta$ is directed in the symmetry plane of the wedge.

Recall, that earlier we observed that the direction of the wedge was irrelevant to the produced geometry. We are therefore free to always choose the wedge to be cut along the movement direction of the defect. This choice has the advantage that we know that  \eqref{eq:wedgeidentify2} only identifies points on the same time slice. This precludes the creation of closed timelike curves by the identification, and allows us to analyse the generated geometry on a single time slice.

In fact, the identification \eqref{eq:wedgeidentify2} also produces zero shift in the $d_0 =d'_0$ direction, so we can analyse the new geometry completely in the $e_1e_2$-plane. This plane is defined by the equations
\begin{equation}
\begin{aligned}
 v_0\cdot x &=0,& d_0\cdot x &=0.
\end{aligned}
\end{equation}
Let
\begin{equation}
\bar\Sigma'_i = \set{\bar\sigma'_i = \mu \bar{e}'_i|\mu\in\RR_{\geq 0}}
\end{equation}
be the intersection of $\Sigma'_i$ and the $e_1e_2$-plane. We can then find $\bar{e}'_i$ by solving
\begin{equation}
\begin{aligned}
 v_0\cdot \sigma'_i(\tau,\lambda,\mu) &=0,& d_0\cdot \sigma'_i(\tau,\lambda,\mu) &=0
\end{aligned}
\end{equation}
for $\tau$ and $\lambda$. The result after normalizing is
\begin{equation}
\bar{e}'_i = \frac{e'_i -(e_i\cdot\eta)v'_0}{\sqrt{1-(e_i\cdot\eta)^2}}.
\end{equation}
The effective angle of the boosted wedge in the $e_1e_2$-plane, $\alpha'$, is then defined through
\begin{equation}
\begin{aligned}
\cos\frac{\alpha'}{2} &\equiv \frac{\bar{e}'_i\cdot\eta}{\abs{\eta}}\\
&=\frac{e_i\cdot\eta}{\gamma\abs{\eta}\sqrt{1-(e_i\cdot\eta)^2}}.
\end{aligned}
\end{equation}
If we observe that $e_i\cdot\eta=\abs{\eta}\cos\tfrac{\alpha}{2}$, we can apply some elementary trigonometry to produce a relation between the angle of the stationary wedge $\alpha$, and the effective angle $\alpha'$ of the boosted wedge in the $e_1e_2$-plane,
\begin{equation}
\tan\frac{\alpha'}{2} = \gamma\tan\frac{\alpha}{2}.
\end{equation}
The Lorentz contraction of the boost causes the effective angle of the wedge to open up. As a result moving defects have larger effective deficit angles than stationary ones with the same mass. 

\begin{figure}[tbp]
\centering\includegraphics[width=\textwidth]{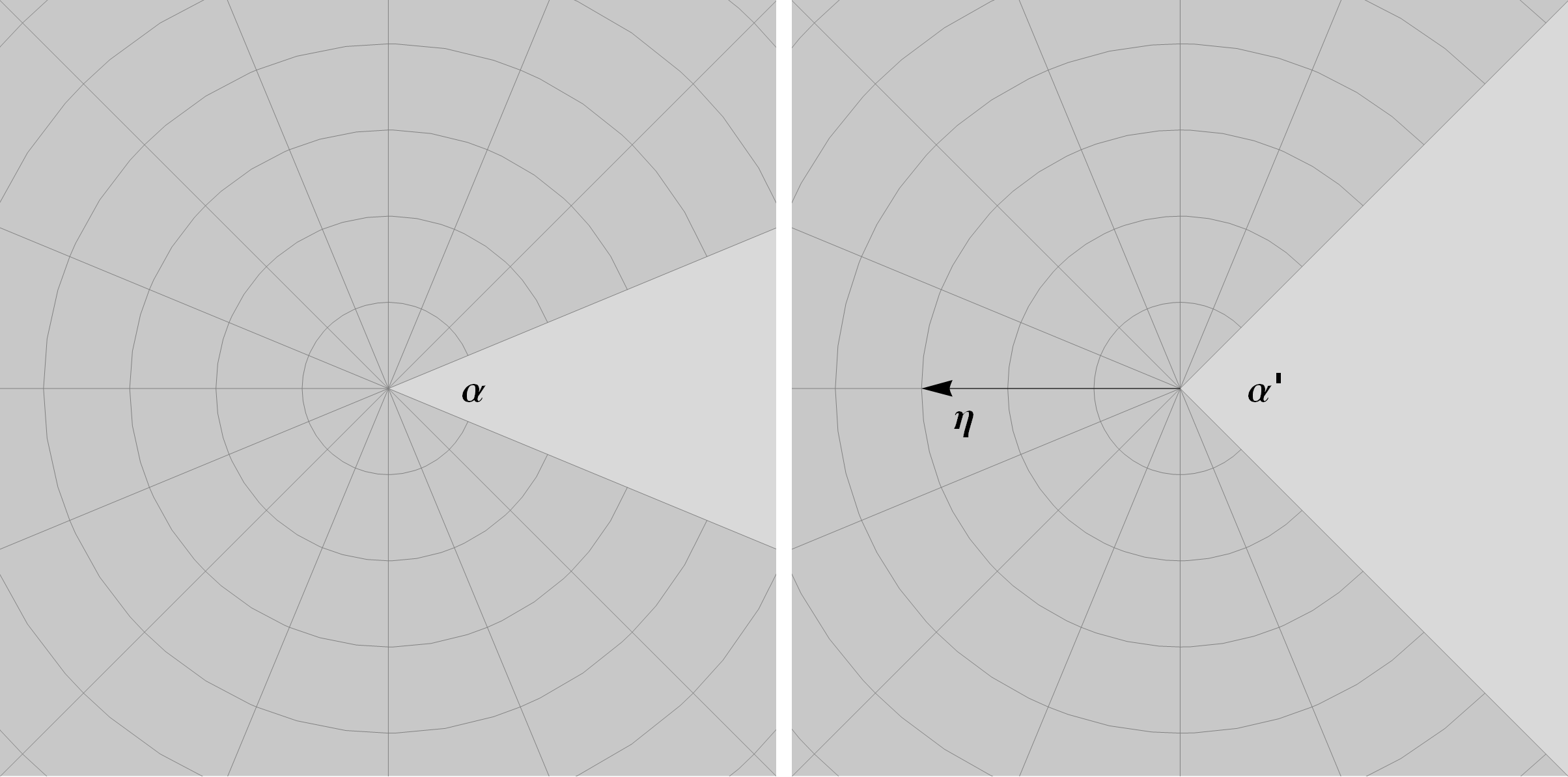}
\caption{On the left: the $e_1e_2$-plane before a boost $\eta$. On the right:the $e_1e_2$-plane after a boost $\eta$.}\label{fig:boostedwedge}
\end{figure}

\section{Parametrizing a defect}\label{sec:paramdefect}
To describe a general defect we need the codimension 2 subspace of the defect, and the deficit angle in its rest frame $\alpha$. A codimension 2 subspace in $d$-dimensional Minkowski space is described by its normal 2-form $\omega_{\mu\nu}$ and its displacement $X^\mu$. Together these define the subspace through the equation
\begin{equation}\label{eq:norm2form}
\omega_{\mu\nu}(x^\mu-X^\mu)=0.
\end{equation}
The normal 2-form is the wedge product of two perpendicular normal vectors. In $d+1$ spacetime dimensions this gives us  $2d-2$ free parameters. Equation  \eqref{eq:norm2form} is invariant under a shift $X^\mu\mapsto X^\mu+ X'^\mu$, with $\omega_{\mu\nu}X'^\mu=0$. The displacement $X^\mu$ therefore contributes another $(d+1)-(d-1)=2$ independent parameters. We therefore need a total of
\begin{equation}
(2d-2) + 2 +1 = 2d+1
\end{equation}
independent  parameters to describe a codimension 2 conical defect. In our case, $d=3$, this is 7 parameters.

The condition that the defect is timelike is captured by requiring that both normals to the codimension 2 subspace are spacelike. In terms of the normal 2-form this means that 
\begin{equation}
\omega_{\mu\nu}\omega^{\mu\nu}=1.
\end{equation}

In our $(3+1)$-dimensional case it is often useful to parametrize the spacetime orientation of the defect by a timelike and a spacelike 4-vector $u^\mu$ and $d^\mu$, instead of the normal 2-form $\omega_{\mu\nu}$. If we choose these such that $u^0 =1$, $d^0=0$ $u^\mu d_\mu=0$, then we can write
\begin{equation}
\begin{aligned}
u^\mu = (1,\vec{v}),\quad
d^\mu = (0,\hat{d}),
\end{aligned}
\end{equation}
 where $\vec{v}$ is the velocity of the defect with respect to the equal time slices and $\hat{d}$ is the direction of the defect line. Moreover, we can use the translation freedom in the displacement such that $X^0 =0$, and write $X^\mu = (0,\vec{X})$ with $\vec{X}$ the position of the defect at $t=0$.

We can therefore describe a defect with mass $\alpha$ by a triple of 3-dimensional vectors $(\vec{X},\hat{d},\vec{v})$ which satisfy
\begin{equation}
\begin{aligned}
\vec{v} &\perp \hat{d}, \\
\vec{X} &\perp \hat{d}, \text{and}\\
\norm{\hat{d}}&=1.
\end{aligned}
\end{equation}
\subsection{Holonomy}
A particularly useful way of characterizing a defect is by the holonomy of a loop around it. The holonomy of a loop is found by parallel transporting a vector around the loop and comparing it to the original. The relation between the original and parallel transported vector will be given by a Lorentz transformation, $Q_{\phantom{\mu}\nu}^\mu$, called the \emph{holonomy}. For a space $M$ with an arbitrary metric the holonomy of a loop $\gamma\colon S^1\to M$ can be calculated using the following formula (see \cite{Carroll:2004AppI} for a derivation),
\begin{equation}
Q_{\phantom{\mu}\nu}^\mu =\pathord \exp\hh{-\int_\gamma \Gamma^\mu_{\sigma\nu} \frac{\md \gamma^\sigma}{\md \lambda} \md \lambda},
\end{equation}
where $\pathord$ indicates that the matrix exponential is path ordered. If the space is flat, then the holonomy does not depend on the details of the path taken. In particular, for any loop that can be contracted to a point without meeting a curvature defect the holonomy is the identity map.

For our piecewise flat model this means that the holonomy of a loop around a defect only detects the number of times the loop wraps around the defect. This means that we can talk about \emph{the} holonomy of a defect, meaning the holonomy of a loop wrapping the defect exactly once.\footnote{There is still an ambiguity here regarding the sign of the wrapping number, corresponding to the direction in which the loop wraps the defect.}  

In equation \eqref{eq:holstatic} we already calculated that the holonomy of a static defect was given by a rotation about the axis of the defect line. The angle of this rotation was found to be equal to the linear mass density of the defect. This holonomy coincides with the map that identifies the two sides of the removed wedge. This is true in general. For each wedge of spacetime there is a unique Lorentz transformation, $Q$,  that maps $\Sigma_1$ to $\Sigma_2$ in such a way that
\begin{equation}
Q\cdot\sigma_1(\tau,\lambda,\mu) = \sigma_2(\tau,\lambda,\mu).
\end{equation}
As this map is used to identify two points on the sides of the wedge, its pullback to the tangent bundle is used to identify the tangent spaces at those points. Therefore, a vector that is parallel transported across the identified sides of the wedge is transformed by this pulled back Lorentz transformation, which may be identified with the original identification since the tangent space of Minkowski space is isomorphic to Minkowski space. Since the rest of the loop is in flat space no further transformation with respect to the background frame occurs in parallel transport. Hence, the holonomy of a loop around a conical defect is equal to the Lorentz transformation associated with the removed wedge.

So what is the holonomy of a moving defect? A defect moving at a subluminal speed can always be transformed into a stationary defect by a boost $B$. As we already observed, a stationary defect has a pure rotation as its holonomy. So, if a moving defect is constructed by identifying the sides $\Sigma_1$ and $\Sigma_2$ of a wedge, then applying $B$ gives the sides $\Sigma'_i = B\cdot\Sigma_i$ of a stationary wedge. The sides $\Sigma'_i$ are identified by a rotation $R$ such that
\begin{equation}
R\cdot\sigma'_1(\tau,\lambda,\mu) = \sigma'_2(\tau,\lambda,\mu).
\end{equation}
This implies that
\begin{equation}
R\cdot B\cdot\sigma_1(\tau,\lambda,\mu) = B\cdot\sigma_2(\tau,\lambda,\mu).
\end{equation}
Consequently, the Lorentz transformation $Q$ that identifies the sides $\Sigma_1$ and $\Sigma_2$ is given by
\begin{equation}
Q= B^{\m 1}\cdot R\cdot B.
\end{equation}
The holonomy of an arbitrary moving defect therefore is a Lorentz transformation that is conjugate to a pure rotation. Such Lorentz transformations are called \emph{rotationlike}.\footnote{Similarly, Lorentz transformations that are conjugate to a pure boost are called \emph{boostlike}.} The angle of the pure rotation is the linear mass density of the defect in its rest frame, and can therefore be associated with the rest mass of the moving defect.

Conversely, given a rotationlike Lorentz transformation $Q$ we can construct a moving defect with holonomy $Q$. Like a pure rotation, a rotationlike Lorentz transformation has two eigenvectors with eigenvalue 1, which span a timelike surface. This surface will be the leading edge of the wedge needed in the construction of the defect. We can now pick an arbitrary spacelike vector $e_1$ perpendicular to this surface and construct $\Sigma_1$ as the span of the leading edge and $e_1$. We define $e_2$ as $e_2 = Q\cdot e_1$. Since $Q$ is an orthogonal transformation and $Q$ leaves the leading edge invariant, $e_2$ will also be perpendicular to the leading edge, and we can construct $\Sigma_2$ as the span of the leading edge and $e_2$. This defines a wedge whose identifying map, by construction, is $Q$. Therefore, if we remove this wedge from Minkowski space we obtain a defect with holonomy $Q$.

The holonomy $Q$ therefore completely encodes all the information about the mass and the movement of a line defect. Following the example of point defects in $2+1$ dimensions,\cite{Matschull:1997du} we could interpret $Q$ as the covariant 4-momentum of the line defect.

Since we are in a locally flat background, we can go a step further than just calculating the parallel transport of vectors. The flatness of the spacetime allows us to use the exponential map to expand each frame of the tangent bundle to a complete coordinate frame that is isometric to the Minkowski frame. Parallel transporting a frame around a loop, can therefore be interpreted as transporting the entire coordinate system around the loop. Consequently, the coordinate system obtained after parallel transport around  a loop, will be related to the original coordinate system by an isometry of the Minkowski frame, a Poincaré transformation. This transformation is called the \emph{Poincaré holonomy} of the loop.  Like the ordinary holonomy it does not depend on the local details of the loop and can be interpreted as a property of the defect.

If we use a coordinate system in which the studied defect passes through the origin of the coordinate system, then the Poincaré holonomy will just give the ordinary holonomy. A defect that passes through any other point may be constructed by applying a shift to the whole system. Going through the same motions as we did for determining the holonomy of a moving defect, we conclude that a moving defect with an arbitrary position produces a rotationlike Poincaré holonomy, i.e. a Poincaré holonomy that is conjugate to a pure rotation by a Poincaré transformation. Conversely, given a rotationlike Poincaré transformation $P$, we can construct a defect with Poincaré holonomy $P$.

The Poincaré holonomy of a loop around a defect depends on the coordinate frame chosen at the initial point of the loop. When choosing a different coordinate frame, the holonomy of the loop is transformed by conjugating with the Poincaré transformation associated to the change of frame. The only truly frame independent property of a defect, therefore is the conjugacy class of its holonomy. For rotationlike Poincaré transformations, the conjugacy classes can be distinguished by the angle of the rotation in the rest frame of the transformation (i.e. the frame where the transformation becomes a pure rotation). We already saw that this angle was proportional to the linear mass density of the corresponding defect in its rest frame. The invariant of a defect given by the conjugacy class of its holonomy can therefore be associated with its rest mass.

There is a subtlety in this association. The rotation angle of a rotationlike Poincaré transformation can only distinguish deficit angles (and therefore mass densities) modulo $2\pi$. This is not a problem if we only allow  defects with a positive deficit angle, since a deficit angle of $2\pi$ corresponds to a spacetime with no volume. We, however, also wish to allow defects with a negative deficit (or surplus) angle, which in principle are not bounded by any value.

The Poincaré holonomy of a defect therefore contains (nearly) all the information needed to identify a defect. At the start of this section we saw that for complete description of a conical defect we needed 7 independent parameters, one for the deficit angle and six to describe an arbitrary codimension 2 subspace. We should be able to reproduce these from the holonomy. The space of rotationlike Poincaré transformations is indeed a 7 dimensional subspace of the 10 dimensional  Poincaré group. So, a rotationlike holonomy has the right number of parameters.

We just saw that the deficit angle can be obtained as the conjugacy class of the holonomy. The codimension 2 subspace, $S$, forming the defect is obtained as the set of fixed points of the Poincaré holonomy $P$,
\begin{equation}
S=\set{x\in\RR^{3,1}| P(x) = x}.
\end{equation}
More specifically, the Poincaré group can be interpreted as the semi-direct product of the Lorentz group with the abelian group of spacetime translations. As such, each Poincaré transformation may be identified with a Lorentz transformation $Q$ and a spacetime translation $T$. The Lorentz transformation $Q$ associated to the Poincaré holonomy $P$ captures the spacetime orientation of the defect, while the displacement is captured by the associated spacetime translation $T$. 

\section{More general defects}\label{sec:generaldefects}
In the preceding section we have seen that there is a close relation between defects and the Poincaré holonomy of a loop around it. Thus far we have considered only defects with a rotationlike holonomy. One may wonder if it is possible to construct defects with more general holonomies. The answer is yes, but not all of these defects are physical.

In the previous sections we constructed a conical defect by removing a wedge from a flat spacetime and identifying the opposing sides. This can be generalized in the following way. Take two hyperplanes $\Sigma_1$ and $\Sigma_2$ intersecting on a codimension 2 surface $E$, remove one of the wedges of spacetime bounded by $\Sigma_1$ and $\Sigma_2$, and identify $\Sigma_1$ and $\Sigma_2$ using a Poincaré transformation $P$ that maps $\Sigma_1$ to $\Sigma_2$. This requires that the leading edge of the wedge $E$ is invariant under $P$. Moreover, if the resulting geometry is to be a regular topological manifold along $E$, then $P$ restricted to $E$ is to be the identity map. That is, the leading edge of the wedge $E$ must consist of fixed points of $P$. This requirement severely restricts the possible transformations $P$.

Without loss of generality we may assume that $E$ passes through the origin, otherwise we may apply a shift to the whole system such that it does. The requirement that the points in $E$ are fixed under $P$ implies that $P$ has no translational part. Moreover, the Lorentz part of the holonomy, $Q$, must have $E$ as an eigenspace with eigenvalue 1.

If $E$ is timelike, as in the case of a timelike conical defect, then $Q$ must be a rotationlike transformation. There are no other timelike defects that are the fixed point of their own holonomy, than the conical defects that we have already discussed. In the case that $E$ is spacelike $Q$ must be a boostlike transformation.  This corresponds to a defect that is completely spacelike, which we do not allow since it would correspond to a non-local degree of freedom.

The final possibility is that $E$ is lightlike, in which case $Q$ is a null rotation. The resulting defect can be interpreted as a line defect propagating at the speed of light. There is no good physical reason to disallow such objects, and in fact they will prove crucial in the construction of gravitational wave solutions in section \ref{ch:gravwave}. These defects will be discussed in the next section (\ref{sec:masslessdefects}).

Not imposing the requirement that $P$ is the identity when restricted to $E$ implies that points in $E$ will get identified with other points of $E$. The result generically is a space that is not a topological manifold along the points of $E$. For us this is enough reason not to consider such defects in our model. But let us briefly comment on what such defects would represent physically.

Lets assume that $E$ is timelike. Without loss of generality we may then take $E$ to be the $tz$-plane. Even if the points of $E$ are not fixed under $P$, $E$ must still be an invariant subspace of $P$. This implies that $P$ restricted to $E$ is a Poincaré transformation of the $tz$-plane; i.e. it is a combination of a boost in the $z$-direction  and shifts in the $t$ and $z$ directions.

If $P$ is a pure shift in the $tz$-plane, it is possible to write down a locally flat metric for the resulting spacetime, 
\begin{equation}\label{eq:stationarydefectmetric}
\md s^2 = -(\md t+\beta \md\theta)^2 +\md r^2 + r^2\md \theta^2 +(\md z+\gamma\md\phi)^2,
\end{equation}
with $\beta$ the shift in the $t$-direction and $\gamma$ the shift in the $z$-direction. Smearing out the defect like we did for ordinary conical defects in section \ref{sec:statdefect} reveals that $\beta$ can be associated with a rotation of the defect, while $\gamma$ is associated to a constant torsion exerted on the defect.

If $\beta>\gamma$, then the loops with constant $t$, $r$, and $z$ become timelike for small $r$. That is, rotating infinitely thin defects generate closed timelike curves, just like rotating point particles in 2+1 dimensions. This can actually be seen directly from the action of the shift on $E$. If the shift is timelike, $E$ becomes compactified in a timelike direction. Similarly, if we consider a $P$ that reduces to a pure boost on $E$, then this boost will identify points on $E$ with a timelike separation. Consequently, there will be closed timelike curves on $E$. By continuity these can be deformed to closed timelike curves in a neighbourhood of $E$ in the quotient manifold.  

We conclude that allowing defects with a holonomy that includes a boost or a timelike shift would introduce unacceptable acausal  features in the model. Even though they would allow the introduction of physically interesting concepts like intrinsic spin to the model. Defects with a spacelike shift in the holonomy seem less harmful, yet they do introduce a weird topology in the neighbourhood of the defect, which seems difficult to reconcile with the interpretation of the defect as a line defect.

\subsection{Massless defects}\label{sec:masslessdefects}
There is no physical reason not to allow defects, where the 2-dimensional surface of the defect is lightlike. Since the defect surface consists of the fixed points of its holonomy, this means that the (Lorentz part of the) holonomy should be a Lorentz transformation that has a lightlike and an orthogonal spacelike eigenvector with eigenvalue one. If in Cartesian coordinates  these are chosen to be $(1,1,0,0)$ and $(0,0,0,1)$, then the most general orientation preserving orthogonal transformation that leaves these vectors invariant is
\begin{equation}\label{eq:nullrotation}
Q^\mu_{\phantom{\mu}\nu} = \begin{pmatrix}
1 + \frac{\alpha^2}{2}	& -\frac{\alpha^2}{2}	& \alpha	& 0 \\
\frac{\alpha^2}{2}		& 1 - \frac{\alpha^2}{2}	& \alpha	& 0 \\
\alpha					& -\alpha				& 1		& 0 \\
0						& 0						& 0		& 1 
\end{pmatrix},
\end{equation}
where $\alpha\in\RR$ is a free parameter. Lorentz transformations of this type are called \emph{null rotations} or \emph{parabolic Lorentz transformations}, and $\alpha$ is called the parabolic angle of the transformation.

All null rotations belong to the same conjugacy class of the Lorentz group. They do not form a closed subgroup of the Lorentz group. In fact, the smallest subgroup of the Lorentz group that contains all null rotations is the Lorentz group itself. The largest subgroups that can be formed from just null rotations are two dimensional abelian groups isomorphic to $\CC$ generated by two commuting nilpotent elements of $\So{3,1}$. 

In general a null rotation can be specified by giving a null vector, $u^\mu$, a spacelike vector, $d^\mu$ perpendicular to $u^\mu$ and one scalar parameter $\alpha$ called the parabolic angle. Since the scaling of $u^\mu$ and $d^\mu$ does not matter this gives 4 independent real parameters. The subset of null rotations does, however, not form a manifold due to a nodal singularity at the identity element. Much like that the lightcone is not a proper submanifold of Minkoswki space.

The subset of null rotations forms the boundary of the subset of rotationlike  transformations in the Lorentz group. This means that any null rotation can be viewed as the limit of a sequence of rotationlike Lorentz transformations. For example the null rotation \eqref{eq:nullrotation} can be viewed as the limit of a sequence of rotations about the $z$-axis boosted in the $x$ direction. If $R_z(\phi)$ is a rotation of $\phi$ about the $z$-axis and $B_x(\eta)$ is a boost in the $x$-direction with rapidity $\eta$, then 
\begin{multline}\label{eq:boostedrotation}
B_x(\eta)\cdot R_z(\phi)\cdot B_x(\m\eta)=\\
\begin{pmatrix}
\cosh^2\eta -\cos\phi\sinh^2\eta
	& \frac{\cos\phi -1}{2}\sinh 2\eta
		& \sin\phi \sinh\eta
			& 0\\
\frac{1-\cos\phi}{2}\sinh 2\eta
	& \cos\phi\cosh^2\eta -\sinh^2\eta
		& \sin\phi \cosh\eta
			& 0\\
\sin\phi \sinh\eta
	& \m\sin\phi \cosh\eta
		& \cos\phi 
			& 0\\ 			 
0
	& 0
		& 0
			& 1\\ 
\end{pmatrix}
\end{multline}
is a rotation about the $z$-axis boosted in the $x$-direction. For any constant value of $\phi$ (not equal to an integer multiple of $2\pi$) the components of this transformation diverge as $\eta$ approaches infinity. To approach a finite limit, $\phi$ must approach zero as $\eta$ goes to infinity. If we write $\phi = \tfrac{\alpha}{\cosh\eta}$, then using that
\begin{align}
\cos(\frac{\alpha}{\cosh\eta}) &= 1 + \frac{\alpha^2}{\cosh^2\eta} + \bigO(\ee^{-4\eta}),\\
\sin(\frac{\alpha}{\cosh\eta}) &= \frac{\alpha}{\cosh\eta} + \bigO(\ee^{-3\eta}),
\end{align}
we find that
\begin{equation}
\lim_{\eta\to\infty} B_x(\eta)\cdot R_z(\frac{\alpha}{\cosh\eta})\cdot B_x(\m\eta) =\begin{pmatrix}
1 + \frac{\alpha^2}{2}	& -\frac{\alpha^2}{2}	& \alpha	& 0 \\
\frac{\alpha^2}{2}		& 1 - \frac{\alpha^2}{2}	& \alpha	& 0 \\
\alpha					& -\alpha				& 1		& 0 \\
0						& 0						& 0		& 1 
\end{pmatrix}.
\end{equation}

This limiting procedure is analogous to applying an Aichelburg-Sexl ultraboost \cite{Aichelburg:1970dh} to a line defect. The rotationlike transformation \eqref{eq:boostedrotation} corresponds to a line defect oriented in the $z$-direction with rest mass density $\phi$ and moving in the $x$-direction with rapidity $\eta$. As $\eta$ increases the kinetic energy of the line defect diverges. The choice $\phi = \tfrac{\alpha}{\cosh\eta}$ precisely ensures that the energy density of the line defect stays constant as $\eta$ increases. We can therefore interpret lightlike defects as massless conical defects moving at the speed of light.

The Aichelburg-Sexl ultraboost of a line defect was studied by Barrabes et al in.\cite{Barrabes:2002hn} They found the metric corresponding to the holonomy \eqref{eq:nullrotation} to be,
\begin{equation}\label{eq:gmasslessdefect}
\md s^2 = -\md t^2 + \md x^2 + \md y^2 +\md z^2 - 2\alpha \abs{y}\delta(x-t)(\md x -\md t)^2. 
\end{equation}

The corresponding energy--momentum tensor can be found by taking the energy momentum tensor for a stationary defect as calculated in equation \eqref{eq:Tstationarystring},
\begin{equation}
T_{\mu\nu} = \begin{pmatrix}
\alpha & 0 & 0 & 0\\
0 & 0 & 0 & 0\\
0 & 0 & 0 & 0\\
0 & 0 & 0 & \m\alpha
\end{pmatrix} \delta(x)\delta(y),
\end{equation}
and boosting that in the $x$-direction. We then find that the energy--momentum tensor corresponding to the holonomy \eqref{eq:boostedrotation} is
\begin{equation}
T_{\mu\nu}(\eta) = \begin{pmatrix}
\phi \cosh\eta & \m\phi\sinh\eta & 0 & 0\\
\m\phi\sinh\eta & \phi \cosh\eta & 0 & 0\\
0 & 0 & 0 & 0\\
0 & 0 & 0 & \m\frac{\phi}{\cosh\eta}
\end{pmatrix} \delta(x-\tanh(\eta) t)\delta(y).
\end{equation}
The choice $\phi = \tfrac{\alpha}{\cosh\eta}$ keeps the $T_{00}$ component constant. The limit as $\eta$ goes to infinity is given by
\begin{equation}
\lim_{\eta\to\infty}T_{\mu\nu}(\eta) = \begin{pmatrix}
\alpha & \m\alpha & 0 & 0\\
\m\alpha & \alpha & 0 & 0\\
0 & 0 & 0 & 0\\
0 & 0 & 0 & 0
\end{pmatrix} \delta(x-t)\delta(y).
\end{equation}
Alternatively, we could have started directly from the metric \eqref{eq:gmasslessdefect}, and calculated the Einstein tensor directly. This was done in \cite{Barrabes:2002hn} and gives the same result.

Note that in the massless limit the tension in the spacelike direction of the defect vanishes. This means that the energy-momentum tensor at a single point of the defect does not have any information on the direction of the defect. This property of massless defects will be very important when we construct gravitational waves in chapter \ref{ch:gravwave}. Because the energy--momentum tensor of massless strings is the same for all defects moving in the same direction, we can get the energy--momentum for perpendicular positive and negative energy defects to cancel each other. 

\section{Multiple defects}\label{sec:multipledefects}
\begin{figure}[tbp]
\centering\includegraphics[width=120mm]{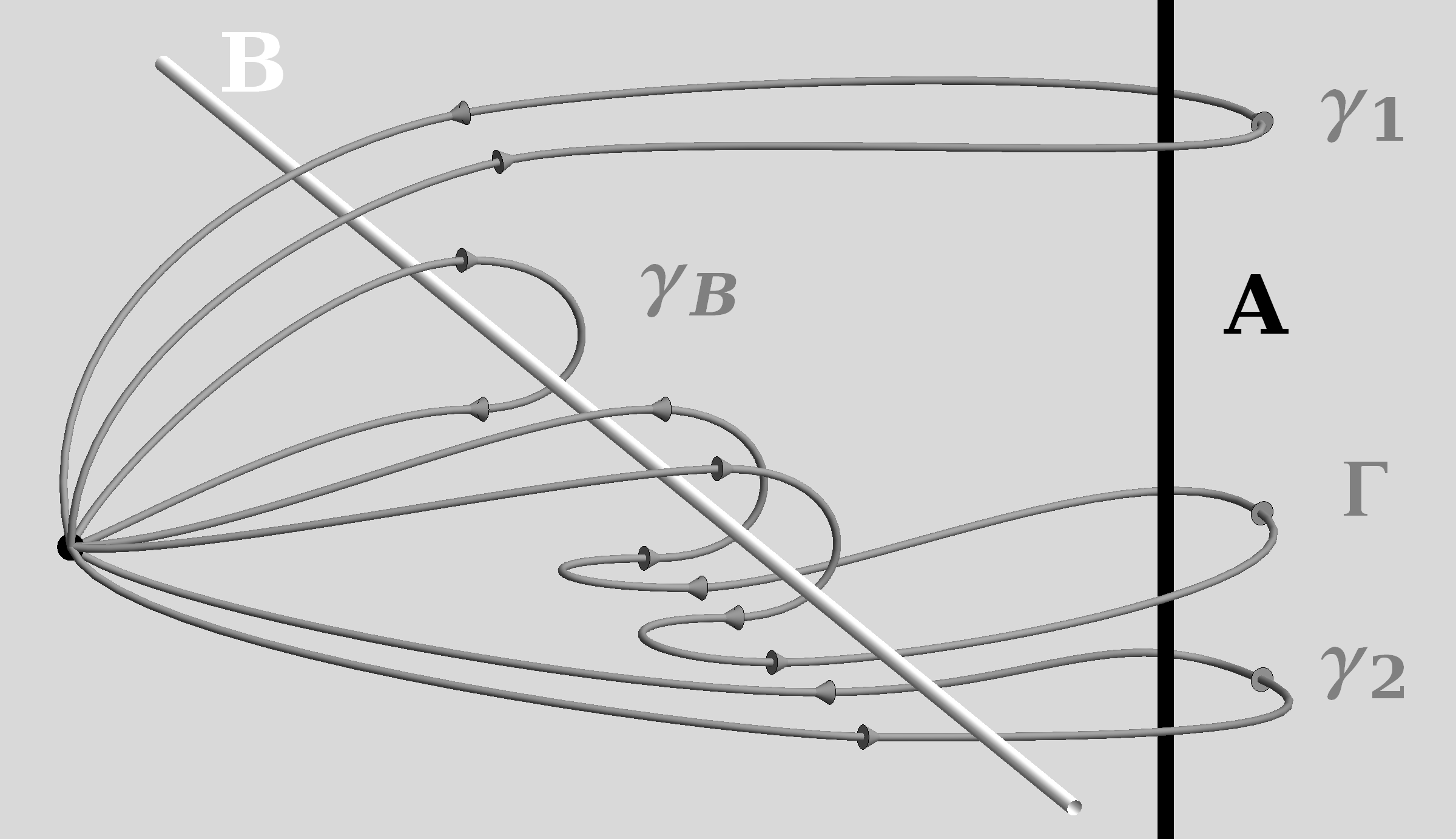}
\caption{The path $\gamma_1$ and path $\gamma_2$ around the defect line $A$ are topologically inequivalent. The path $\Gamma$ shows that $\gamma_1$ can be deformed to the sequence of paths $\gamma_B$ then $\gamma_2$ then $\gamma_B^{-1}$.}\label{fig:holoframes}
\end{figure}

Geometries with multiple defects can be generated by removing multiple wedges from Minkowski spacetime. This introduces a new subtlety when describing the defects in terms of their holonomies. In a spacetime with multiple defects, there are multiple topologically distinct loops around a defect line. Consider the situation depicted in figure \ref{fig:holoframes} (for visual clarity the removed wedges have been suppressed). There are two topologically  distinct loops from the point $p$ around the defect line labelled $A$. The loop can either pass above the line labelled $B$, as the path labelled $\gamma_1$ does, or it can pass below the defect $B$ like loop $\gamma_2$. The loop $\gamma_2$ cannot be deformed into loop $\gamma_1$ without crossing defect $B$. Since the holonomy of a loop changes when a path is deformed across a defect, the holonomy of path $\gamma_2$ is different from the holonomy of $\gamma_1$.

We can however relate the holonomies of the paths $\gamma_1$ and $\gamma_2$, by observing that the loop $\Gamma$ formed by first taking the loop $\gamma_B$ around defect $B$, then taking loop $\gamma_2$ around $A$, and finally taking $\gamma_B$ around $B$ in the opposite direction, can be continuously deformed to path $\gamma_1$ without crossing any of the defects. Consequently, the holonomy of $\Gamma$ is equal to the holonomy of $\gamma_1$. Since the holonomy of $\Gamma$ is the product of the holonomies of the individual loops, we conclude that if we denote the holomies of the individual loops  $Q_{\gamma_1}$, $Q_{\gamma_2}$, and $Q_{\gamma_B}$, then they are related as through,
\begin{equation}
\begin{aligned}
Q_{\gamma_1} &= Q_{\Gamma}\\
&=Q_{\gamma_B^{-1}}Q_{\gamma_2}Q_{\gamma_B}\\
&=Q_{\gamma_B}^{-1}Q_{\gamma_2}Q_{\gamma_B},
\end{aligned} 
\end{equation}
where in the last line we used that the holonomy of a loop followed backwards is the inverse of the holonomy of the loop. In particular, we see that $Q_{\gamma_1}$ and $Q_{\gamma_2}$ belong to the same conjugacy class. Therefore if either one is rotationlike, so is the other.

This relation can be expressed in formal way for a general configuration. Let $X$ be a spacetime containing an arbitrary configuration of defects, let $X^{(2)}\subset X$ by the 2-dimensional subset of all defects, and let $p\in X\setminus X^{(2)}$ be a point not on any of the defects. Two loops in $X$ starting at $p$ can be continuously deformed into each other without crossing any of the defects if and only if they are homotopic in $X\setminus X^{(2)}$. The set of classes of topologically equivalent loops together with the operation of concatenating loops therefore gives the fundamental group $\pi_1\left[X\setminus X^{(2)},p\right]$. Since there is no curvature on $X\setminus X^{(2)}$ all homotopic loops have the same holonomy. We therefore have a map 
\begin{equation}
 \map{Q}{\pi_1\left[X\setminus X^{(2)},p\right]}{\ISO{T_p X}},
\end{equation}
that assigns to each homotopy class of loops $\gamma$ its holonomy $Q_\gamma$, an element of the Poincaré group at the point $p$; $\ISO{T_p X}$. The consistency requirement of the example above generalizes to the requirement that the assignment $Q$ is a group homomorphism. That is, for any two (equivalence classes of) loops $\gamma_1$ and $\gamma_2$,
\begin{equation}\label{eq:holonomygroupresp}
Q_{\gamma_1 \cdot \gamma_2} = Q_{\gamma_1}Q_{\gamma_2}.
\end{equation}

The holonomy of each simple loop (i.e. a loop enclosing a single defect once) must  be rotationlike in order for all the defects to be timelike. Equation \eqref{eq:holonomygroupresp} tells us that the holonomy of a loop enclosing multiple defects is the product of the holonomies of simple loops around the individual defects. Consequently, because the subspace of rotationlike Poincaré transformations is not closed under multiplication, the holonomy of a general loop will not be rotationlike. In fact, rotationlike Poincaré transformations generate the whole Poincaré group (in the sense that the smallest subgroup of the Poincaré group containing all rotationlike transformations is the Poincaré group itself). This means that a loop taken around a suitable set of rotationlike defects can have any holonomy. 

A homomorphism form $\pi_1\big[X\setminus X^{(2)},p\big]$ to the Poincaré group can be defined by specifying an element of the Poincaré group for each generator of $\pi_1\big[X\setminus X^{(2)},p\big]$. If $X$ is contractible , then $\pi_1\big[X\setminus X^{(2)},p\big]$ is generated by a set of simple loops around the defect.\footnote{If the fundamental  group  of $X$ is not trivial, the assignment $Q$ also includes global ``topological'' degrees of freedom. As was the case in 2+1 dimensional pure gravity.} Since the fundamental group of $X\setminus X^{(2)}$ is torsion-free, the number of generators will be equal to the first Betty number of $X\setminus X^{(2)}$. In the special case that $X^{(2)}$ consists of a disconnected collection of infinite defects\footnote{Configurations with finite connected defects will be considered in the next section.} the number of generators will exactly equal the number of defects. Since each of the simple loops must be assigned a rotationlike holonomy this gives $7n$ degrees of freedom for $n$ defects.

The assignment $Q$ of holonomies to equivalence classes of loops depends on the choice of base point $p$ and the choice of Poincaré frame at that point. Another choice of point and frame will lead to an assignment $Q$ that differs by conjugation with an element of the Poincaré group. The choice of the base point and its frame can therefore be used to eliminate up to 10 degrees of freedom. For example a configuration with two defects has $2\cdot 7 -10 =4$ degrees of freedom. Note that the choice of base point can never eliminate the frame independent degrees of freedom given by the conjugacy class of each generator.

\section{Finite defects and junctions}\label{sec:junctions}
\begin{figure}[tbp]
\centering\includegraphics[width=120mm]{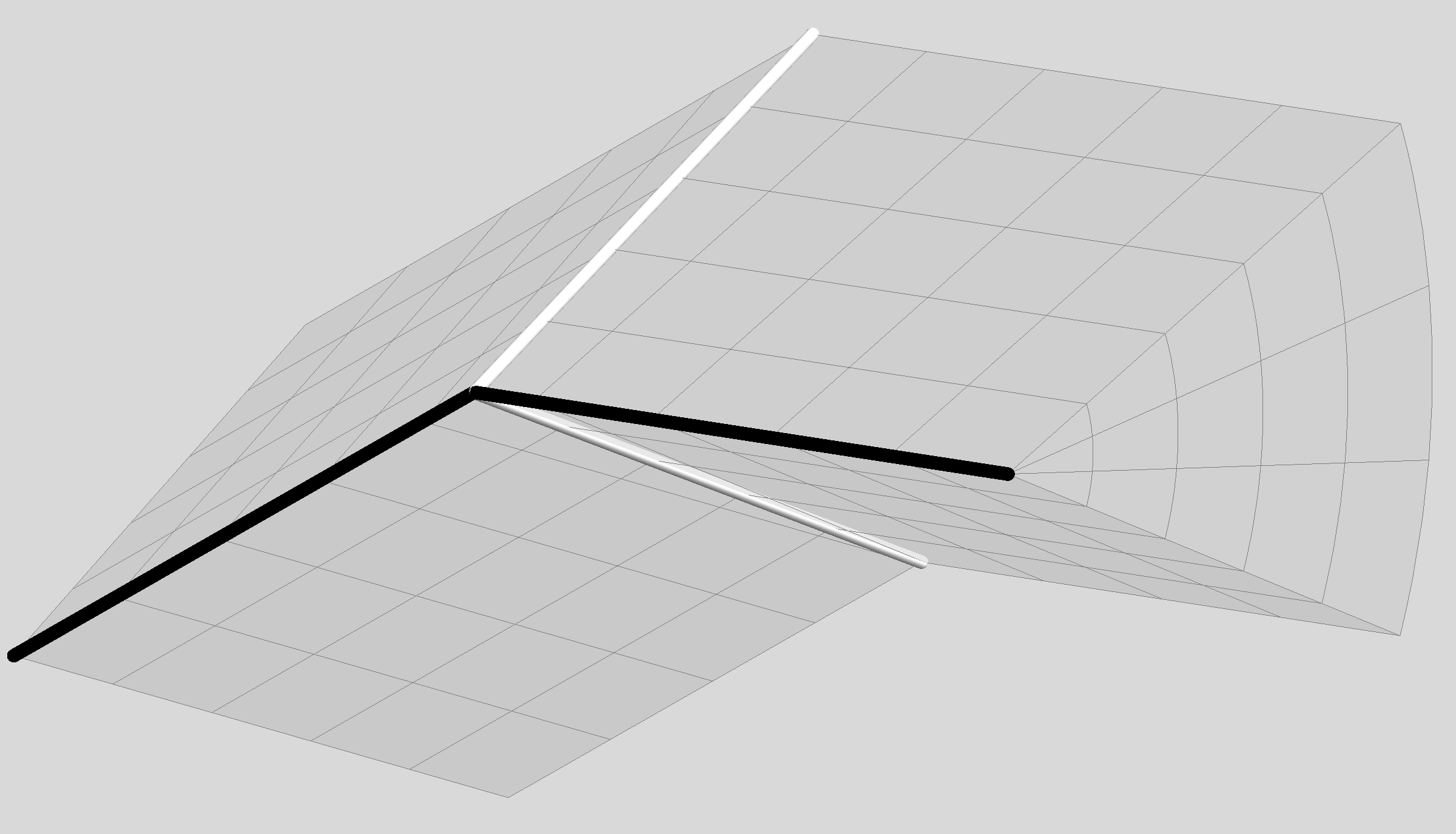}
\caption{If two defects meet at an angle, then there must exist a third defect (in white) at that junction.}\label{fig:twodefectjunction}
\end{figure}
Thus far we have only considered defects of infinite extent. More generally we can also consider finite defects, i.e. defects that have a finite length either in their spacelike direction of in their timelike direction (or both). To get a feeling for these objects let us first consider finite conical defects in 3-dimensional Euclidean space, where they are 1-dimensional lines. In a locally flat background it is impossible for a defect line to simply end in mid space. So, finite defect lines have to end on other objects. In our piecewise flat model the only available objects to end on are other finite defects.

If two finite defect lines end at one point there are two options:
\begin{enumerate}
\item Both defects have the same deficit angles and same direction. I.e. the defects form a single longer line defect.
\item There is at least one more defect ending at the same point.
\end{enumerate}
The second situation is shown in figure \ref{fig:twodefectjunction}. The two defect lines (shown in black) meet at an angle. As a result the line where the deficit angles meet (shown in red) cannot be flat, but instead forms a third line defect.

The same situation can occur in 3+1 dimensions. Conical defects cannot end simply in mid air, but instead must end on a codimension  3  submanifold called a \emph{junction}, where at least two other bounded conical defects must end. The junction is a line that is shared between the defects ending there. Therefore, if $Q_1,\dots,Q_n$ are the holonomies of defects ending at a junction, then the points of the junction must satisfy,
\begin{equation}\label{eq:junctionfixedspace}
Q_1 x = Q_2 x= \cdots = Q_n x = x.
\end{equation}
We can distinguish three types of junction based on their metric signature: timelike subluminal junctions, lightlike null junctions, and spacelike superluminal junctions.

\begin{figure}[t]
\centering\includegraphics[width=\textwidth]{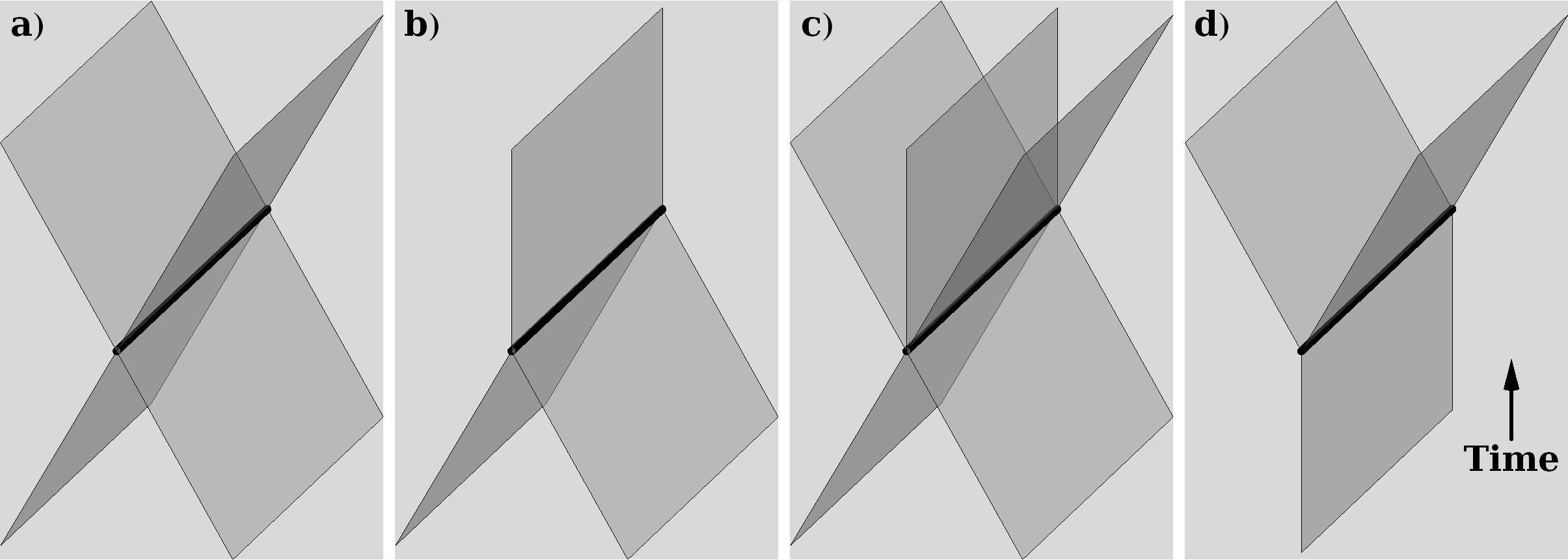}
\caption{Four different types of superluminal junction. The black line is a spacelike junction, the grey sheets are defects ending at the junction. From left to right:...\newline a) ... two incoming and two outgoing defects, constant number of degrees of freedom.\newline
 b) ... less outgoing than incoming defects. Information is lost.\newline
 c) ... more outgoing than incoming defects. Information is created, but the incoming defects do predict the location of the junction.\newline
 d) ... just one incoming defect. Information is spontaneously created. Not compatible with local causality.}\label{fig:junctiontypes}
\end{figure}

Of these the superluminal junctions appear troublesome. If three or more timelike defects end on a spacelike line, then there exists a Lorentz frame in which this line lies in a constant time slice. In this frame, the defects ending at this junction may be divided into two classes depending on whether they extend to the future or the past of this time slice (see figure \ref{fig:junctiontypes}). If there is a mismatch between  the number of defects in the future and the past of the junction, then the junction implies a change in the number of local degrees of freedom.

In principle, this does not need to be the problem. If the number of defects decreases at the junction, this would be an explicit example of information loss. Although a bit peculiar, this might simply be a feature of the model. More troublesome, is the case where the number of defects increases at the junction. In that case there is a spontaneous non-local creation of information at the junction, which seems to violate local causality. Yet, if there are at least two defects in the past of the junction, then the junction describes the frontal collision of two (or more) collinear  defects. It may yet be that the proper way to continue such a singular collision involves the creation of new defects. (In chapter \ref{ch:collisions} we will see that in general collisions are accompanied by the creation of new defects.)

The situation becomes more dire if there is just one defect (or none at all) in the past of the junction.  In that case the geometry to the past of the junction has no prior information about the junction at all. The junction simply appears instantaneously at all points of the junction. It is impossible to reconcile such an event with local causality. Since local causality was one of the primary assumptions of our model, appearance of such junctions would indicate a possibly fatal inconsistency of the model. Much of chapter \ref{ch:collisions} will deal with trying to avoid the appearance of such superluminal  junctions.

\begin{figure}[tbp]
\centering\includegraphics[width=100mm]{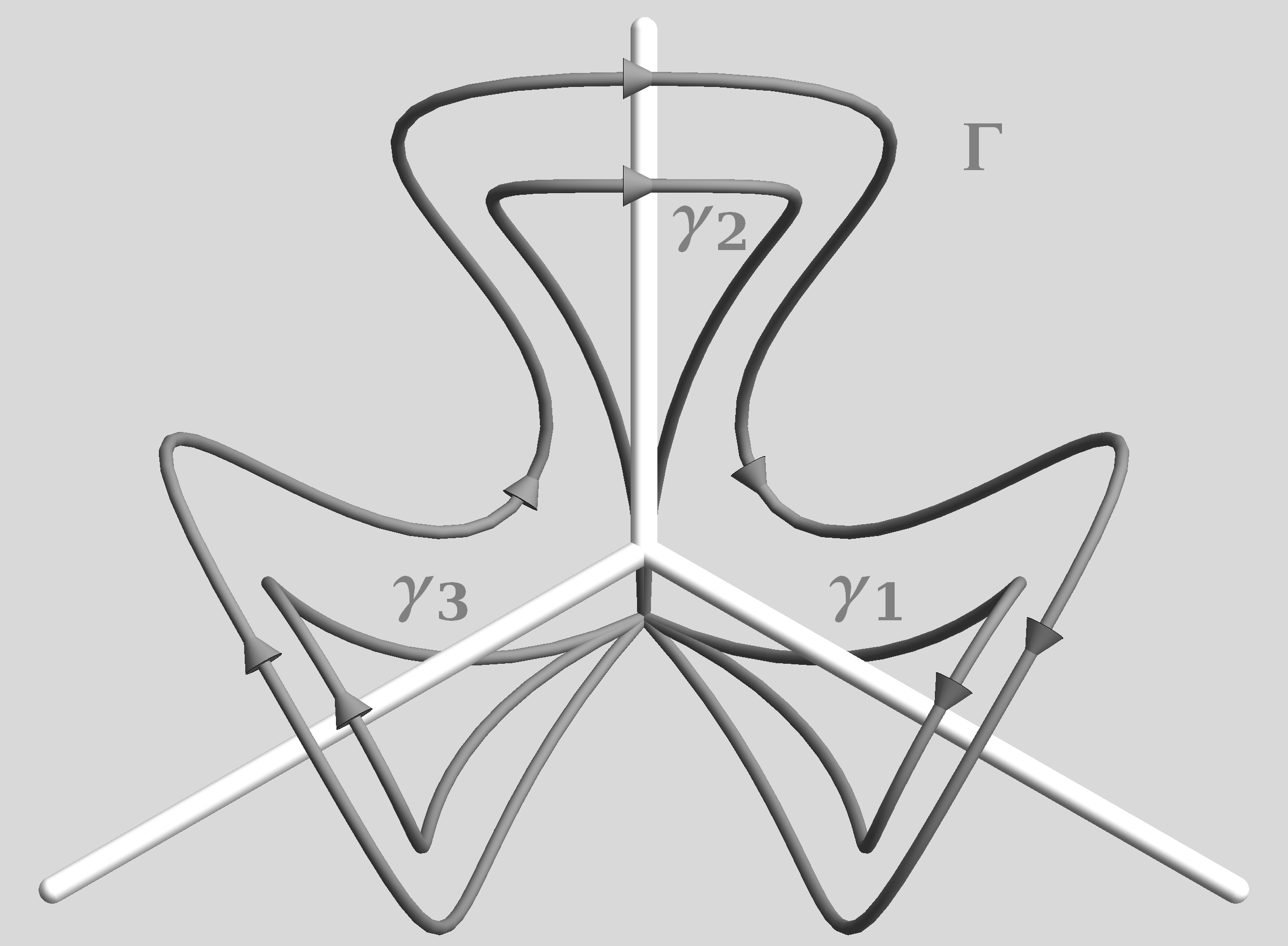}
\caption{The loop $\Gamma$ is homotopic to both the concatenation of paths $\gamma_1\cdot\gamma_2\cdot\gamma_3$ and to the trivial loop. The holonomies assigned to the loops $\gamma_i$ must thus satisfy $Q_{\gamma_1}Q_{\gamma_2}Q_{\gamma_3}=\Id$.}\label{fig:junction}
\end{figure}

Another issue we saw in the example of line defects in 3-dimensional Euclidean space was that not just any set of defects can connect at a junction. In the example, the third line defect was determined in terms of the other two line defects. In 3+1 dimensions, there exists a similar consistency condition for general junctions. This condition is a result of the requirement that the space around the junction is locally flat. It is most easily expressed  in terms of the holonomy. Recall that the assignment of holonomies to each loop is required to be a group homomorphism with respect to the concatenation of loops. The topology of a junction is such that one can find simple loops around each of the ``legs'' of the junction, which can be concatenated to a single contractible loop (see figure \ref{fig:junction} for an example with a 3-valent junction). This implies that the product of the holonomies of these loops must be the identity element. I.e. if the loops are labelled, $\gamma_1,\dots,\gamma_n$, then the corresponding holonomies satisfy
\begin{equation}\label{eq:junctioncondition}
Q_{\gamma_1}\cdots Q_{\gamma_n} = \Id.
\end{equation}
 This equation is referred to as the \emph{junction condition} of the holonomies.

A second restriction comes from the requirement that all defects share a common 1-dimensional invariant subspace. This means that all holonomies must satisfy equation \eqref{eq:junctionfixedspace}. This condition is not completely independent of the junction condition \eqref{eq:junctioncondition}. If the junction condition  \eqref{eq:junctioncondition} is satisfied for $n$ defects, then if $n-1$ holonomies share a common invariant line then so does the last one. 

%


\section{Piecewise flat manifolds}\label{sec:PWFmanifolds}
The description of a configuration of conical defects in terms of the Poincaré holonomies of loops is very powerful. In particular, it gives a succinct formulation of the consistency conditions at a junction. However, this method also has a few drawbacks. We already remarked that the holonomy does not quite uniquely describe a defect configuration if we include defects with negative curvature.

A second problem is that  the path dependence of the holonomy means that the local geometry of a defect is encoded in a non-local way depending on the path chosen to reach the defect. Keeping track of the chosen paths becomes increasingly inconvenient for larger configurations of defects.

An alternative approach focuses on the geometry of the locally flat patch of spacetime rather that the curvature of the defects. In this approach the geometry of a configuration of defects is described as a \emph{piecewise flat manifold}. 

\begin{definition}
A \emph{piecewise flat manifold} is a manifold $X$ with a CW complex structure and a flat (pseudo)-Riemannian metric $g^\alpha$ on each $n$-cell $e^n_\alpha$. A geometry on $X$ is defined by the following requirements
\begin{enumerate}
\item The metric $g^\alpha$ is smooth on the interior of $e^n_\alpha$ and piecewise smooth on the boundary $\partial e^n_\alpha$.
\item  The attachment map $\phi_\alpha\colon S^{n-1}_\alpha \to X^{(n-1)}$ of each $n$-cell $e^n_\alpha$ to the $(n-1)$-skeleton  $X^{(n-1)}$ is an isometry.
\item For each $(n-1)$-cell $e^{n-1}_\beta$ in the image of $\phi_\alpha$, the inverse image $\phi_\alpha^{\m 1}(e^{n-1}_\beta)$ has zero extrinsic curvature in $e^n_\alpha$.
\end{enumerate}
\end{definition}
These conditions ensure that each $n$-cell is an $n$-dimensional polytope. An $n$-cell can be subdivided by adding  an $(n-1)$-cell dividing it in two pieces. Repeated subdivisions can be used to ensure that the cells satisfy some additional properties. For example, one could subdivide the cells until all cells are convex polytopes, or ultimately until all cells are simplices.

Subdivision into a simplicial complex has the additional advantage that a flat geometry on the interior of a $n$-simplex is uniquely determined by the lengths of the 1-simplices in its boundary. This leads to a more usual definition of a piecewise flat manifold in terms of a simplicial complex.\cite{CMS:1984} Here we opt for the more general definition above since it gives more flexibility.

If we take the manifold to be four dimensional and require that the metrics on the 4-cells have Lorentzian signature, then $X$ can be viewed as a piecewise flat spacetime consisting of polytope building blocks glued together along their edges.

Because the attachment maps are isometries, for each $(n-1)$-cell $e^{n-1}_\beta$ in the boundary of a $n$-cell $e^{n}_\alpha$ there is an inverse isometry $i_{\beta}^{\alpha}\colon e^{n-1}_\beta\to e^{n}_\alpha$.  Moreover, because $X$ is a manifold, each 3-cell is included in exactly two 4-cells.

Each 4-cell is isometric to a piece of Minkowski space.\footnote{In the pathological cases where this is not possible, it is possible to subdivide the 4-cells into cells that are isometric to a subset of Minkowski space.} We can fix a frame in a 4- cell  $e^4_\alpha$ by choosing a specific embedding $\iota_\alpha\colon e^4_\alpha\to \RR^{3+1}$. Given two 4-cells $e^4_\alpha$ and $e^4_\beta$ that share a 3-cell $e^3_{\alpha\beta}$ in their common boundary and framings $\iota_\alpha$ and  $\iota_\beta$, there is a unique Poincaré transformation $P_{\beta\alpha}$ that makes the following diagram commute:
\[\begindc{\commdiag}[75]
\obj(0,1)[m1]{$\RR^{3+1}$}
\obj(2,1)[m2]{$\RR^{3+1}$}
\obj(0,0)[ea]{$e^4_\alpha$}
\obj(1,0)[eab]{$e^3_{\alpha\beta}$}
\obj(2,0)[eb]{$e^4_\beta$}
\mor{m1}{m2}{$P_{\beta\alpha}$}
\mor{ea}{m1}{$\iota_{\alpha}$}
\mor{eb}{m2}{$\iota_{\beta}$}
\mor{eab}{ea}{$i^{\alpha}_{\alpha\beta}$}[-1,0]
\mor{eab}{eb}{$i^{\beta}_{\alpha\beta}$}
\enddc\]
The map $P_{\alpha\beta}$ relates the change of frame as you move from cell $\alpha$ to cell $\beta$.

A general 2-cell is in the boundary of a finite number of 4-cells, say $\alpha_1,\dots,\alpha_k$. We can follow the change of frame as we follow a loop $\gamma$ around the 2-cell by composing the Poincaré transformations $P_{\alpha_i\alpha_j}$. The Poincaré holonomy of the loop can therefore be calculated as
\begin{equation}
P_\gamma = P_{\alpha_1\alpha_k}P_{\alpha_k\alpha_{k-1}}\cdots P_{\alpha_2\alpha_1}.
\end{equation}
Note that this holonomy only depends on the 2-cell, around which the loop was taken and the frame chosen in the 4-cell where the loop was started.

We can use piecewise flat manifolds to describe general configurations of defects. In such a description the 2-cells correspond to conical defects, the 1-cells correspond to junctions, and the 0-cells correspond to collisions of defects.

In a generic piecewise flat manifold any 2-cell will have a non-trivial holonomy including spacelike 2-cells (i.e. 2-cells with a Euclidian metric signature). Such a 2-cell would correspond to a spacelike conical defect, which is prohibited by the local causality principle of our model. One could attempt to enforce this principle by disallowing piecewise flat manifolds with spacelike 2-cells. This however seems overly restrictive. Instead we opt to allow spacelike 2-cells only when they have a trivial holonomy. These cells are treated as ``virtual defects'' necessary to facilitate the piecewise flat description.

The advantage of using a piecewise flat manifold structure to describe a configuration of defects is that it encodes the geometry in a local cell-by-cell way. Moreover, if two configurations share the same description as a piecewise flat manifold they are geometrically indistinguishable, and should physically be considered the same configuration.

The downside is that there are many piecewise flat structures that describe the same configuration of defects. Not only is there the choice of framing of the 4-cells (which can be considered as a kind of local gauge fixing), the process of subdivision allows the creation of many equivalent piecewise flat structures. A related issue is that the 4-cells are compact which means that we even need an infinite number of cells to describe Minkowski space. This last issue can easily be resolved by a slight generalization of the concept of a piecewise flat manifold by allowing half-open cells that extend to infinity. In the rest of this thesis we will generally use this generalized concept for simplicity.

\section{Other piecewise flat approaches to gravity}\label{sec:otherapproaches}
The use of piecewise flat structures in gravity dates back to 
Tulio Regge's seminal 1961 paper ``General Relativity without Coordinates''.\cite{Regge:1961} His approach --- now known as Regge calculus --- is based on the observation that any (pseudo)Riemannian manifold may be approximated by an appropriately fine piecewise flat simplicial complex. The lengths of the 1-simplices, which determine the flat metrics on the other simplices, are taken to be the fundamental variables.

Regge calculus has been successfully used both as an approximation scheme for numerically attacking problems in classical general relativity (see \cite{williams:1992} for a review and bibliography), and as the basis for quantum mechanical approaches to gravity (see \cite{RW:2000} for a review). In many respects the piecewise flat gravity model studied in this thesis is similar to Regge calculus. A key difference, however, is that in Regge calculus the piecewise flatness of the geometry is a discretization of the classical geometry. As such the conical defects on the codimension 2 simplices are not viewed as physical excitations of the model. Consequently, no restrictions are enforced on the spacetime signature with which the defects may occur.

Another difference is that quantum gravity approaches based on Regge calculus such as the original Ponzano-Regge model \cite{Ponzano:1968} and the Turaev-Viro model \cite{Turaev:1992hq} in 3 dimensions and the Barrett-Crane model \cite{Barrett:1997gw} in 4 dimensions typically assume an Euclidean signature. Although the model under consideration here does not currently address any issues of quantization, it is inherently Lorentzian in nature.

The spin foam models considered in the context of loop quantum gravity can be viewed as a Lorentzian generalization of the Euclidean quantum gravity models based on Regge calculus. Moreover, loop quantum gravity uses the holonomies of spatial loops as the fundamental variables, much like the holonomy description of a configuration of conical defects. This begs the question whether there is some relation with the holonomy description of the model studied here. In particular, one could imagine that this model could appear as the classical limit of loop quantum gravity. This question was addressed  by Eugenio Bianchi.\cite{bianchi2009} He tried to reproduce the kinematical state space of loop quantum gravity by standard path integral quantization techniques. His conclusion was that restricting the path integral to locally flat metrics was too strong of a restriction. Instead to obtain the kinematical state space of loop quantum gravity one needs to integrate over the wider class of locally-flat connections.

Another piecewise flat approach to quantum gravity based on simplicial complexes is causal dynamical triangulations.\cite{Ambjorn:1998xu,Ambjorn:2000dv,Ambjorn:2001cv} Unlike Regge calculus, causal dynamical triangulations does not use the edge lengths of the simplices as variables, but rather fixes all edge lengths to a constant, and relies on the sum of different triangulations to produce dynamics. Even more so than Regge calculus this model views the piecewise flat structure as a discretization tool with the theory proper being defined only in the limit that the edge lengths go to zero. As such in causal dynamical triangulations the appearance of spacelike 2-simplices with a non-trivial holonomy is not viewed as problematic and is in fact unavoidable for the model to work.

In spirit, the world crystal model proposed by Kleinert  \cite{Kleinert:2003za,Kleinert2008} seems to be the most similar to the model considered here. He too considers a model of gravity where the fundamental degrees of freedom are curvature defects. In his model however the defects are not constrained to be flat planes in spacetime, but are allowed to have an extrinsic curvature. As consequence they must be accompanied by a non-zero gravitational field. The resulting spacetime is not actually locally flat. In that aspect that model crucially differs from our model, where local flatness is a fundamental principle.

\ifx\fullTeX\undefined
\bibliographystyle{../bib/utcaps}
\bibliography{../bib/thesis}
\end{document}
\fi

\cleardoublepage 
\ifx\fullTeX\undefined
\documentclass[11pt,a4paper]{article}
\usepackage[utf8]{inputenc} 
\usepackage[english]{babel}
\usepackage{graphicx}
\usepackage{hyperref}
\usepackage{color}
\usepackage{amsmath}
\usepackage{amsthm}
\usepackage{amssymb}
\usepackage{bbm}
\usepackage{pictexwd,dcpic}

\graphicspath{{../pics/}}

\title{Collisions}
\author{Maarten van de Meent}
\date{\today} 

\begin{document}
\maketitle
\else
\chapter{Collisions}\label{ch:collisions}
\fi
A new issue in 3+1 dimensions is that we must consider the collision of two defects. Generically, two codimension 2 planes have an intersection of codimension 4. Consequently, conical defects in 2+1 dimensions generically have no intersections. That is, in 2+1 dimension we can safely assume that two point particles never collide. In 3+1 dimensions, however, the world sheets of two line defects generically have one common point. This point indicates an event where the two line defects collide.\footnote{Note that for an arbitrary time slice considered as the present this point may lie in the future or in the past.}

\begin{figure}[tbp]
\centering\includegraphics[width=\textwidth]{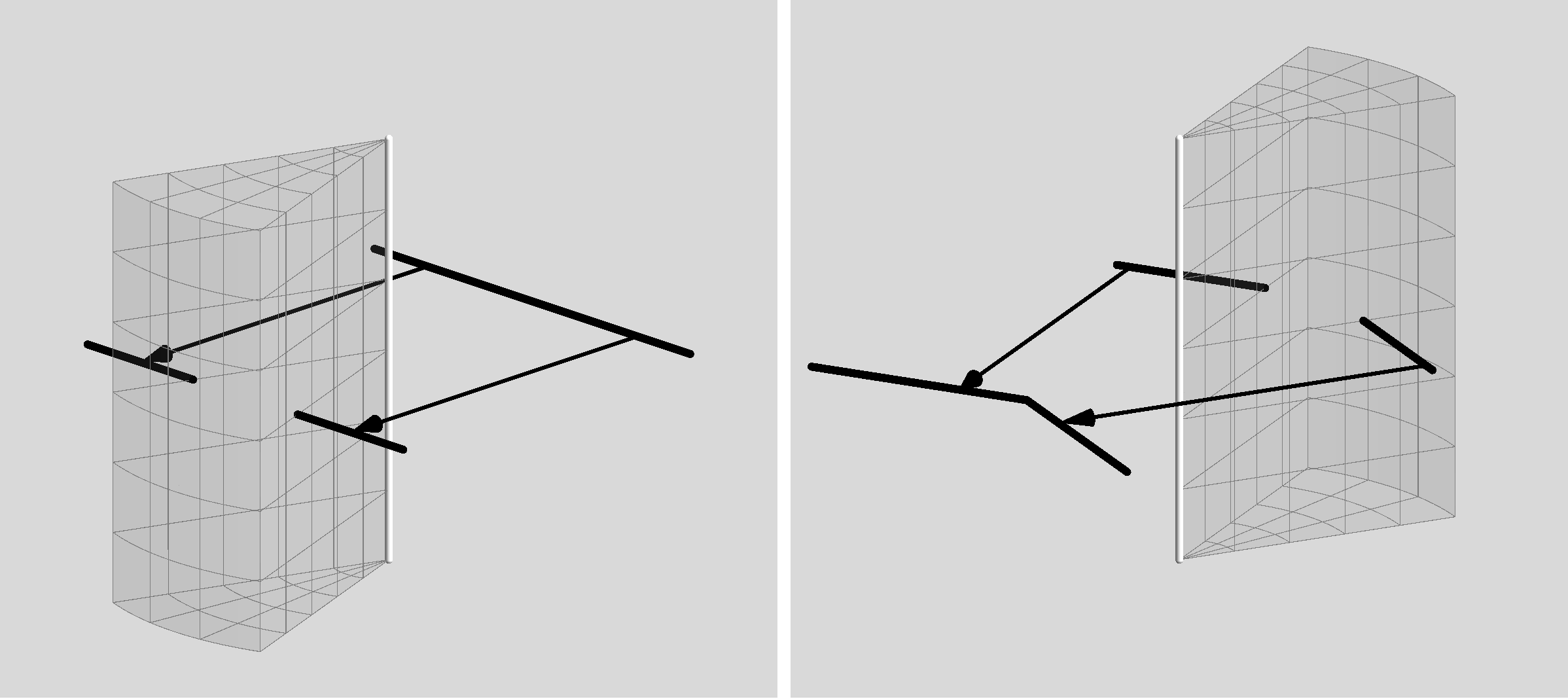}
\caption{The black defect cannot pass through the white defect without incurring a kink.}\label{fig:defectshallnotpass}
\end{figure}

Consequently, we must add a prescription of what happens when two line defects collide to our a model. Because the holonomies carried by the line defects do not commute, the line defects cannot simply pass through each other. This can be understood geometrically if we consider the situation in figure~\ref{fig:defectshallnotpass}, which shows the collision of two line defects. On the left the deficit angle of the white defect is drawn away from the approaching black defect. On the right the same situation is drawn with the deficit angle in the opposite direction. The black defect now has a kink after passing through the white defect. In section \ref{sec:junctions} we saw that in our piecewise flat model a kink in a defect implies that there is at least one other defect meeting at that point.

Our piecewise flat model must somehow prescribe what happens after such a collision. For this it needs to provide an intermediate configuration of conical defects that continues the geometry after the collision.\footnote{For the sake of brevity of expression we will simply refer to such an intermediate configuration as a ``continuation''.} In this chapter we will discuss how these ``continuations'' can be constructed for general collisions.

\section{Collision parameters}
We wish to study a general collision of two conical defects $A$ and $B$. For any collision we can find a locally flat neighbourhood that contains no other defects than $A$ and $B$ (prior to the collision). For the study of general collisions it is therefore sufficient to study the collisions of infinite defects.

Two defects have 14 free parameters. To determine how many of these can be fixed by the choice of Poincaré frame, we proceed as follows. Let us use the translational freedom to fix the spacetime point of the collision at the origin. This has the advantage that the translational parts of the holonomies of both defects vanish and we only have to worry about the Lorentz part of the holonomies. We can then use two rotations to fix the orientation of the defect $B$ (say along the z-axis), and use two boosts to fix the velocity of $B$ (say we make $B$ stationary). We now have only two degrees of freedom in our choice of frame, a rotation and a boost in the direction of the orientation of $B$. We can use the boost to make the velocity of $A$ perpendicular to the defect $B$. The remaining rotation can then be used to completely fix the direction of that velocity (say along the $x$-axis).

\begin{figure}[tb]
\centering\includegraphics[width=12cm]{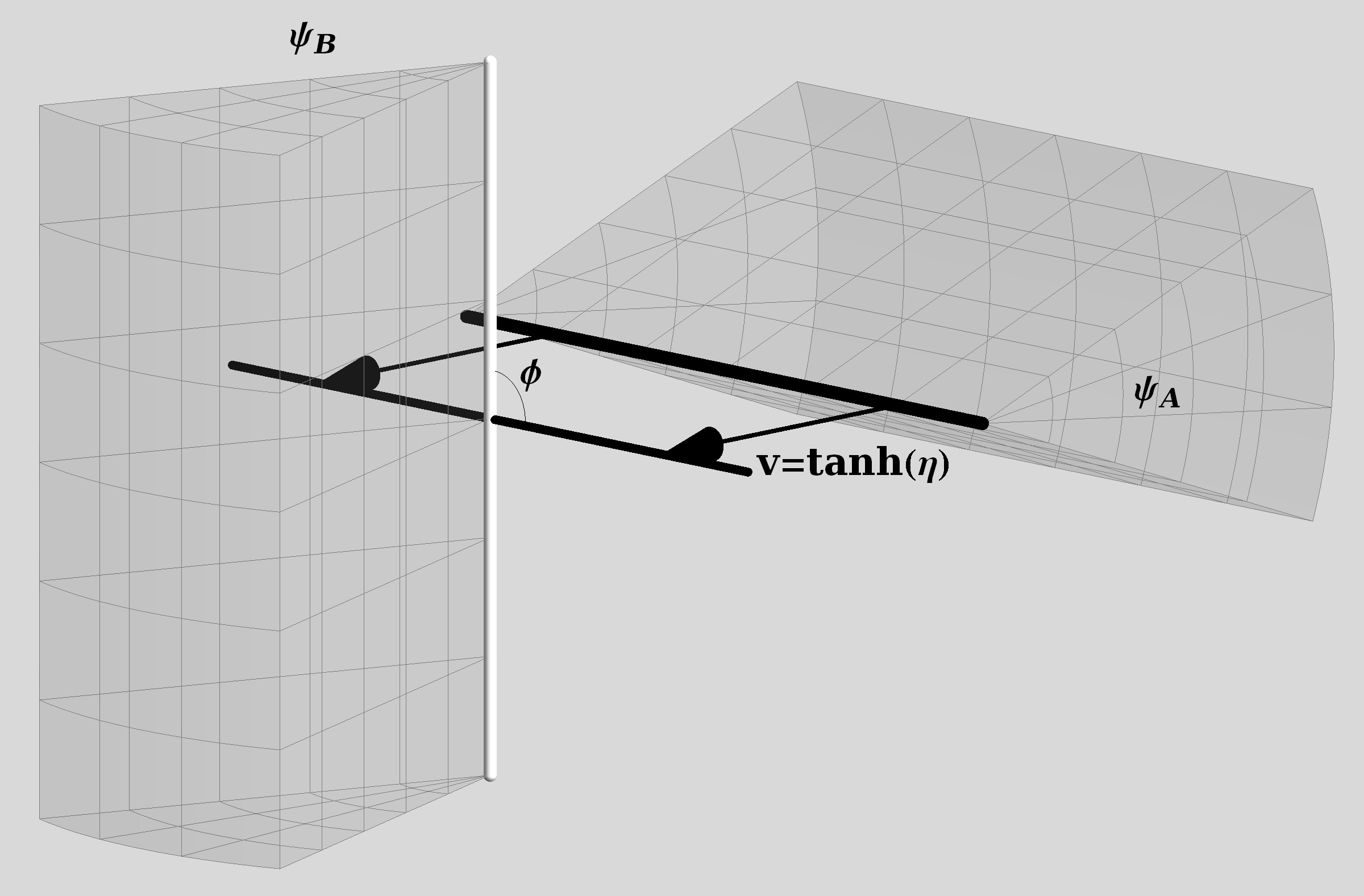}
\caption{The parameters of a general collision of two defects: $\psi-A$, $\psi_B$, $v$ ($=\tanh\eta$), and $\phi$.}\label{fig:collisionparameters}
\end{figure}

Thus we find that there are 4 remaining degrees of freedom: the conjugacy classes of the two defects given by the size of their deficit angles in their rest frame $\psi_A$ and $\psi_B$, the difference in orientation of the defects given by the angle $\phi$ between the two defects at the instant of collision, and the relative velocity between the two defects given by the difference in rapidity $\eta$.

Note that there are different ways in which the frame can be fixed. In the example above we chose the frame where defect $B$ was stationary. We will call this the rest frame of $B$. Similarly we could have chosen the rest frame of $A$. These frames are often convenient when doing calculations with the holonomy. For geometrical calculations, it is often useful to make a more symmetrical choice of frame where both defects have an equal velocity. We will refer to that frame as the ``center of velocity frame''. In this frame the difference in velocity will not be equal to that in the rest frames, the difference in rapidity however is the same in all three frames.
 
\section{Orthogonal collisions}\label{sec:orthogonalcollisions}
We first consider a specific class of collisions where the colliding defects are orthogonal at the point of impact, i.e. $\phi=\pi/2$. These types of collisions turn out to allow an especially simple continuation, yet they already introduce some of the issues that we encounter when constructing continuations for general collisions.

\begin{figure}[tbp]
\centering\includegraphics[width=\textwidth]{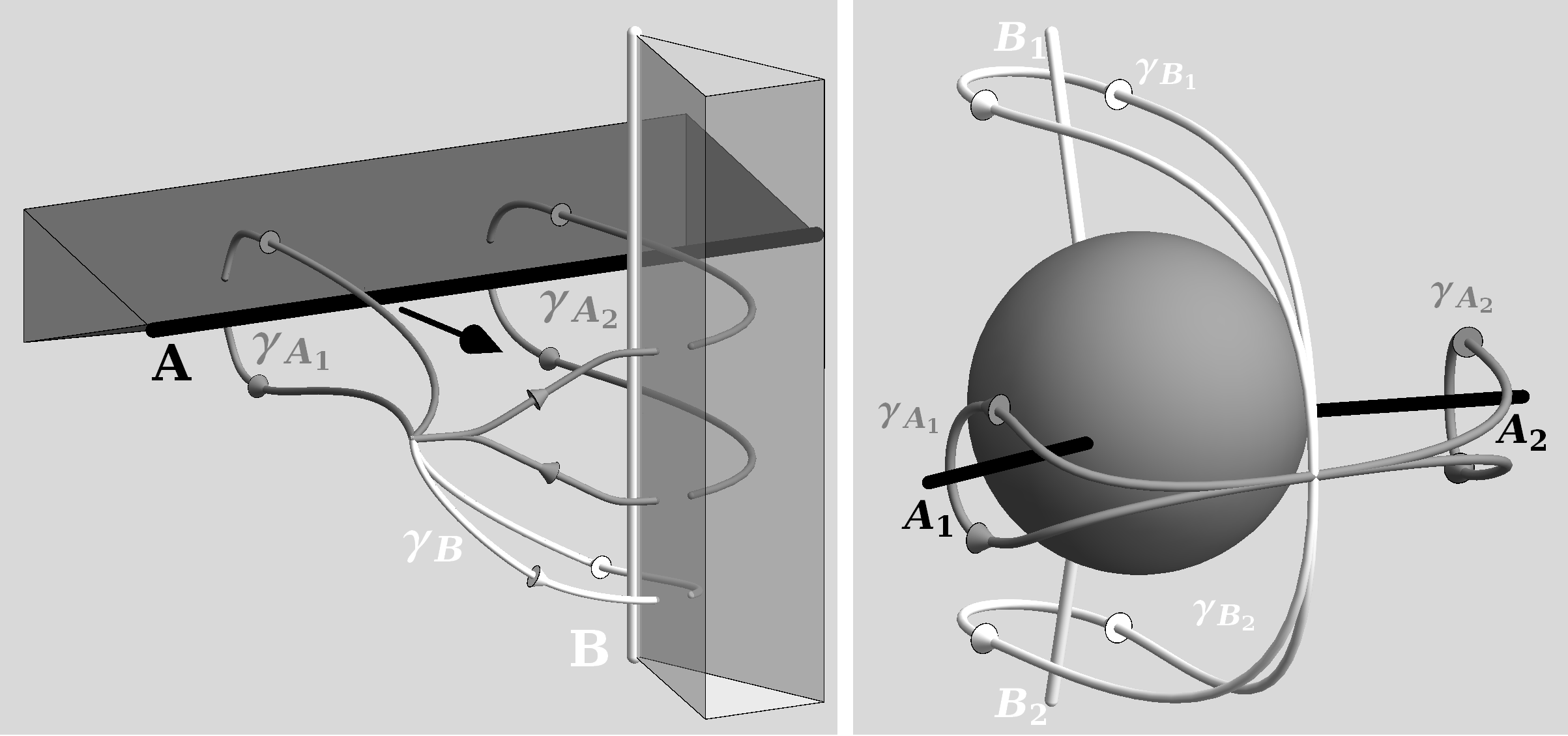}
\caption{The loops relating the holonomies of the defects before the collision (on the left) to the holonomies of the semi-infinite defects after the collision (on the right).}\label{fig:collisionholonomies}
\end{figure}

As defects $A$ and $B$ collide they effectively cut each other in four semi-infinite defects (which we label $A_1$, $A_2$, $B_1$, and $B_2$ as shown in figure \ref{fig:collisionholonomies}). The task of constructing a continuation for the collision consists in finding a network of finite defects that interpolate between the ends of these semi-infinite defects.

Figure \ref{fig:collisionholonomies} shows how the holonomies of the semi-infinite defects are related to the holonomies of the original defects. The loops $\gamma_{A_1}$ and $\gamma_{A_2}$ around the semi-infinite defects $A_1$ and $A_2$ after the collision are homotopic to the loops with the same labels before the collision. The loop $\gamma_{A_2}$ before the collision is homotopic to the concatenation $\gamma_{B}^{\m 1}\cdot\gamma_{A_1}\cdot\gamma_{B}$. Consequently, for the holonomies we find
\begin{equation}\label{eq:orthcollisioncond1}
Q_{A_2} = Q_{B}^{\m 1}Q_{A_1}Q_{B}.
\end{equation}
The loops $\gamma_{B_1}$ and $\gamma_{B_2}$ after the collision are both homotopic to the loop $\gamma_B$ before the collision, and therefore have the same holonomy. Consequently, we can rewrite equation \eqref{eq:orthcollisioncond1} as
\begin{equation}\label{eq:orthcollisioncond2}
Q_{A_1}Q_{B_1}Q_{A_2}^{\m 1}Q_{B_2}^{\m 1}=\Id.
\end{equation}
This equation expresses that the concatenation of loops $\gamma_{A_1}\cdot \gamma_{B_1}\cdot\gamma_{A_2}^{\m 1}\cdot\gamma_{B_2}^{\m 1}$ is contractible, and thereby that the defects formed two separate networks (i.e. two distinct defects) before the collision.

In the rest frame of the original $B$ defect, the defect $A_1$ is spanned by the 4-vectors
\begin{equation}
\begin{aligned}
u_1^\mu = (1,v,0,0),\\
d_1^\mu = (0,0,1,0).\\
\end{aligned}
\end{equation}
The holonomy of the $B$ defect in this frame is a rotation about the $z$-axis, $R_z(\psi_B)^\mu_{\phantom{\mu}\nu}$. Consequently, the semi-infinite defect $A_2$ is spanned by the 4-vectors,
\begin{equation}
\begin{aligned}
u_2^\mu &= R_z(\psi_B)^\mu_{\phantom{\mu}\nu} u_1^\nu & &= (1,v \cos\psi_B,v \sin\psi_B,0),\\
d_2^\mu &= R_z(\psi_B)^\mu_{\phantom{\mu}\nu} d_1^\nu & &= (0,-\sin\psi_B, \cos\psi_B,0).\\
\end{aligned}
\end{equation}
Consequently, the semi-infinite defects $A_1$ and $A_2$ share a common junction spanned by the 4-vector
\begin{equation}
j^\mu = (1,v,v\tan\frac{\psi_B}{2},0).
\end{equation}
A similar junction is shared by the semi-infinite defects $B_1$ and $B_2$.

\begin{figure}[tbp]
\centering\includegraphics[width=10cm]{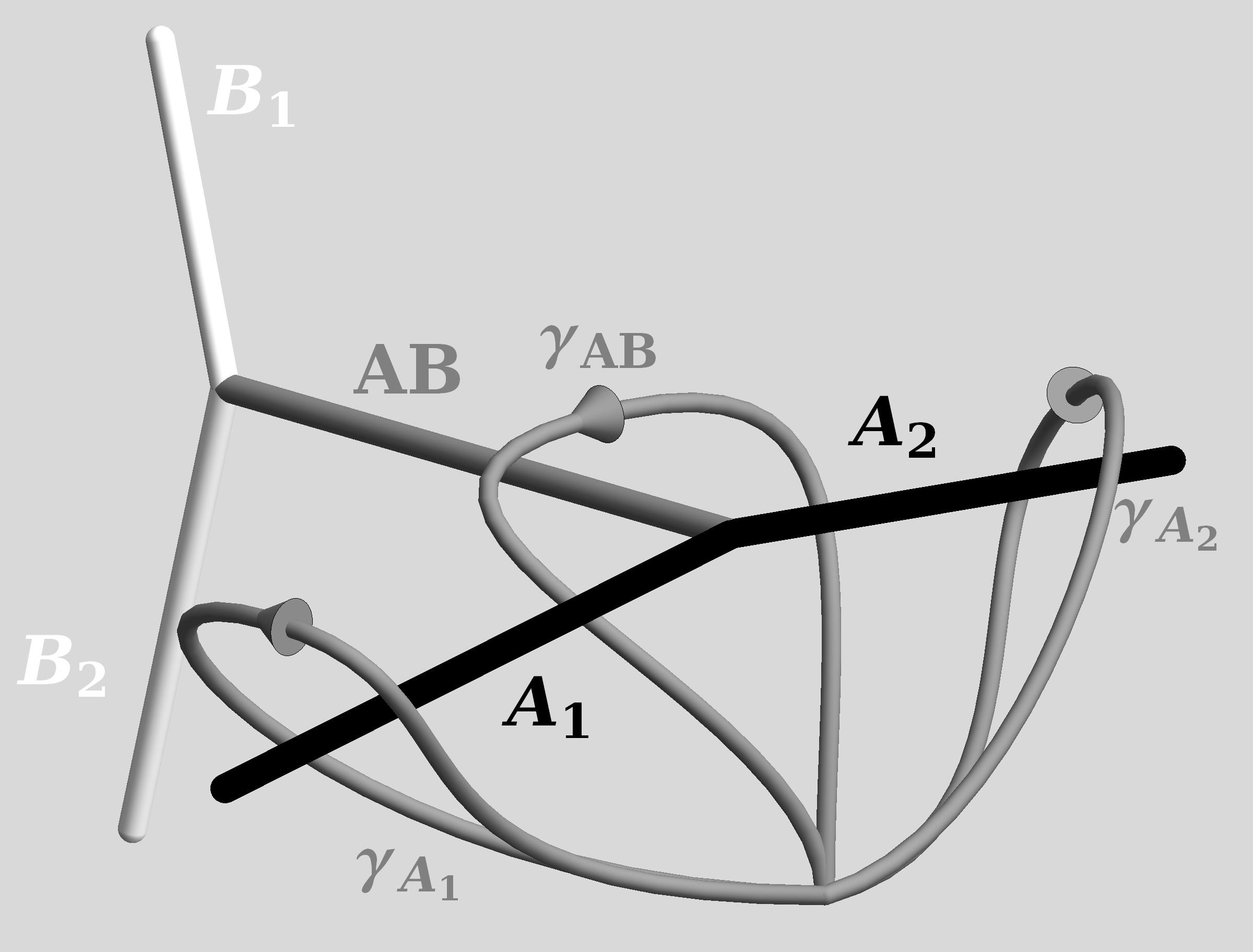}
\caption{An orthogonal collision allows a continuation consisting of just a single intermediate defect connecting the kinks in the defects $A$ and $B$. The loops in the diagram show how the holonomy of this new defect is related to the holonomies of the colliding defects.}\label{fig:orthogonalcollision}
\end{figure}

This suggests that a simple continuation is possible with a single finite defect $AB$ connecting the junctions in the $A$ and $B$ defects. The holonomy of this defect is given by the junction condition of the junction in the $A$ defect (see figure \ref{fig:orthogonalcollision})\footnote{Or equivalently the junction condition in the $B$ defect.} 
\begin{equation}
Q_{AB} = Q_{A_1}Q_{A_2}^{\m 1} = Q_A Q_B^{\m 1} Q_A^{\m 1} Q_B.
\end{equation}
We still need to check if the resulting defect is physically acceptable, i.e.
whether its holonomy is rotationlike. Since the junction is a common line for all defects meeting there, if the junction is timelike so are all incident defects. Consequently, a sufficient (although not necessary) condition for defect $AB$ to be timelike is that the junction in the $A$ defect is timelike. 

The norm of $j^\mu$ is
\begin{equation}
j^\mu j_\mu = v^2(1+\tan^2\frac{\psi_B}{2}) -1.
\end{equation}
So the junction is spacelike when 
\begin{equation}\label{eq:1defectjunctionlimit}
v^2 > \cos^2\frac{\psi_B}{2}.
\end{equation}
Consequently, for orthogonal collisions with high velocities and/or high deficit angles the junctions become superluminal. In section \ref{sec:junctions}, we observed that such junctions were undesirable for our model. However, both the $A_1$ and $A_2$ defect lie in the past of this junction. It is therefore of the least problematic kind, i.e. the fact that this junction is tachyonic needs not be seen as a violation of local causality.

A more pressing worry is that a superluminal junction leaves room for the possibility that the intermediate defect $AB$ itself is spacelike. This can be checked by finding the conjugacy class of the holonomy $Q_{AB}$ of the intermediate defect. If this is rotationlike, the intermediate defect is timelike, and if this is boostlike, the intermediate defect is spacelike. A direct way to find out to what conjugacy class $Q_{AB}$ belongs is to calculate the trace of $Q_{AB}$ in the $\PSL{2,\CC}$ representation of the Lorentz group (see appendix \ref{app:PSL2C}). In the rest frame of the $B$ defect the $\PSL{2,\CC}$ representations of $Q_A$ and $Q_B$ are
\begin{equation}
\begin{aligned}
Q_A &= \begin{pmatrix}
\cos\frac{\psi_A}{2}-\sin\frac{\psi_A}{2}\sinh\eta 
	& \sin\frac{\psi_A}{2}\cosh\eta \\
-\sin\frac{\psi_A}{2}\cosh\eta
	&\cos\frac{\psi_A}{2}+\sin\frac{\psi_A}{2}\sinh\eta 
\end{pmatrix},\\
Q_B &= \begin{pmatrix}
\ee^{\ii\frac{\psi_B}{2}} & 0\\
0 & \ee^{\m \ii\frac{\psi_B}{2}}
\end{pmatrix}.
\end{aligned}
\end{equation}
The trace of the holonomy of the intermediate $AB$ defects is consequently given by
\begin{equation}
\begin{aligned}
\tr Q_{AB} & = \tr Q_A Q_B^{\m 1} Q_A^{\m 1} Q_B\\
 &= \frac{1-2v^2+\cos\psi_A +\cos\psi_B - \cos\psi_A\cos\psi_B}{1-v^2},
\end{aligned}
\end{equation}
with $v=\tanh\eta$. A Lorentz transformation in $\PSL{2,\CC}$ is rotationlike if and only if its trace is real and the absolute values of the trace is smaller than 2. Consequently, the intermediate defect $AB$ is rotationlike only when
\begin{equation}\label{eq:orthotracebound}
v^2 < 1 - \frac{(1-\cos\psi_A)(1-\cos\psi_B)}{4}.
\end{equation}
Since the right hand side of this inequality is smaller than one for non-zero values of $\psi_A$ and $\psi_B$, there are high velocity orthogonal collisions for which the intermediate defect becomes spacelike.

\begin{figure}[tbp]
\centering\includegraphics[width=12cm]{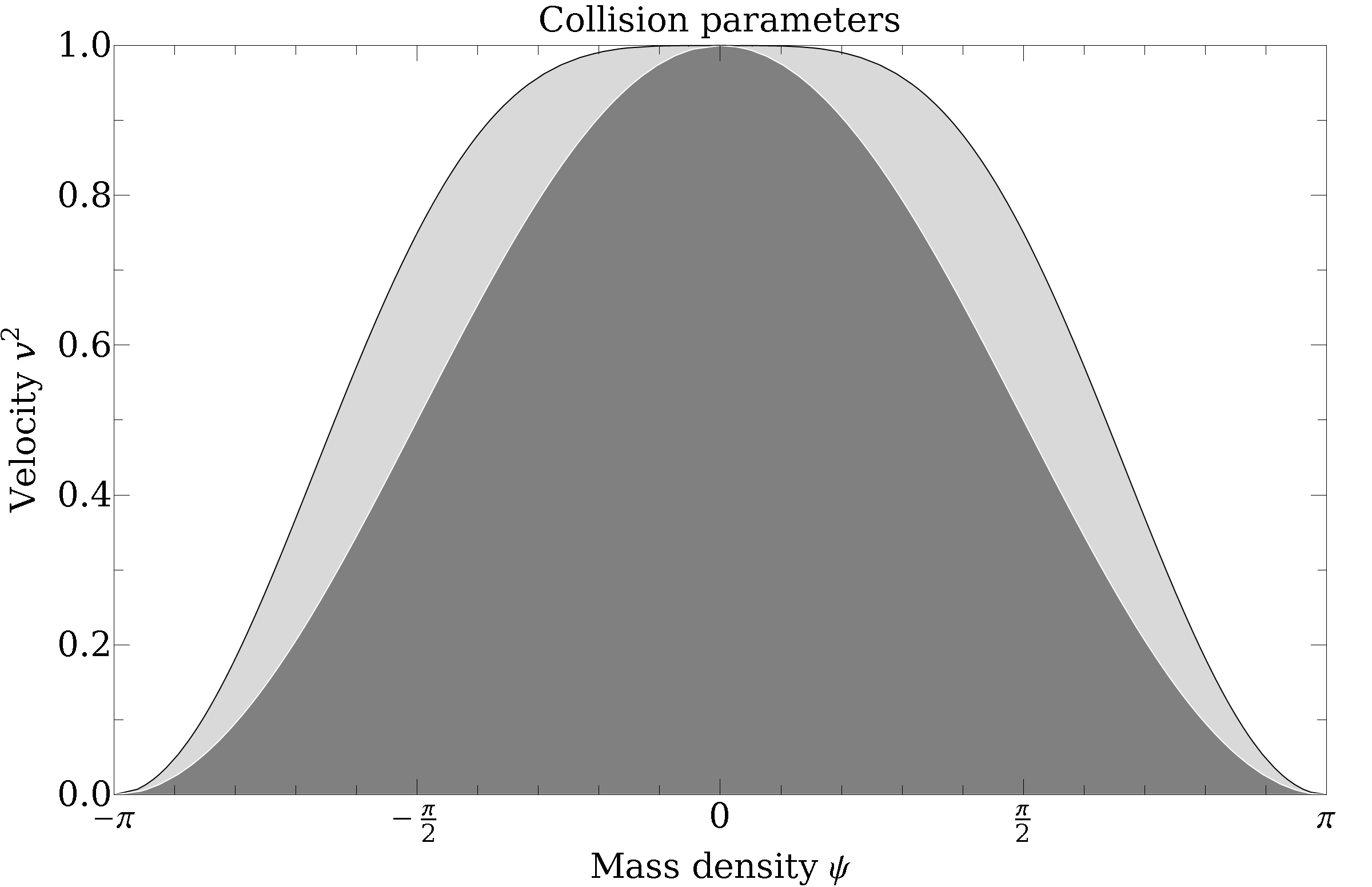}
\caption{The regions of the parameter space of orthogonal collisions of defects with equal mass $\psi$, where the junctions are timelike (in dark gray) and where the intermediate defect itself is timelike (In light gray).}\label{fig:orthobounds}
\end{figure}

We can compare the region of the collision parameter space where subluminal junctions become impossible to the region where the intermediate defect becomes spacelike --- as is done for a collision of equal mass defects in figure \ref{fig:orthobounds}. As we argued, the former is contained within the latter; The junctions being timelike is a sufficient condition for the intermediate defect to be spacelike. The region where the intermediate defect is timelike is significantly larger, showing that requiring the junctions to be timelike is not a necessary condition for the intermediate defect to be timelike. The region however still is bounded.

Apparently, there is a large region of the parameter space of collisions where the single defect continuation is not consistent within our model. In particular, collisions with a high velocity cannot be continued in this way. Now, as we will see in the next section, the single defect continuation is not even an option for non-orthogonal collisions. We therefore need to consider more complicated continuations involving more intermediate defects. These continuations may also help us deal with high velocity orthogonal collisions.

\section{Non-orthogonal collisions}
\begin{figure}[tbp]
\centering\includegraphics[width=\textwidth]{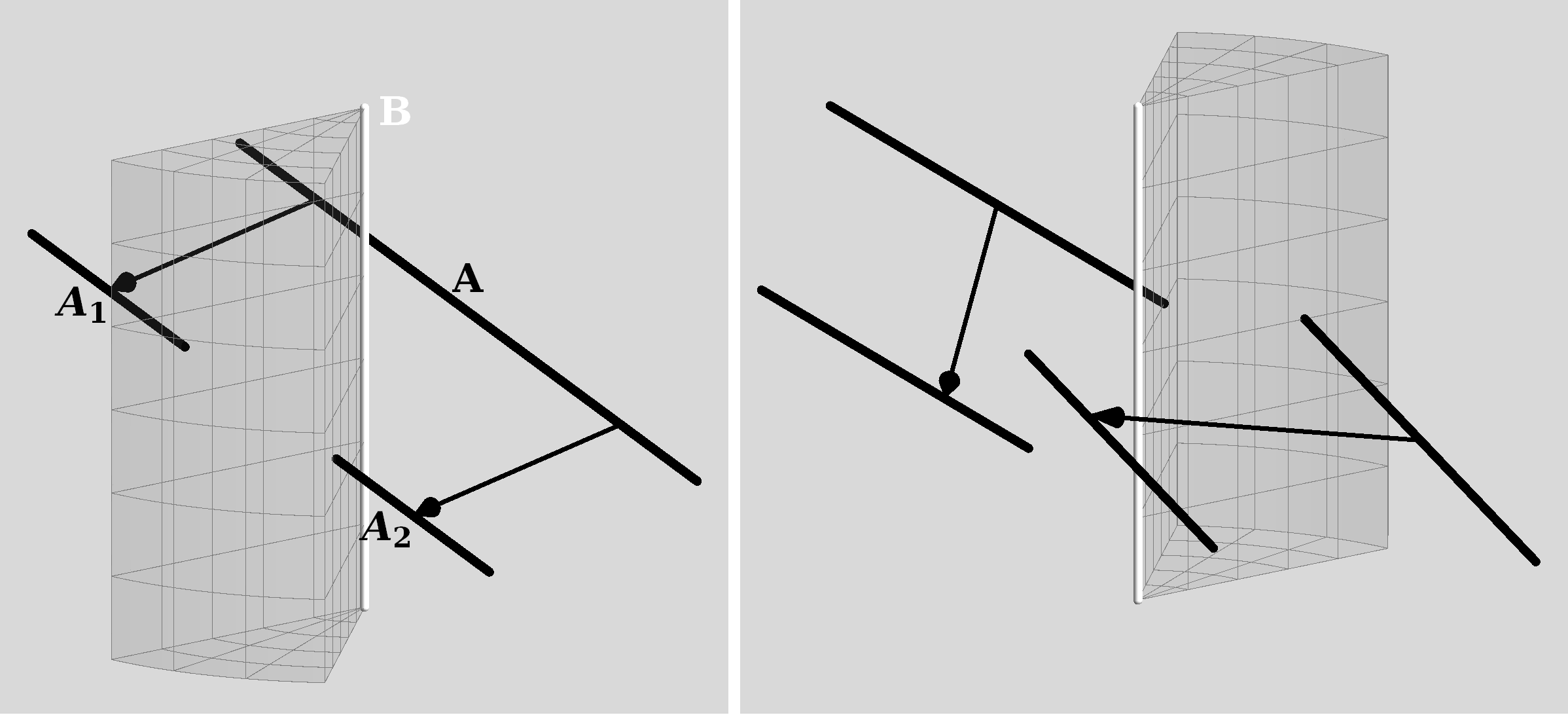}
\caption{A slanted collision. The semi-infinite defect lines $A_1$ and $A_2$ do not meet up after the collision. (Compare to figure \ref{fig:defectshallnotpass}.) }\label{fig:nonorthogonalcollision}
\end{figure}

When the defects $A$ and $B$ are not orthogonal at the collision point (i.e. $\phi\neq \pi/2$), then the semi-infinite defect lines $A_1$ and $A_2$ do not share a common junction after the collision (see figure \ref{fig:nonorthogonalcollision}). Consequently, a continuation with just one intermediate defect is not possible. We must therefore consider more complicated continuations involving multiple intermediate defects.

\begin{figure}[p]
\centering\includegraphics[width=7.5cm]{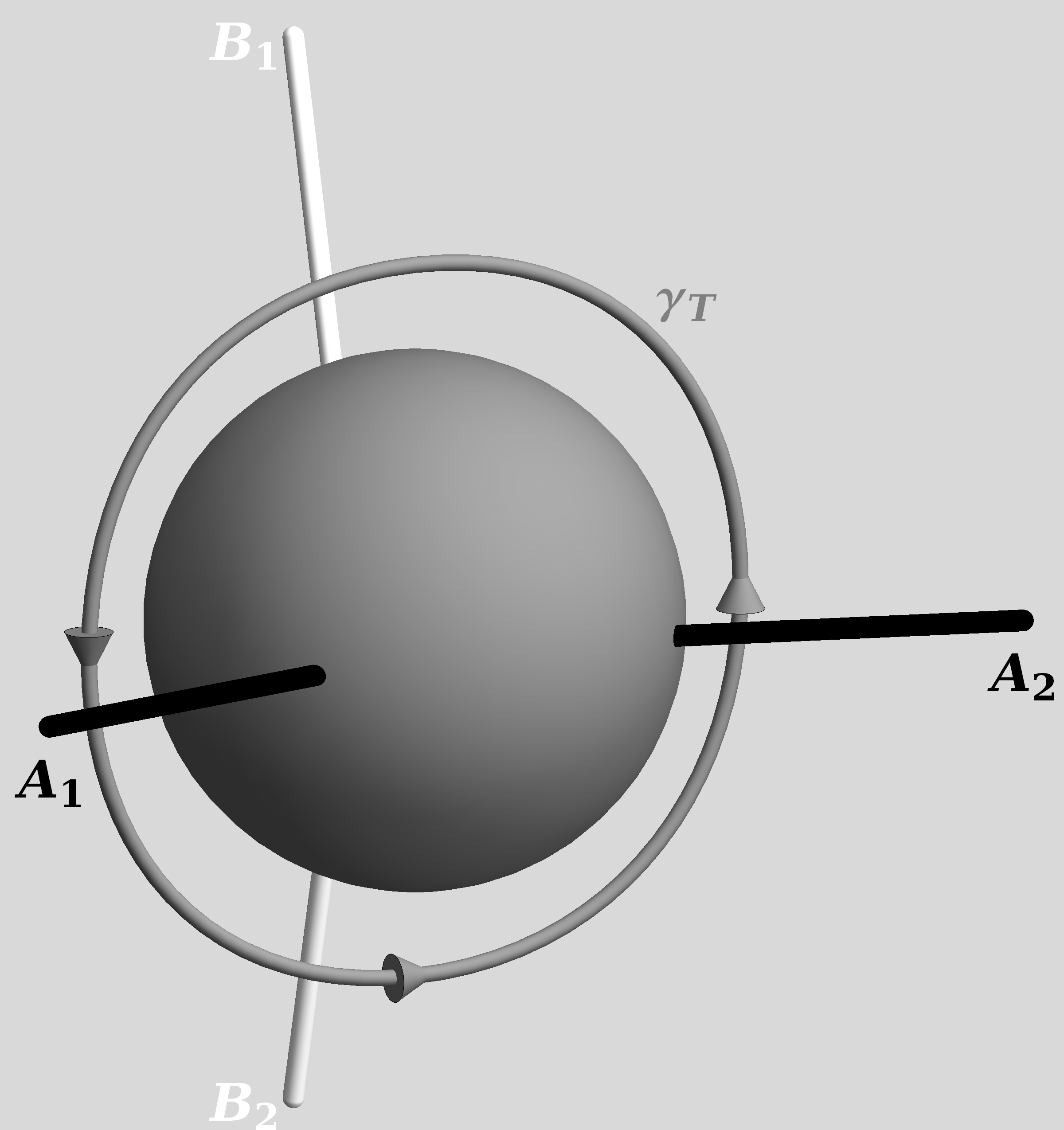}
\caption{A loop $\gamma_T$ transverse to the collision point.} \label{fig:transverseholonomy}
\end{figure}

Another way to see that a continuation with a single defect line is impossible is to calculate the holonomy, $Q_T$, of a loop transverse to the collision (see figure \ref{fig:transverseholonomy}). This is the holonomy of a defect that would connect the colliding defects. In the previous section we found that it is given by
\begin{equation}
Q_{T} = Q_{A_1}Q_{A_2}^{\m 1} = Q_A Q_B^{\m 1} Q_A^{\m 1} Q_B.
\end{equation}
For non-orthogonal $Q_A$ and $Q_B$ the trace of this holonomy is a complex number (we will spare the reader the actual expression). Consequently, it cannot be the holonomy of a single defect line.

\begin{figure}[p]
\centering\includegraphics[width=8cm]{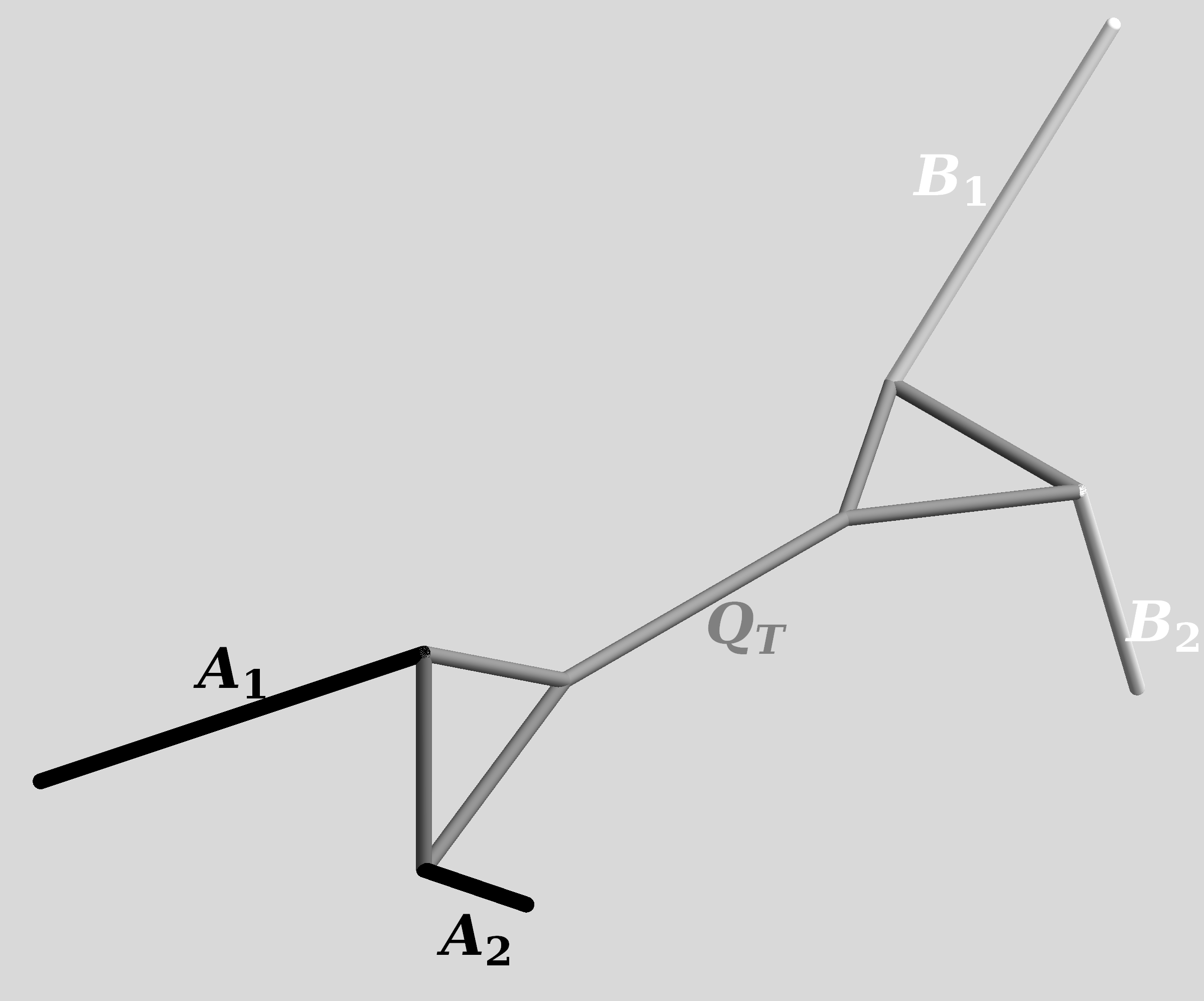}
\caption{Continuations like this with triangle resolved junctions are not allowed for non-orthogonal collision because the middle defect would still need to have the loxodromic holonomy $Q_T$.}\label{fig:triangleresolution}
\end{figure}

This result is, in fact, slightly stronger than the previous result. If the only problem was that the defects $A_1$ and $A_2$ did not meet up in a single junction we could have tried to resolve the issue by replacing the junctions with triangle loops such as depicted in figure \ref{fig:triangleresolution}. However, for such a continuation the single defect connecting the two triangles would still need to have the holonomy of the transverse loop $\gamma_T$. Such continuations are therefore also impossible. We need continuations that involve at least two intermediate defects running between the two original defects $A$ and $B$.

\section{Quadrangle continuations}\label{sec:quadsols}
\begin{figure}[bt]
\centering
\includegraphics[width=\textwidth]{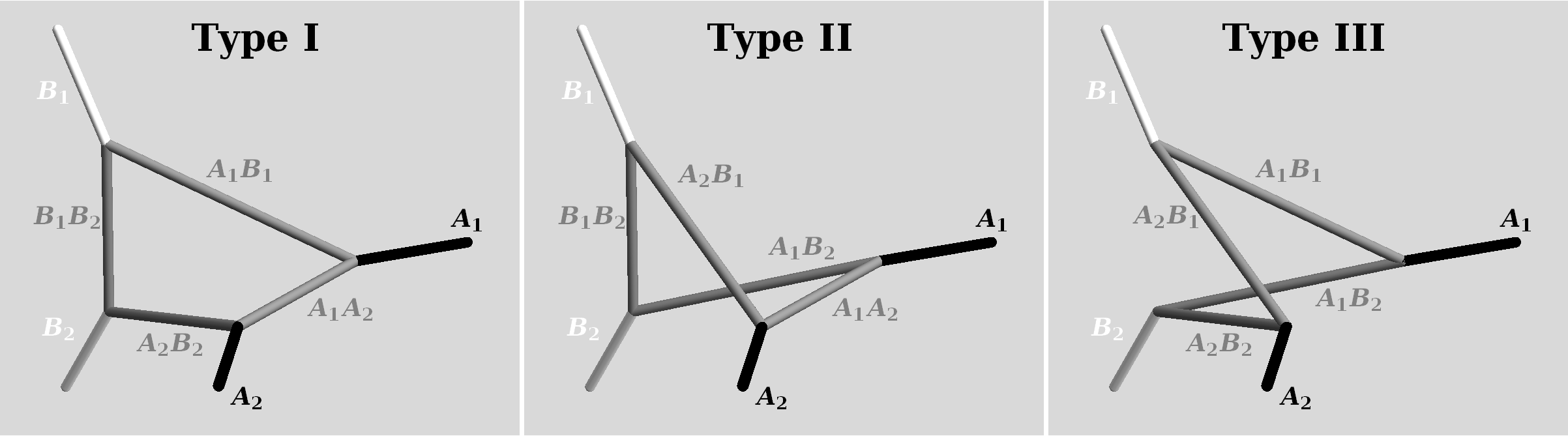}
\caption{There are three different types of quadrangle configurations that provide a continuation of a collision.}\label{fig:squares}
\end{figure}

The simplest possible continuation involving at least 2 defects connecting the original defects $A$ and $B$ is a square loop of four intermediate defects connecting the four ends of the semi-infinite defects $A_1$, $A_2$, $B_1$, and $B_2$. There are four topologically distinct loops of this type possible as shown in figure \ref{fig:squares}.

We choose loops around the external semi-infinite defects in such a way that equation \eqref{eq:orthcollisioncond2} relating their holonomies becomes
\begin{equation}\label{eq:extcond}
Q_{B_2}Q_{A_2}Q_{B_1}Q_{A_1}=\Id.
\end{equation}
This can be achieved by taking the loops as defined in fig \ref{fig:orthogonalcollision} and reversing the directions of the $\gamma_{A_2}$ and $\gamma_{B_2}$ loops. This choice is convenient in the sense that all loops are in the positive direction with respect to the outwards direction of each external semi-infinite defect.

A set of four intermediate defects introduces $4 \times 7 =28$ new degrees of freedom in the form of four holonomies. We are however not completely free to choose the parameters of the internal defects. First of all, all internal defects must pass through the collision point at the time of collision. This fixes the positions of the internal defects. Since we choose the collision point as the origin of our coordinate system, this means that the holonomies of the intermediate defects maybe  expressed as pure Lorentz transformations. This reduces the number of free parameters from 28 to $4 \times 5 = 20$.

The internal defects must also satisfy the junction conditions at each of the four junctions. For example, for a type III configuration (see figure \ref{fig:squares}) the junction conditions become (for a suitably chosen set of loops),
\begin{equation}
\begin{aligned}\label{eq:squarejunctions}
Q_{A_1B_1} &= Q_{A_1} Q_{A_1B_2}\quad & \quad
Q_{A_2B_1} &= Q_{B_1} Q_{A_1B_1} \\
Q_{A_2B_2} &= Q_{A_2} Q_{A_2B_1} &
Q_{A_1B_2} &= Q_{B_2} Q_{A_2B_2}.
\end{aligned}
\end{equation}
Each of these equations imposes 6 algebraic conditions on the parameters of the internal defects. The equations are however not independent. The condition \eqref{eq:extcond} implies that if any three of the above equations is satisfied, the remaining fourth equation is automatically satisfied as well. The junction conditions therefore provide $3\times 6 = 18$ independent algebraic conditions on the parameters of the internal defects. Consequently, we expect that for each type of quadrangle continuation, there is a two-dimensional set of configurations that satisfies all conditions.

In this section we will go through the steps necessary to find these configurations in terms of two parameters, the rapidity with which two of the junctions move along the external defect they connect to. As an explicit example we will take a type III configuration, because as it turns out the answer takes its simplest form in this case. Type I and II configurations can be solved by following the same steps, but the intermediate results are more involved and will not be shown here.

To do the calculations it is convenient to represent the holonomies by  $\PSL{2,\CC}$ matrices as is explained in appendix \ref{app:PSL2C}. This has a number of advantages. First, a $2\times2$ complex valued matrix has only 8 real parameters, as compared to the 16 real parameters of a $4\times4$ real valued matrix. This helps reduce the number of redundant equations. Second, the condition the holonomy of the defect lines needs to be rotation-like takes a particular easy form in terms of the trace.

As is discussed in appendix \ref{app:PSL2C}, a necessary and sufficient condition for a $\PSL{2,\CC}$ matrix, $Q$, to correspond to a rotationlike Lorentz transformation is
\begin{equation}\label{eq:tracecond}
\begin{aligned}
\im\tr Q &= 0 \\
\abs{\tr Q} &\leq 2.
\end{aligned}
\end{equation} 
So, in addition to the junction conditions \eqref{eq:squarejunctions}, the holonomies of the intermediate defects are also subject to the conditions
\begin{equation}
\begin{aligned}\label{eq:imtrace}
\im\tr Q_{A_1B_1} &= 0\quad & \quad \im\tr Q_{A_2B_1} &= 0 \\
\im\tr Q_{A_2B_2} &= 0 & \im\tr Q_{A_1B_2} &= 0,
\end{aligned}
\end{equation}
and
\begin{equation}
\begin{aligned}\label{eq:retrace}
\abs{\tr Q_{A_1B_1}} &\leq 2\quad & \quad \abs{\tr Q_{A_2B_1}} &\leq 2\\
\abs{\tr Q_{A_2B_2}} &\leq 2\quad & \quad \abs{\tr Q_{A_1B_2}} &\leq 2.
\end{aligned}
\end{equation}

We can parametrize the space of configurations by taking one of the intermediate holonomies ($Q_{A_1B_1}$ in a type III configuration) to be a general $\GL{2,\CC}$ matrix,
\begin{equation}
Q_{A_1B_1} = \begin{pmatrix}
a + \ii b & c + \ii d \\
g + \ii h & e + \ii f.
\end{pmatrix}
\end{equation}
The junction conditions \eqref{eq:squarejunctions} can then be used to obtain the holonomies of the other intermediate defects. Since the holonomies of the external defects all have determinant equal to one, the junction conditions ensure that if we require that $\det(Q_{A_1B_1})=1$, the determinants of the other intermediate holonomies are one as well.

The conditions \eqref{eq:imtrace}, provide four linear constraints on the eight real parameters. The inequalities \eqref{eq:retrace} then constrain the resulting four-dimensional parameter space to a hypercube. The complex condition $\det(Q_{A_1B_1})=1$ provides two further real constraints. These, however are quadratic in the parameters. Therefore, the question whether these are compatible with the linear inequalities \eqref{eq:retrace} is somewhat delicate.

The question can be simplified by taking a slightly different route. First we go to the frame where the external defect $B_1$ is stationary and oriented along the z-axis.\footnote{The requirement that $B_1$ is stationary does not uniquely define a frame, since there is the liberty to boost the frame in the z-direction. Here and onwards, when we say "the frame were defect $X$ is stationary", we mean the frame that can be reached from the center of velocity frame by a boost along the propagation direction of the incident defects.} In this frame,
\begin{equation}
Q_{B_1} =\begin{pmatrix}
\ee^{\ii\psi_B/2} & 0 \\
0 & \ee^{-\ii\psi_B/2}
\end{pmatrix}.
\end{equation}
The equation $\im\tr Q_{A_1B_1}=0$ implies that $f = -b$. The combination of the junction condition $Q_{A_2B_1} = Q_{B_1} Q_{A_1B_1}$ with the reality condition on the trace of $Q_{A_2B_1}$ then implies that
\begin{equation}
\begin{aligned}
0 &= \im\tr  Q_{B_2} Q_{A_2B_2}\\
&= \im(\ee^{\ii\psi_B/2}(a+\ii b)+ \ee^{-\ii\psi_B/2}(e-\ii b))\\
&= (a-e)\sin\psi_B/2, 
\end{aligned}
\end{equation}
from which we can infer that $e = a$. The condition that the determinant of $Q_{A_1B_1}$ is real then implies that,
\begin{equation}
Q_{A_1B_1} =\begin{pmatrix}
a+\ii b & \pm\ee^{-\mu}(c+\ii d) \\
-\ee^{\mu} (c-\ii d) & a-\ii b
\end{pmatrix}.
\end{equation}

In the case that the $\pm$ is a $+$, we can conjugate $Q_{A_1B_1}$ with a boost in the z-direction of rapidity $\mu$ to obtain a unitary matrix. That is, in the z-boosted frame both $Q_{B_1}$ and $Q_{A_1B_1}$ are pure rotations. Therefore in this frame both $B_1$ and $A_1B_1$ are stationary defects, consequently the junction of these defects is also stationary. The parameter $\mu$ is thereby identified as the rapidity of the $B_1$ junction along the $B_1$ defect, which we label $\mu_{B1}$.

In the other case where $\pm$ is a $-$, it is impossible to obtain a unitary matrix from $Q_{A_1B_1}$ by boosting it in the z-direction. It is therefore impossible to find a frame where the junction on the $B_1$ defect is stationary. The junction must consequently move at a superluminal velocity. Note that this does not necessarily mean that the defects $A_1B_1$ and $A_2B_1$ move at superluminal velocities.

We thus see that requiring that the junction moves at a subluminal velocity is a sufficient (although not necessary) condition for the incident holonomies to satisfy \eqref{eq:imtrace} and \eqref{eq:retrace}. By requiring that $Q_{A_1B_1}$ takes the form,
\begin{equation}\label{eq:staticform}
Q_{A_1\bar{B}_1} =\begin{pmatrix}
a+\ii b & \ee^{-\mu_{B1}}c\,\ee^{\ii\chi}) \\
-\ee^{\mu_{B1}}c\,\ee^{-\ii\chi} & a-\ii b
\end{pmatrix},
\end{equation}
where the bar on the $B_1$ indicates that the holonomy is observed in the frame where the $B_1$ defect is stationary,  we can impose the conditions \eqref{eq:imtrace} and \eqref{eq:retrace} for the two incident holonomies $Q_{A_1B_1}$ and $Q_{A_2B_1}$, and the requirement that determinant of the intermediate holonomies is real. If we change to a frame where the external defect diametrically opposite $B_1$ ($B_2$ in the case of a type III configuration) is stationary, we can require that one of the incident intermediate holomies $Q_{A_1B_2}$ or $Q_{A_2B_2}$ is of the form \eqref{eq:staticform} to enforce the other half of the conditions \eqref{eq:imtrace} and \eqref{eq:retrace}. This leaves only one equation, the requirement that the determinant is one. For a matrix of the form \eqref{eq:staticform}, this takes the simple form,
\begin{equation}
a^2 + b^2 + c^2 = 1.
\end{equation}

Enforcing all conditions will leave 2 free parameters. A logical choice would be the rapidities of the two diametrically opposite junctions, $\mu_{B1}$ and $\mu_{B2}$, since we already needed to require that these junctions move at subluminal velocities to ensure that the inequalities \eqref{eq:retrace} are met. However, as it turns out these parameters are not independent; fixing one automatically fixes the other. As far as this author has been able to find, there is no simple argument that explains why this should be, other than explicitly doing the calculations. This is what we will now do for a configuration of type III.

To do the calculations we must first choose frames for the external defects. What sets configurations of type III apart from the other two types is that the diametrically opposite junctions $B_1$ and $B_2$ lie on the two ends of what was a single defect before the collision. This allows us to choose a frame in which both ends are stationary simultaneously.

The holonomies of the external defects with this choice are
\begin{equation}\label{eq:extassign}
\begin{aligned}
Q_{B_1} &= R_z(\psi_B) = \begin{pmatrix}
\ee^{-\ii \psi_B/2} & 0 \\
0 & \ee^{\ii \psi_B/2}
\end{pmatrix}\\
Q_{B_2} &= R_z(-\psi_B) = \begin{pmatrix}
\ee^{\ii \psi_B/2} & 0 \\
0 & \ee^{-\ii \psi_B/2}
\end{pmatrix}\\
Q_{A_1} &= B_x(\eta)R_x(\phi)R_z(\psi_A)R_x(\phi)^{\m 1}B_x(\eta)^{\m 1}\\
 &= \begin{pmatrix}
\cos\tfrac{\psi_A}{2}-\ii\cosh(\eta+\ii\phi)\sin\tfrac{\psi_A}{2} & \ii\sinh(\eta+\ii\phi)\sin\tfrac{\psi_A}{2} \\
-\ii\sinh(\eta+\ii\phi)\sin\tfrac{\psi_A}{2} & \cos\tfrac{\psi_A}{2}+\ii\cosh(\eta+\ii\phi)\sin\tfrac{\psi_A}{2}
\end{pmatrix}\\
Q_{A_2} &= R_z(\psi_B)B_x(\eta)R_x(\phi)R_z(\psi_A)R_x(\phi)^{\m 1}B_x(\eta)^{\m 1}R_z(\psi_B)^{\m 1}\\
 &= \begin{pmatrix}
\cos\tfrac{\psi_A}{2}-\ii\cosh(\eta+\ii\phi)\sin\tfrac{\psi_A}{2} & -\ii\,\ee^{-\ii\psi_B}\sinh(\eta+\ii\phi)\sin\tfrac{\psi_A}{2} \\
\ii\,\ee^{\ii\psi_B}\sinh(\eta+\ii\phi)\sin\tfrac{\psi_A}{2} & \cos\tfrac{\psi_A}{2}+\ii\cosh(\eta+\ii\phi)\sin\tfrac{\psi_A}{2}
\end{pmatrix}.
\end{aligned}
\end{equation}
One can readily check that these holonomies indeed satisfy \eqref{eq:extcond}. We further choose 
\begin{equation}
Q_{A_1\bar{B}_1} =\begin{pmatrix}
a+\ii b & \ee^{-\mu_{B1}}c\,\ee^{\ii\chi}) \\
-\ee^{\mu_{B1}}c\,\ee^{-\ii\chi} & a-\ii b
\end{pmatrix},
\end{equation}
such that conditions \eqref{eq:imtrace} and \eqref{eq:retrace} are automatically satisfied for $Q_{A_1\bar{B}_1}$ and $Q_{A_2\bar{B}_1}$. We can then obtain $Q_{A_1B_2}$ by using \eqref{eq:squarejunctions};
\begin{equation}
Q_{A_1B_2} = Q_{A_1}^{-1} Q_{A_1B_1}.
\end{equation}
We now need to check that $Q_{A_1B_2}$ has the form \eqref{eq:staticform} in the frame where $B_2$ is stationary. Normally this would require a conjugation with an appropriate Lorentz transformation, but in the special frame we choose both $B_1$ and $B_2$ are stationary, so 
\begin{equation}
Q_{A_1\bar{B}_2} = Q_{A_1}^{-1} Q_{A_1\bar{B}_1}.
\end{equation}
Since $\det  Q_{A_1} =1$, the determinant of $Q_{A_1\bar{B}_2}$ is automatically real. Therefore, to ensure that $Q_{A_1\bar{B}_2}$ has the form \eqref{eq:staticform} it is enough to require that
\begin{equation}
Q_{A_1\bar{B}_2}[1,1]^* = Q_{A_1\bar{B}_2}[2,2],
\end{equation}
where $Q[i,j]$ denotes the entry of the matrix $Q$ at position $(i,j)$. This complex-valued equation, can be written as two linear real-valued equations for the real and imaginary part. These equations can be solved for $a$ and $b$,
\begin{equation}
\begin{aligned}
a &= c (\cos\chi\frac{\sinh\mu_{B1}}{\tanh{\eta}}-\sin\chi\frac{\cosh\mu_{B1}}{\tan{\phi}})\\
b &= c (\cos\chi\frac{\sinh\mu_{B1}}{\tanh{\eta}}+\sin\chi\frac{\cosh\mu_{B1}}{\tan{\phi}}).
\end{aligned}
\end{equation}
Substituting this in the equation for the determinant of $Q_{A_1B_1}$,
\begin{equation}
\begin{aligned}
\det Q_{A_1B_1} &= a^2 +b^2 +c^2\\
&= c^2 (1+\frac{\cosh^2 \mu_{B1}}{\tan^2\phi}+\frac{\sinh^2 \mu_{B1}}{\tanh^2\eta}),
\end{aligned}
\end{equation}
we find that setting the determinant to one fixes $c$ to be 
\begin{equation}
c = \frac{1}{\sqrt{1+\frac{\cosh^2 \mu_{B1}}{\tan^2\phi}+\frac{\sinh^2 \mu_{B1}}{\tanh^2\eta}}}.
\end{equation}
We now have two free parameters $\mu_{B1}$ and $\chi$. We would like to replace $\chi$ with $\mu_{B2}$. We can extract $\mu_{B2}$ from $Q_{A_1\bar{B}_2}$ by taking,
\begin{equation}
\begin{aligned}\label{eq:case3muB2}
\ee^{2\mu_{B2}} &= -\frac{Q_{A_1\bar{B}_2}[1,2]}{Q_{A_1\bar{B}_2}[2,1]^*}\\
&=\frac{(A+h)m_{B_1} + h(h A-1)}{(A-h)m_{B_1} - h(h A+1)},
\end{aligned}
\end{equation}
where
\begin{equation}
 A = \tan\frac{\psi_A}{2},\quad 
 h = \frac{\sinh\eta}{\sin\phi},\quad\text{and}\quad
 m_{B_1} = \tanh{\mu_{B1}}.
\end{equation}
The dependence on $\chi$ has completely dropped out; $\mu_{B2}$ is completely determined by $\mu_{B1}$ and the collision parameters $\eta$, $\phi$, and $\psi_A$. We can therefore not replace the free parameter $\chi$ with $\mu_{B2}$. If we want the second free parameter to be a junction rapidity, we will have to use the rapidity of the $A_1$ or $A_2$ junctions.

The rapidity $\mu_{A1}$ may be obtained by transforming $Q_{A_1B_1}$ to the frame where $A_1$ is stationary. From the assignment of the external holonomies \eqref{eq:extassign} we see that this may be achieved by conjugating with  $B_x(\eta)R_x(\phi)$,
\begin{equation}
Q_{\bar{A}_1B_1} = R_x(\phi)^{\m 1}B_x(\eta)^{\m 1}Q_{A_1\bar{B}_1}B_x(\eta)R_x(\phi).
\end{equation}
Extracting $\mu_{A1}$ and solving for $\chi$ we find,
\begin{equation}\label{eq:xsol1}
\tan\chi = -\frac{h}{m_{A_1}},
\end{equation}
where $m_{A_1}=\tanh{\mu_{A1}}$. The same steps may be repeated to obtain $\mu_{A2}$ from $Q_{\bar{A}_2B_1}$, and we find 
\begin{equation}\label{eq:xsol2}
\tan\chi = \frac{h-B m_{A_2}}{B h + m_{A_2}},
\end{equation}
where $m_{A2}=\tanh{\mu_{A2}}$. Combining expressions \eqref{eq:xsol1} and \eqref{eq:xsol2} we find 
\begin{equation}\label{eq:type3A1A2}
m_{A_1}m_{A_2} - \frac{h}{B}(m_{A_1}+m_{A_2}) -h^2=0.
\end{equation}
If we introduce $m_{B_2} = \tanh{\mu_{B2}}$, then equation \eqref{eq:case3muB2} can be rewritten to take a similar form,
\begin{equation}\label{eq:type3B1B2}
m_{B_1}m_{B_2} - \frac{h}{A}(m_{B_1}+m_{B_2}) -h^2=0.
\end{equation}
The space of quadrangle continuations of type III may be parametrized by two numbers, one of the pair $(m_{A_1}, m_{A_2})$ and another of the pair $(m_{B_1}, m_{B_2})$. Physically, these pairs of numbers represent the velocities of opposing junctions on the quadrangle in the center of velocity frame. If one of the velocities of a pair is known, then the other is given as a function of the collision parameters $\eta$, $\phi$, $\psi_A$ and $\psi_B$.

Similar relations between the velocities of opposing junctions may be found for configurations of type I and II. They may be obtained by following the same steps as outlined above for a type III configuration. The intermediate results are somewhat lengthy so we just give the final relations. For a configuration of type I one finds,
\begin{align}
\label{eq:type1A1B2}
m_{A_1}m_{B_2} 
+h\frac{\hat{h}A +B }{\hat{h}-A B}m_{A_1} 
+h\frac{A +\hat{h}B}{\hat{h}-A B} m_{B_2}
-h^2\frac{1-\hat{h} A B}{\hat{h}-A B} &=0\\
\label{eq:type1A2B1}
m_{A_2}m_{B_1} 
+h\frac{\hat{h}A +B }{\hat{h}-A B}m_{A_2} 
+h\frac{A +\hat{h}B }{\hat{h}-A B} m_{B_1}
-h^2\frac{1-\hat{h} A B}{\hat{h}-A B} &=0,
\end{align}
and for a type II configuration,
\begin{align}
\label{eq:type2A1B1}
m_{A_1}m_{B_1} 
+h\frac{\hat{h}A -B }{\hat{h}+A B}m_{A_1} 
-h\frac{A -\hat{h}B }{\hat{h}+A B} m_{B_1}
+h^2\frac{1+\hat{h} A B}{\hat{h}+A B} &=0\\
\label{eq:type2A2B2}
m_{A_2}m_{B_2} 
+h\frac{\hat{h}A -B }{\hat{h}+A B}m_{A_2} 
-h\frac{A -\hat{h}B}{\hat{h}+A B} m_{B_2}
+h^2\frac{1+\hat{h} A B}{\hat{h}+A B} &=0,
\end{align}
where in both sets of equations we used the shorthand, $\hat{h}=\cosh(\eta)/\cos(\phi)$.

Comparing the set of relations for each configuration, we notice a couple of things. First, all configuration types share the property that the velocities of opposing junctions on the quadrangle are related only to each other, and are independent of the velocities of the junctions on the other diagonal. This decoupling was already noticed by 't Hooft in \cite{hooft:2008} based on numerical calculations, and  here is confirmed analytically.

Second, as we already noted, configurations of type III are a bit special. Equation \eqref{eq:type3A1A2} relating the velocities of the $A_1$ and $A_2$ junctions, is completely independent of $\psi_A$, while equation  \eqref{eq:type3B1B2} relating the velocities of the $B_1$ and $B_2$ junctions, is completely independent of $\psi_B$. This property makes equations \eqref{eq:type3A1A2} and \eqref{eq:type3B1B2} take a very simple form, and is undoubtedly related to the fact that in a configuration of type III opposing junctions in the quadrangle lie on ends of what was originally one defect.

Third, in each set of equations --- \eqref{eq:type3A1A2} and \eqref{eq:type3B1B2}, \eqref{eq:type1A1B2} and \eqref{eq:type1A2B1}, and  \eqref{eq:type2A1B1} and \eqref{eq:type2A2B2} --- there is a simple substitution rule that relates one equation to the other. For example, if in equation \eqref{eq:type2A1B1} we replace $m_{A_1}$ with $m_{A_2}$, $m_{B_1}$ with $m_{B_2}$, we obtain equation \eqref{eq:type2A2B2}. 

Finally --- and most disturbingly --- for some values of the collision parameters the equations do not allow solutions where the velocities of  the junctions are all subluminal (i.e. where $m_i^2 < 1$ for all junctions). This can be seen by looking at the dependence on $h$. In each equation the $h^2$ term is independent of the  $m_i$'s. Since the $m_i$'s are bounded, the lower order $h$ terms, which do depend on the  $m_i$'s, cannot cancel the $h^2$ term for large values of $h$.

Consequently, for large enough values of $h$ (in relation to the other  collision parameters) there are no quadrangle solutions of any type with only subluminal junctions. Moreover, unlike the superluminal  junctions in the one defect continuation of an orthogonal collision, the superluminal junctions of the quadrangle continuation have two defects in their future (the new intermediate defects) and only one in its past (of the external defects.). In a frame where such a superluminal junction is contained in a single time slice, it will appear as if the colliding defect instantaneously splits in two at the moment of collision, even at far away points of the defect whose causal past did not include any hint of the impending collision. This is totally at odds with local causality principle at the base of our model. 

Recall that $h$ was defined as$\tfrac{\sinh\eta}{\sin\phi}$. A large value of $h$ is therefore associated with a large value of $\eta$ or a value of $\phi$ that is close to zero. So, as was the case for the one defect continuation of an orthogonal collision, it are the high velocity collisions that cannot be adequately continued by a quadrangle configuration of defects. In addition, we find that collisions where the orientations of the colliding defects are close to collinear, are also problematic.

\begin{figure}[tbp]
\centering\includegraphics[width=12cm]{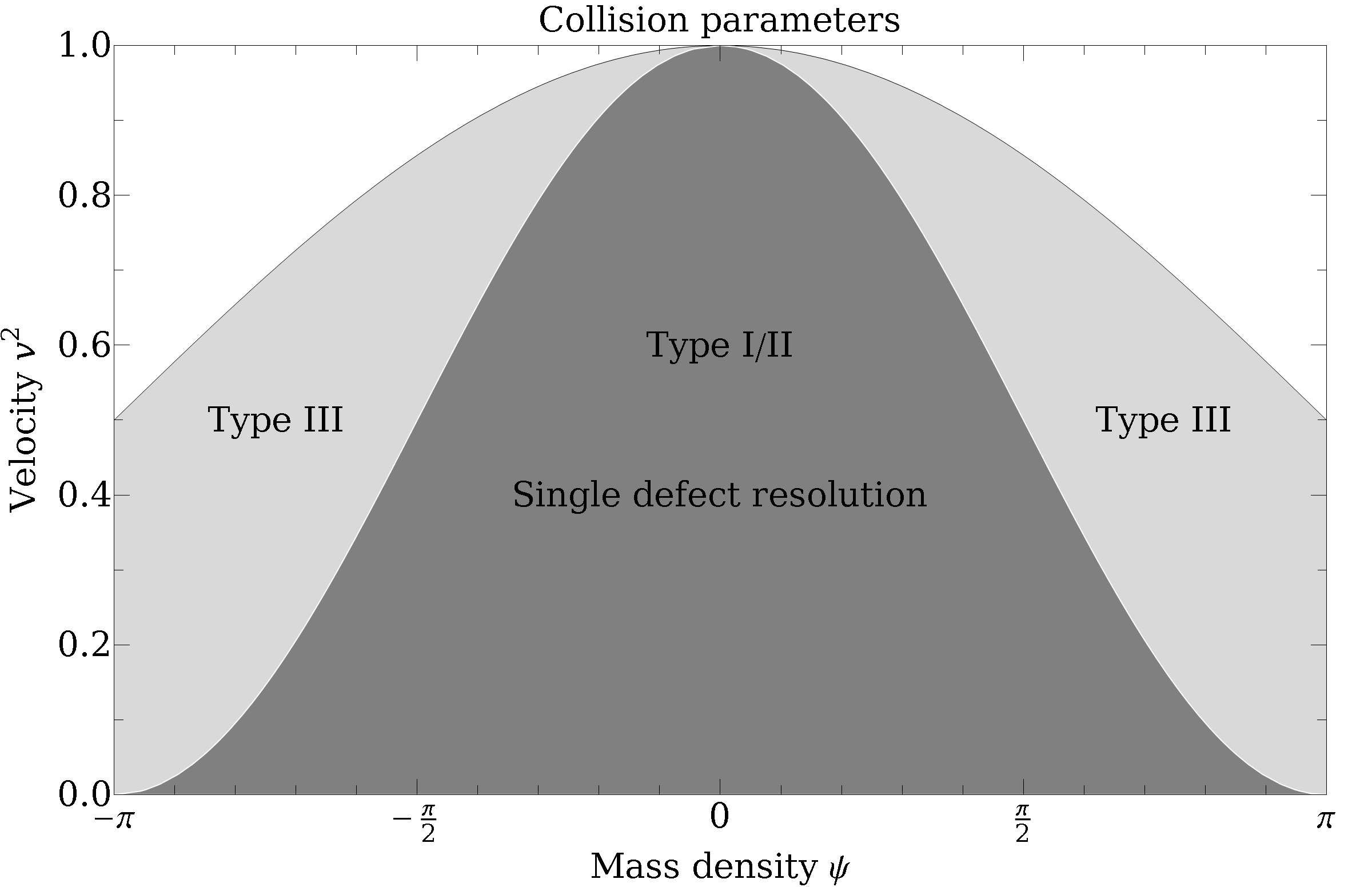}
\caption{The region of the parameter space of orthogonal collisions of defects with equal mass $\psi$, for which a continuation can be found consisting  of quadrangle or single defect configurations with only subluminal junctions.}\label{fig:junctionbounds}
\end{figure}

In the case of orthogonal collisions the quadrangle continuations are however an improvement over the single defect continuation we discussed in section \ref{sec:orthogonalcollisions}. Figure \ref{fig:junctionbounds} plots the regions of the collision parameter space for orthogonal collisions of defects with equal mass, that can be continued by the three types of quadrangle continuation or the single defect continuation. We see that quadrangle continuations of type I and II offer no improvement over the single defect continuation for orthogonal collisions. Quadrangle continuations of the III type with only subluminal junctions, however, can be found for a much larger range of collisions.

The appearance of superluminal junctions of the opening type is a serious challenge to the consistency of the principles of our piecewise flat model of gravity. It seems that the local causality principle should disallow this type of junction, as it disallows superluminal defects. We may still wonder however if allowing superluminal junctions would allow quadrangle continuations for all collisions. 

If we do not require all junctions to be timelike, we are no longer guaranteed that the intermediate defects are timelike as well. In the discussion above we saw that for a general collision we can only fix the velocities of two junctions that are adjacent on the quadrangle. If we fix these to be subluminal, then the three intermediate defects that are incident to these two junctions are timelike. For certain high velocity or nearly collinear collisions, the remaining two junctions will both be superluminal. Consequently there is no guarantee that the intermediate defect connecting these two junctions, is timelike. However, a finite defect connecting two spacelike junctions may still itself be timelike.

We want to know if there are collisions for which at least one of the defects of any quadrangle continuation is spacelike. It is to be expected that the ``best'' result is obtained when all junctions are allowed to become spacelike. The way to check whether a defect is timelike is to check whether the absolute value of the trace of its holonomy (in the $\PSL{2,\CC}$ representation) is smaller than 2.

\begin{figure}[tbp]
\centering\includegraphics[width=12cm]{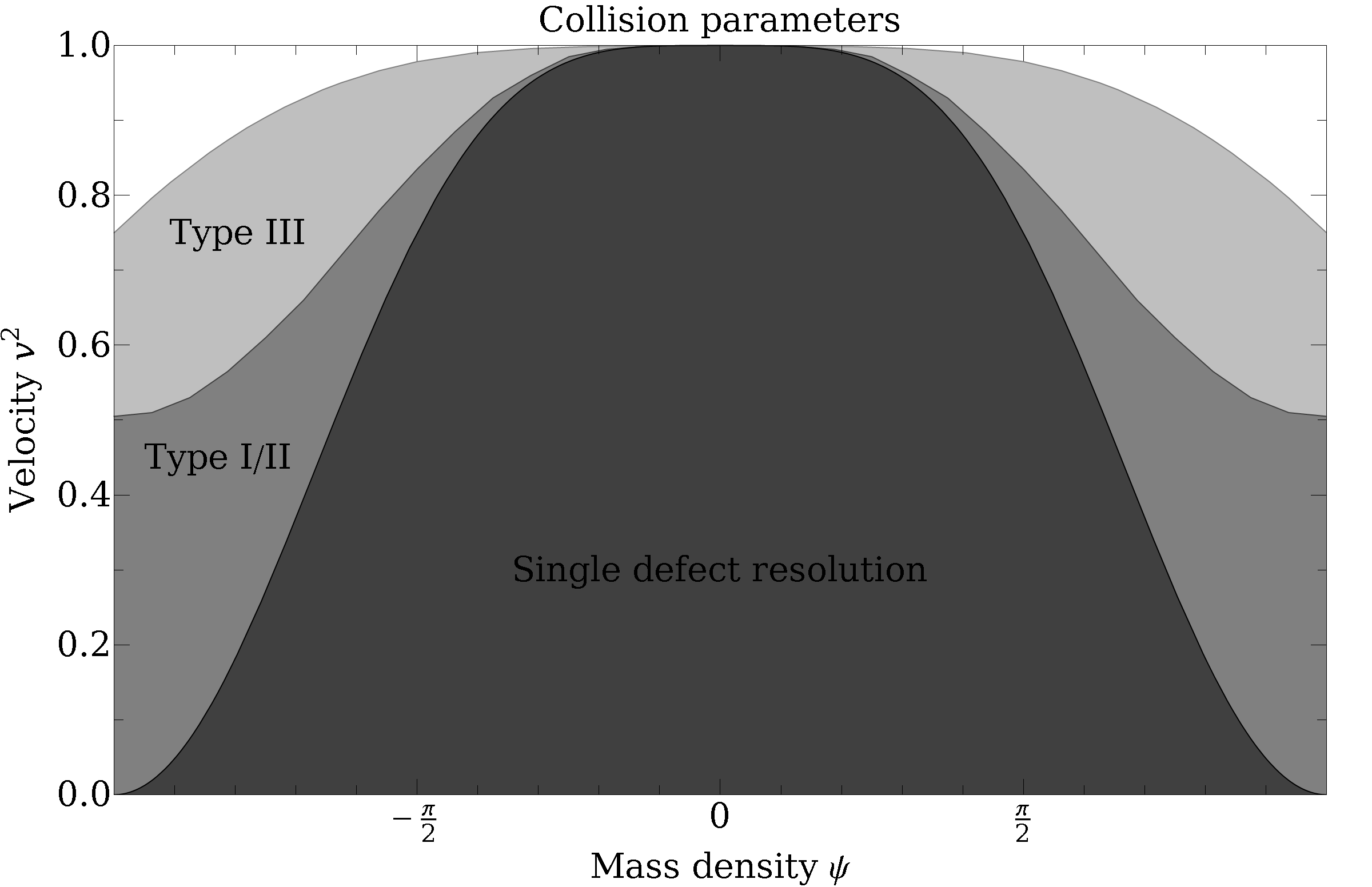}
\caption{The region of the parameter space of orthogonal collisions of defects with equal mass $\psi$, for which a continuation can be found consisting of a  quadrangle or single defect configuration with only non-spacelike defects.}\label{fig:tracebounds}
\end{figure}

Consequently, to test if a continuation with only timelike defects for a certain collision is possible, we need to calculate the holonomies of the intermediate defects and check if there exist values in the two dimensional parameter space of quadrangle continuations (of a fixed type) where the absolute value of each trace is smaller than two. This problem does not seem to be tractable analytically. It can however be checked numerically. Figure \ref{fig:tracebounds} shows the result for orthogonal collisions of defects with equal mass density. 

Compared to the single defect configuration the quadrangle configurations allow continuation of a significantly larger portion of the parameter space of collisions. The quadrangle continuations of type III seem most effective in finding continuations for orthogonal collisions with a high relative velocity.\footnote{Type I and II quadrangle configurations are more suited for continuing slanted collisions. For collisions that are close to collinear these types will be most effective.} However, for any non-zero value of the mass density of the defects, there exist velocities for which none of the types of quadrangle continuations has only timelike defects.

So, even if we allow superluminal junctions, quadrangle configurations cannot be used to continue all collisions within our model. We do however observe that for orthogonal collisions, quadrangle continuations provide a substantial improvement over the single defect continuation considered in section \ref{sec:orthogonalcollisions}. This suggests that similar improvements may be made by considering more complicated continuations. It could be that one can find a continuation for any collision by considering an arbitrarily complicated intermediate configurations. To test this we consider a more complicated type of continuation in the next section.

\section{Tetrahedral continuations}
\begin{figure}[tb]
\centering\includegraphics[width=100mm]{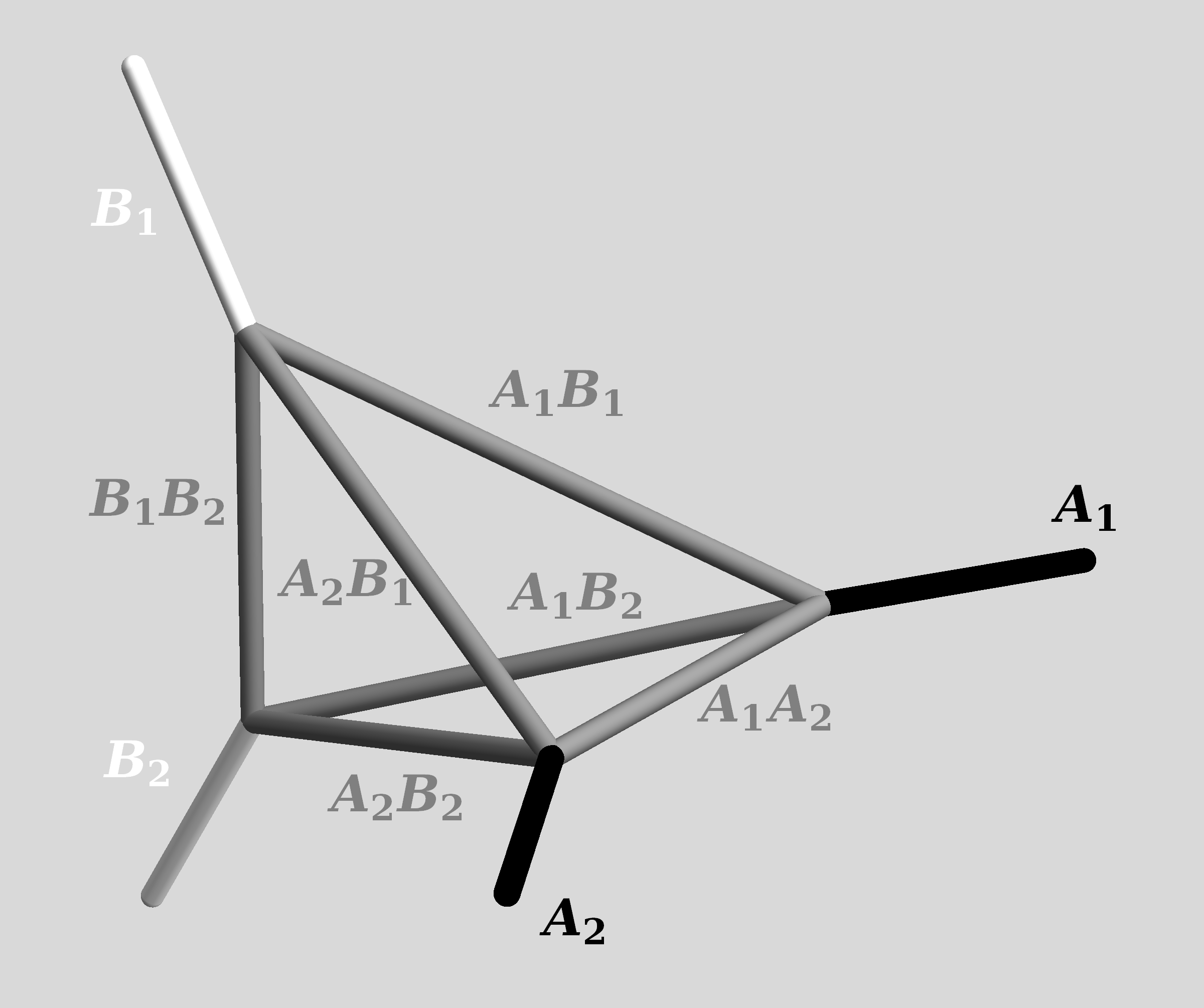}
\caption{Connecting all the vertices yields a tetrahedral configuration of internal strings.}\label{fig:tetrahedron}
\end{figure}

The quadrangle configurations considered in the previous section basically consisted of four vertices moving away from the collision  along the external semi-infinite defects, which were connected by four internal finite defects. This meant we needed to choose which vertices should be connected by finite defects with the three possible distinct choices leading to the three possible types of quadrangle continuation. We will now consider the more general case where each vertex is connected to each other vertex. The result is a tetrahedral configuration of finite defects as shown in figure \ref{fig:tetrahedron}.

The quadrangle continuations of type I, II, and III can be considered to be special cases of this configuration where the deficit angles of two of the internal finite defects become zero. For example, a quadrangle continuation of type III is obtained by setting the deficit angles of the defects labelled $A_1A_2$ and $B_1B_2$ in figure \ref{fig:tetrahedron} to zero.

The six new internal defects originating from the collision point give $6\cdot 5=30$ degrees of freedom.\footnote{Since the defects originate from collision their positional degrees of freedom are fixed by the fact that they must pass through the collision point.} The algebraic junction conditions \eqref{eq:junctioncondition} provide six conditions for each vertex. These are however not all independent. Since the holonomies of the external semi-infinite defects are related through \eqref{eq:orthcollisioncond2} the  junction conditions for the four junctions give 18 independent conditions on the internal defects. The geometric condition that all defects incident to junction must share that junction as a common line, provides a further 2 conditions per 4-valent vertex. Consequently, we expect that a tetrahedral configuration of internal defects has four free parameters.

This agrees with our above observation that we can obtain the quadrangle configurations (each having two free parameters) by setting two deficit angles to zero. In the case of the quadrangle configurations we took two of the velocities of the vertices along the external defects as our free parameters. Since we have four vertices moving along four external defects and four free parameters it is tempting to take these velocities (or their corresponding rapidities) as our free parameters. This choice makes it rather easy to explicitly impose that all vertices are subluminal.

\subsection{Newtonian limit}
Solving the system of 26 conditions on the 30 degrees of freedom of the internal defects is rather complex. It is therefore instructive to first solve these conditions in the Newtonian limit, i.e. the limit that the velocities of all defects are much smaller than the speed of light and all deficit angles (linear mass densities) are much smaller than one.

For low velocities a Lorentz transformation can be approximated by a Galilean transformation. That is, represented as a $4\times 4$ matrix, a Lorentz transformation $Q$ may be written,
\begin{equation}
Q \approx
\begin{pmatrix}
1 & 0 \\
\vec{v} & R(\vec{\phi})
\end{pmatrix},
\end{equation}
where $\vec{v}\in \RR^3$ is the boost velocity and $R(\vec{\phi})\in SO(3)$ is a rotation of $\abs{\vec\phi}$ degrees around the $\vec\phi$ axis. So the holonomy $Q(\vec\phi,\vec{v})$ of a line defect with orientation $\vec\phi$ and deficit angle $\norm{\vec\phi}$ moving with velocity $\vec{v}$ is given by
\begin{align}
Q(\vec\phi,\vec{v}) &\approx 
\begin{pmatrix}
1 & 0 \\
-\vec{v} &\Id
\end{pmatrix}
\begin{pmatrix}
1 & 0 \\
0 & R(\vec{\phi})
\end{pmatrix}
\begin{pmatrix}
1 & 0 \\
\vec{v} & \Id
\end{pmatrix}\\
&\approx \begin{pmatrix}
1 & 0 \\
(\Id-R(\vec{\phi}))\vec{v} & R(\vec{\phi})
\end{pmatrix}\\
&\approx \begin{pmatrix}
1 & 0 \\
-\vec{\phi}\times\vec{v} & R(\vec{\phi})
\end{pmatrix},
\end{align}
where in the last line we used that for small $\norm{\vec\phi}$,  $R(\vec{\phi})\cdot\vec{x} \approx \Id +\vec\phi\times\vec{x}$.

The junction conditions \eqref{eq:junctioncondition} therefore become
\begin{align}
\Id &= Q_1 \cdots Q_n \\
&\approx
\begin{pmatrix}
1 & 0 \\
-\vec{\phi}_1\times\vec{v}_1 & R(\vec{\phi}_1)
\end{pmatrix}
\cdots
\begin{pmatrix}
1 & 0 \\
-\vec{\phi}_n\times\vec{v}_n & R(\vec{\phi}_n)
\end{pmatrix}\\
&\approx \begin{pmatrix}
1 & 0 \\
-\vec{\phi}_1\times\vec{v}_1-\ldots-\vec{\phi}_n\times\vec{v}_n & R(\vec{\phi}_1+\dots+\vec{\phi}_n)
\end{pmatrix}.
\end{align}
Consequently, in the Newtonian limit the junction conditions become
\begin{align}
 \vec{\phi}_1+\dots+\vec{\phi}_n &= 0,\quad\mathrm{and}\label{eq:forcecond}\\
\vec{\phi}_1\times\vec{v}_1+\ldots+\vec{\phi}_n\times\vec{v}_n &= 0.\label{eq:torquecond}
\end{align}

Now consider a tetrahedral configuration of defects as in figure \ref{fig:tetrahedron}, and let $m_i$ be the velocity of the vertex $i$ along external defect $i$ with orientation $\vec{\phi}_i$ and velocity $\vec{v}_i$. Since the velocities are non-relativistic the total velocity of the junction $i$, $\vec{w}_i$,  is then given by 
\begin{equation}
 \vec{w}_i = \vec{v_i} + m_i\frac{\vec{\phi}_i}{\norm{\vec{\phi}_i}}.
\end{equation}
From the velocities of the junctions we can obtain the orientation $\hat{\phi}_{i,j}$ and velocities $\vec{v}_{i,j}$ of the defect connecting junction $i$ and $j$ as follows,
\begin{align}
 \hat{\phi}_{i,j} &= \frac{\vec{w}_i-\vec{w}_j}{\norm{\vec{w}_i-\vec{w}_j}}, \quad\mathrm{and}\\
\vec{v}_{i,j} &= \frac{\vec{w}_i+\vec{w}_j}{2}.
\end{align}
We have now solved 24 of the 30 degrees of freedom of six internal defects in terms of the velocities of the junctions along the external defects $m_i$. 
The  remaining unknown parameters are the deficit angles $\alpha_{i,j}$ of the internal defects. Equation \ref{eq:forcecond} tells us that we can find the $\alpha_{i,j}$ by decomposing  the vector $-\vec{\phi}_j$ on the base formed by the orientations $\hat\phi_{i,j}$ of the three internal defects connected to the external defect $i$. Equation \eqref{eq:torquecond} is automatically satisfied as well.

Since we independently obtain values for $\alpha_{i,j}$ and $\alpha_{j,i}$ it may appear that the answer is overdetermined. However, up till now we have ignored the relation \eqref{eq:extcond} for the external defects. These give two relations among the external defect parameters
 \begin{align}
 \vec{\phi}_{A_1}+\vec{\phi}_{A_2}+\vec{\phi}_{B_1}+\vec{\phi}_{B_2}&= 0,\quad\mathrm{and}\\
\vec{\phi}_{A_1}\times\vec{v}_{A_1} + \vec{\phi}_{A_2}\times\vec{v}_{A_2} + \vec{\phi}_{B_1}\times\vec{v}_{B_1} + \vec{\phi}_{B_2}\times\vec{v}_{B_2} &= 0.
\end{align}
It is a straightforward exercise in vector algebra to show these conditions imply that $\alpha_{i,j}= -\alpha_{j,i}$. The deficit angle of each internal defect is uniquely determined.

We find that --- in the Newtonian limit --- fixing the velocities of the junctions along the external defects indeed fixes all the internal parameters of a tetrahedral configuration. We can thus use these velocities as the four free parameters of the tetrahedral continuation. It is expected that this is true as well in the exact case, at least for small values of $\psi_A$, $\psi_B$, and $\eta$. For larger values of these parameters one may expect that not all choices for the velocities of the junctions yield valid configurations. Ultimately, it may even happen that for some values of the collision parameters there are no valid configurations for subluminal choices of these velocities, as we will see.

\subsection{Exact piecewise flat continuations}
\label{sec:cellresolution}
In the preceding sections we have studied the single defect and quadrangle continuations using the algebraic relations given by the holonomies of the defects. In the case of a tetrahedral continuation the algebraic relations between the 30 (or 36 if you start out with general Lorentz transformations) free parameters become very complex. Already in the case of the quadrangle continuations, solving the relations required significant assistance from computer algebra software like Mathematica. Despite some serious attempts to directly solve these algebraic relations, that approach has proven impractical.

The source of the complications here (in part) lies in the fact that the junction conditions at one junction influence the conditions at other junctions in a non-trivial way. The non-local nature of the algebraic description of the defects in terms of their holonomy really starts to become an hindrance in solving the relations. In section \ref{sec:PWFmanifolds} we formulated an alternative way of describing a configuration using the piecewise flat geometry that it generated. The advantage of that approach was that it encoded the geometry in an essentially local way. In this section we will use this approach to find a general tetrahedral continuation of a collision of defects.

\begin{figure}[p]
\centering\includegraphics[width=0.99\textwidth]{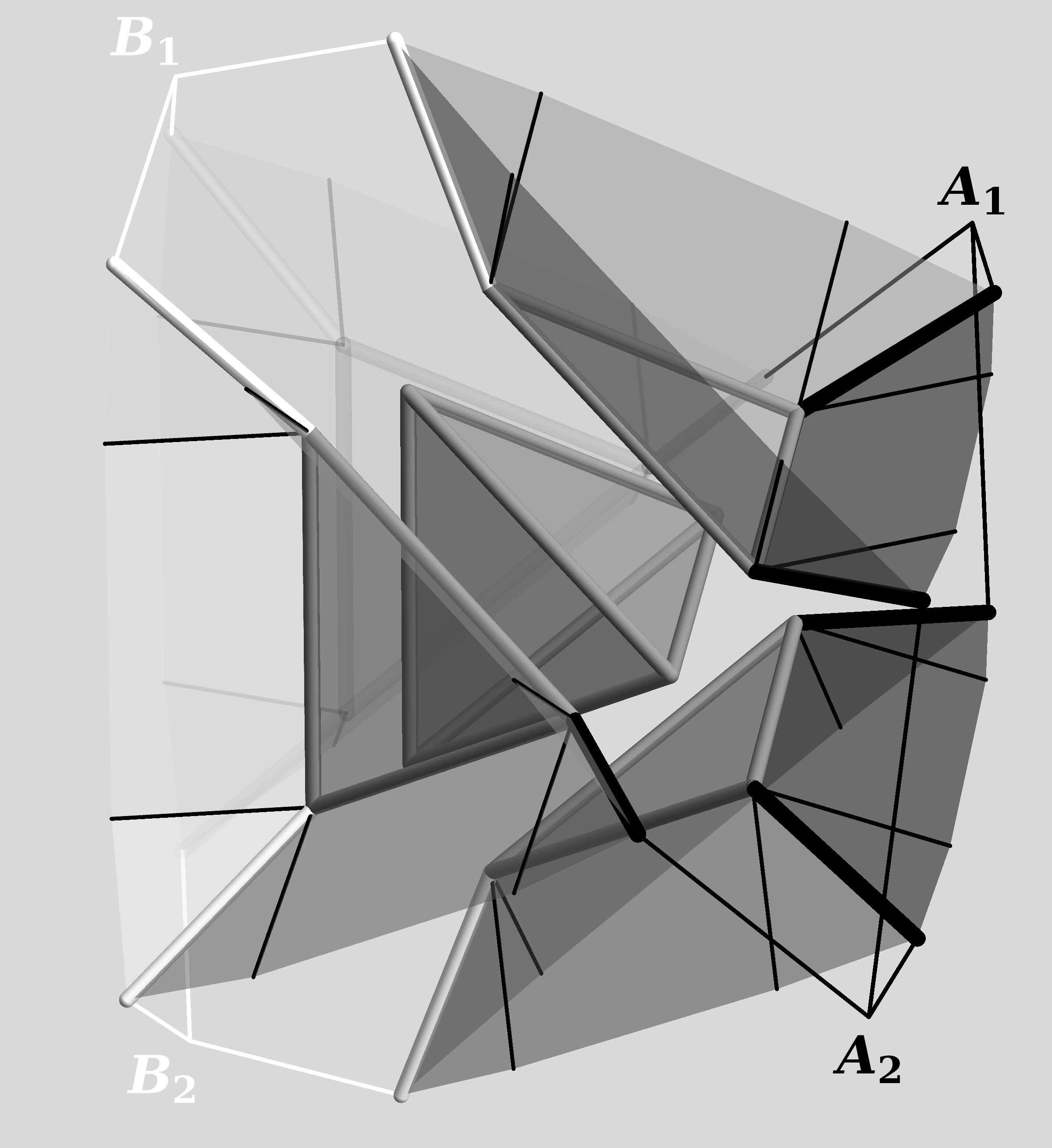}
\caption{The tetrahedral configuration can be divided in five $4$-cells. Here shown on a time slice after the collision, such that each $4$-cell is represented by a $3$-dimensional cell. There are four external cells extending to infinity, and one internal tetrahedral cell. Surfaces (representing $3$-cells) with matching colour are mapped into each other through a Lorentz transformation. The black lines on the external surfaces are virtual defects that subdivide the external surfaces and carry no holonomy. These are necessary because the external defects generically do not lie in the same plane.} \label{fig:tetracells}
\end{figure}

The general idea will be to split the geometry in five 4-cells, as shown in figure \ref{fig:tetracells}. One 4-cell outside of each face of the tetrahedron, and a fifth 4-cell as the inside of the tetrahedron.

Far away from the collision (well outside its lightcone) the geometry of the spacetime will be unaffected by the collision and will simply evolve linearly from the geometry before the collision. If we cut out the inner region containing the collision and its continuation, then this outer region contains four semi-infinite defects, each a half of one of the colliding defects. We will use these defects to divide the outer region in four (half open) 4-cells.

The semi-infinite external defects will become (half open) 2-cells in the boundary of these outer 4-cells. For this we first need to choose an inner boundary for these 2-cells. For each external defect we choose a line contained in the defect passing through the collision point. This line will become the 1-cell boundary of the half-open 2-cell formed by the external defect. It will also serve as one of the junctions of the tetrahedral configuration.

Each choice of line corresponds to one free parameter given by the velocity of the junction moving outwards along the external defect. The lines need to be chosen to move away from the junction fast enough such that the geometry in their vicinity can be obtained as a linear continuation from the geometry before the collision.\footnote{We will later discuss what restrictions this puts on the construction.}
 
Now, to construct one of the outer 4-cells we choose three external defects. Since the junctions on these defects all pass through the collision point, they lie in a common hyperplane. The convex hull of the junctions in that hyperplane will form the 3-cell at the inner boundary of the 4-cell. The boundary of this 3-cell is formed by three 2-cells, each spanned by two of the junctions. These 2-cells will become the new intermediate defects in the tetrahedral configuration.

A pair of external semi-infinite defects does not, in general, lie in one hyperplane. We can therefore not connect each pair of external 2-cells by a single 3-cell. Instead we need to choose a number of extra 2-cells starting at each junction, such that we can define 3-cells that form the boundary between the external 4-cells, as shown in figure  \ref{fig:tetracells}.
 
The spacetime geometry of the 4-cell and its boundary can be found by linearly continuing  the geometry from before the collision. Doing this for each set of three external defects, we obtain the four external 4-cells complete with a flat metric on each boundary cell. Consequently, we have the geometry of the entire 3-skeleton of the tetrahedral configuration.

The missing piece to be added is a 4-cell with a piecewise flat metric, which is to be attached to the boundary formed by the four interior boundary 3-cells of the external 4-cells. The easiest way to proceed is to observe that if we choose a common constant time slicing on the exterior cells --- this can be done, for example, by continuing a constant time slicing from the center of velocity frame --- then the geometry of the boundaries of the internal 4-cell on any two time slices are congruent to each other up to a scaling factor depending linearly on time. It is therefore enough to find the geometry of one time slice of the internal 4-cell.

The problem of filling a tetrahedron with an internal flat metric given a flat metric on the boundary is well-known in Euclidean geometry. It is known that this is possible if and only if the boundary metrics satisfy the (generalized) triangle inequalities.\cite{Schoenberg1935} That is, if and only if the sum of the areas of any three of the boundary triangles is larger than the area of the remaining triangle.

A priori, it is not clear that these triangle inequalities will be satisfied for any choice of the collision parameters and junction velocities. However, even if it is not satisfied, it is still rather easy to construct a piecewise flat metric on the interior of the tetrahedron, as we will show presently.

\begin{figure}[tb]
\centering\includegraphics[width=70mm]{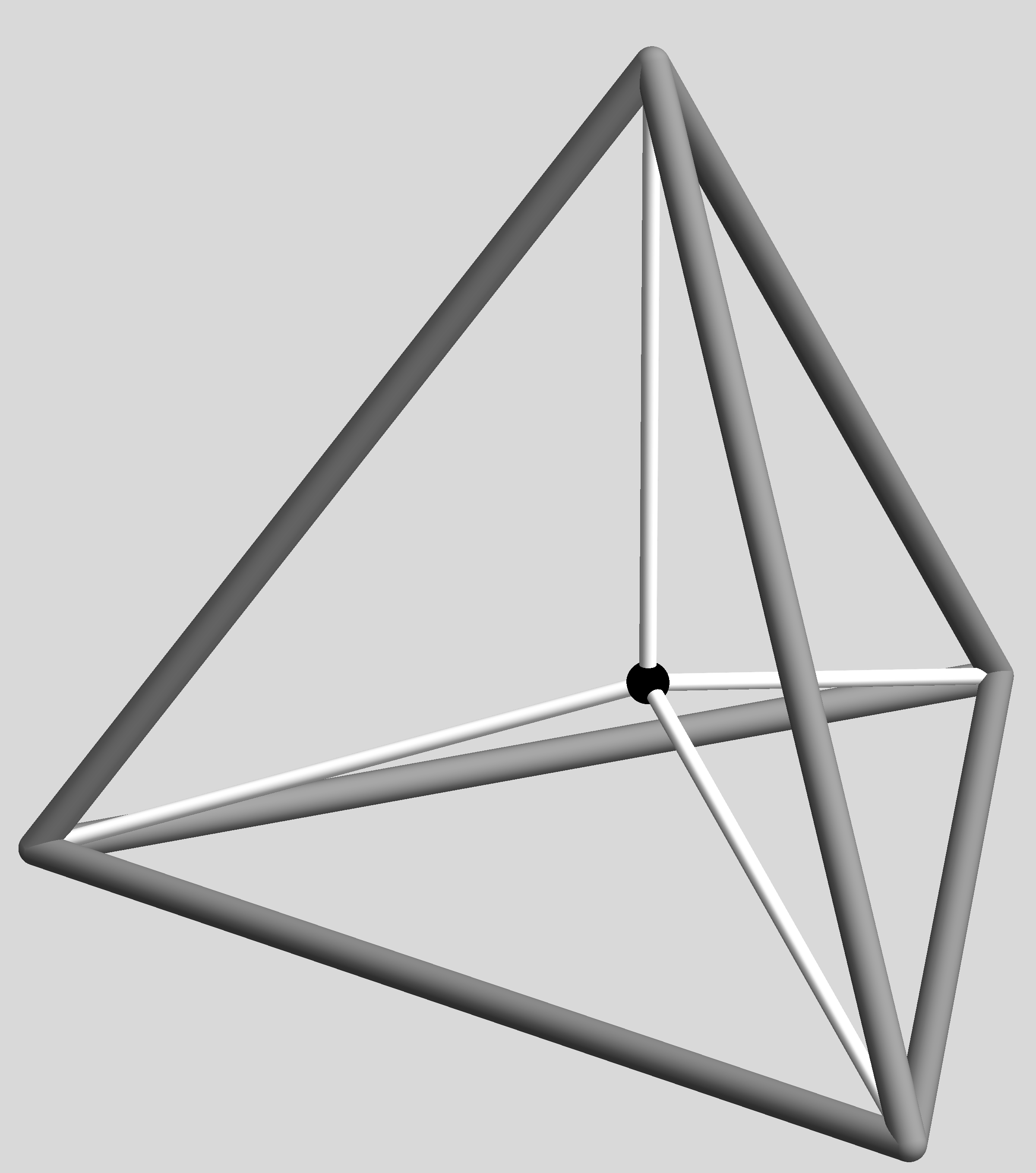}
\caption{A tetrahedron can be subdivided into four smaller tetrahedra by adding a single vertex, four 1-cells, six 2-cells, and four 3-cells.}\label{fig:subdivide}
\end{figure}

To construct an internal piecewise flat metric in the tetrahedron, we will subdivide it in four pieces (see figure \ref{fig:subdivide}). We build a cell complex (in fact a simplicial complex) on the interior in the following way. We add a single vertex, which we connect to each of the four vertices on the boundary by 1-cells. Each of the triangles formed by two of the new internal 1-cells and one of the 1-cells on the boundary is filled by a new 2-cell, and each of the tertahedra formed by three of the new 2-cells and one of the boundary 2-cells is filled by a new 3-cell.

For each of the new cells we will have to specify a flat metric that is compatible with the metrics on its boundary. Any metric on a 1-cell is flat and is specified by a single parameter; its length. Consequently, the four new internal 1-cells give us four free parameters. To construct a flat metric on a 2-cell we need that the metrics on its boundary 1-cells satisfy the (2D) triangle inequalities. This can easily be satisfied for all the new 2-cells. One particular way is to choose all four lengths equal and larger than the length of the longest 1-cell on the boundary.\footnote{This obviously is not the most conservative choice one could make, but it is convenient for the rest of the argument.}
 
To construct a flat metric on a 3-cell, we again need the metrics on its boundary 2-cells to satisfy the (3D) triangle inequalities. The choice above guarantees that three of the four inequalities are satisfied. 

To see this, call the length of the new internal 1-cells $a$ and the length of the three 1-cells on the outer boundary of any particular internal 3-cell $b_1$, $b_2$, and $b_3$. Because they are on the boundary of a given triangle they satisfy
\begin{equation}
 b_i + b_j \geq b_k,
\end{equation}
with $i,j,k\in \set{1,2,3}$, and without loss of generality we can take the labelling such that $b_1 \leq b_2 \leq b_3$. The area $A_i$ of the internal 2-cell incident to the 1-cell with length $b_i$ is thus equal to
\begin{equation}
A_i = \frac{1}{2}b_i\sqrt{a^2- \frac{1}{4}b_i^2}.
\end{equation}
Since $a> b_i$ for all $i$, $b_1 \leq b_2 \leq b_3$ implies that $A_1 \leq A_2 \leq A_3$. Furthermore we have
\begin{align}
 A_1 + A_2 &=  \frac{1}{2}b_1\sqrt{a^2- \frac{1}{4}b_1^2} + \frac{1}{2}b_2\sqrt{a^2- \frac{1}{4}b_2^2} \\
  &\geq \frac{1}{2}(b_1+b_2)\sqrt{a^2- \frac{1}{4}b_3^2}\\
  &\geq \frac{1}{2}b_3\sqrt{a^2- \frac{1}{4}b_3^2} \\
  &\geq A_3.
\end{align}
As a result any sum involving two of the areas of the internal 2-cells will be larger than the area of the remaining 2-cell.

The remaining triangle inequality is that the sum of the areas of the internal 2-cells is larger  than the area of the boundary 2-cell. Since the areas of the internal 2-cells can be made arbitrarily large by increasing $a$, it is possible to also satisfy this fourth triangle inequality for the new 3-cells.\footnote{A sufficient condition for this is that $a$ is larger than the radius of the circumscribed circle of each of the boundary triangles.}

Hence we can construct a piecewise flat metric on the interior of the tetrahedron. Consequently, the geometry of the missing 4-cell  can be taken to be such a piecewise flat metric expanding linearly with time. That is, we can expand our original cell complex by adding four new internal 4-cells each with the geometry of a tetrahedron growing linearly with time (each corresponding to one of the internal 3-cells in our construction of the piecewise flat metric).

The new 1-cells in the construction of the piecewise flat metric, correspond to four new internal 2-cells that, generically, carry non-trivial holonomy. These must therefore be interpreted as new internal defects.

\subsection{Limitations}\label{sec:limits}
In the last section we have shown how to construct a tetrahedral continuation of a collision of defects using (piecewise flat) geometrical techniques. The construction involved the choice of the outward trajectory of the four junctions, providing four new free parameters as expected.\footnote{And possibly four additional parameters for the piecewise flat metric inside the tetrahedron.} The construction, however, does not automatically guarantee that the junctions and the new internal defects are all timelike.

The velocities of the junctions are in principle free parameters. If we choose all junction to be timelike, then, since the junctions are the boundaries of the internal defects, the internal defects are timelike as well. However, our experience with the quadrangle continuations has shown that such a choice may not always be available. We therefore need to examine what the available range of possible velocities for the junctions is.

\begin{figure}[p]
\centering
\includegraphics[width=0.95\textwidth]{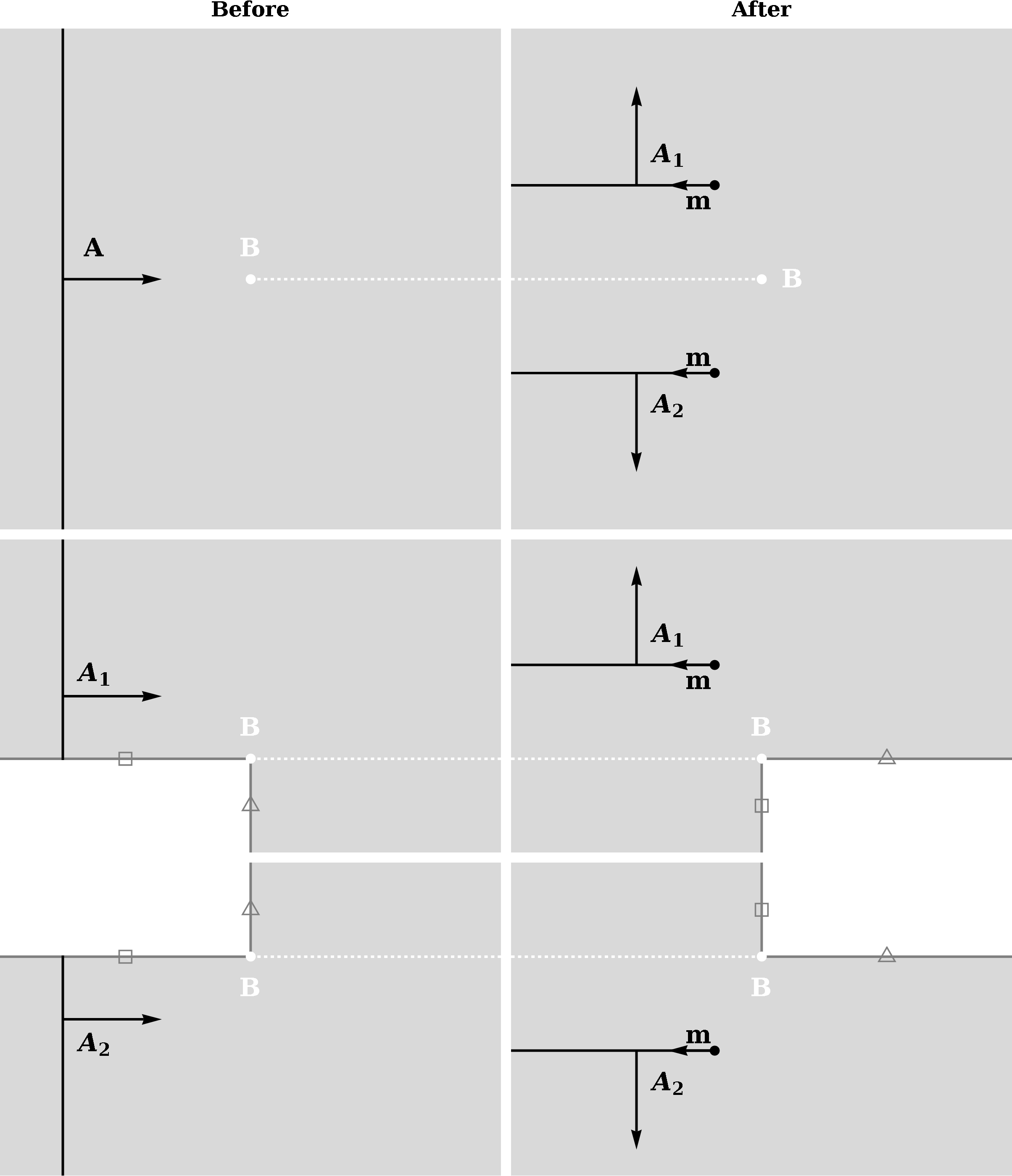}
\caption{Defect $A$ is scattered off the stationary defect $B$, which has a surplus angle of $\pi$. (Projected along the orientation of $B$.) The dashed line indicates the cut leading to the surplus area. On the bottom row the same process is depicted, but now as two separate pieces of space that are glued together along boundary of the white area. (The lines marked with a triangle are identified with each other, idem for the lines marked with a square.) After the collision there exist no straight lines going from $A_1$ to $A_2$. }\label{fig:largesurplus}
\end{figure}

The first issue that we need to consider is that due to the non-euclidean geometry induced by the curvature of the colliding defects it may occur that for a certain choice of the velocities of the junctions, there may not be a plane connecting a pair of the junctions. This typically happens when the curvature of the defects is negative.

For example, figure \ref{fig:largesurplus} shows the 2-dimensional projection of the scattering off a defect of a stationary defect with surplus angle equal to $\pi$. After the collision, it is impossible to connect the junctions labelled $A_1$ and $A_2$ with a straight line, making it impossible to construct a tetrahedral configuration. In the chosen example this is true for any outward moving choice of the junction velocities ($m>0$). Junctions that move inward ($m<0$) can still be connected in that case. If we choose the surplus angle even bigger, say $2\pi$, then it is impossible to connect the junctions for any choice of the junction velocities.

\begin{figure}[tbp]
\centering\includegraphics[width=\textwidth]{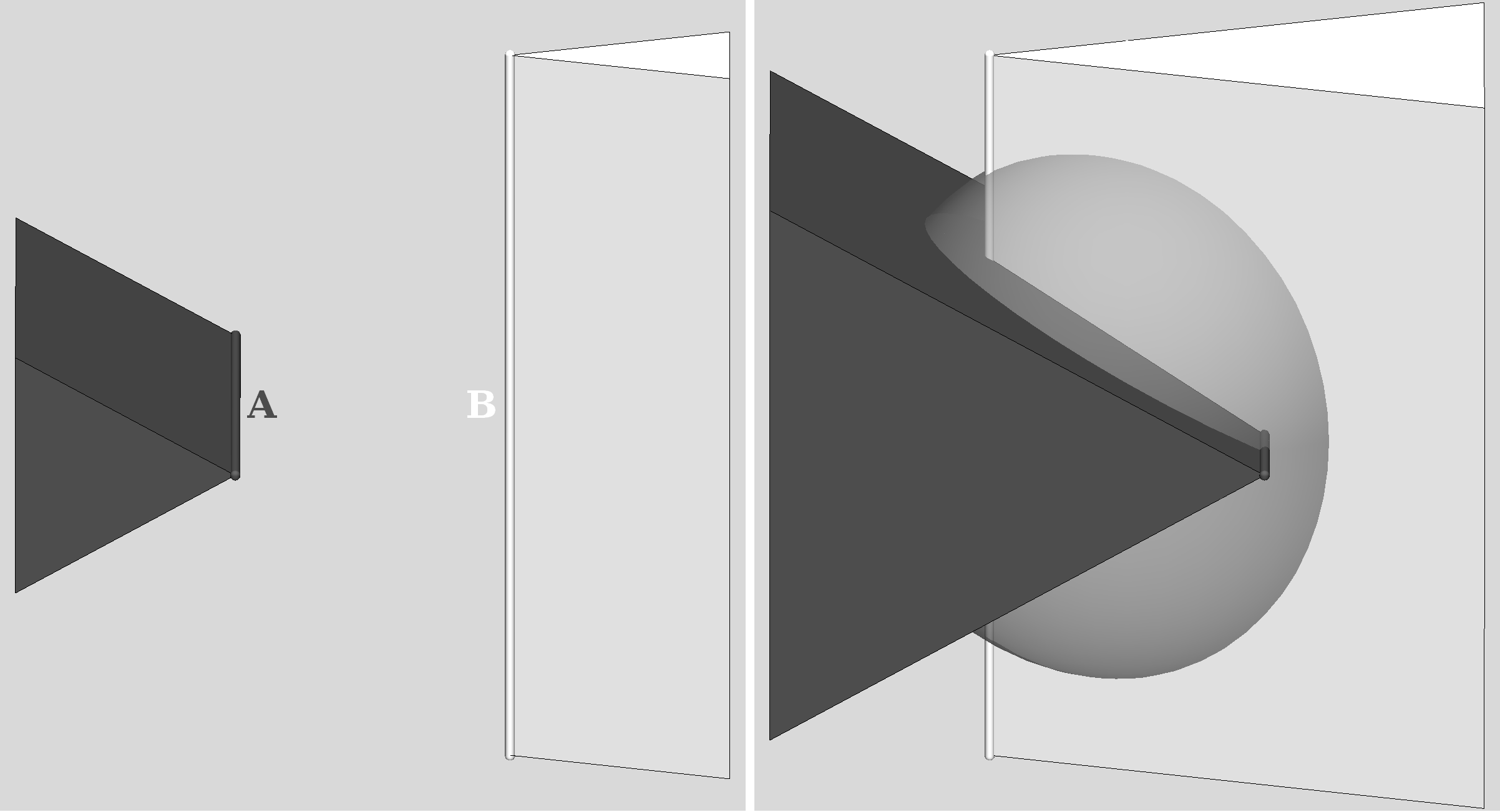}
\caption{Two defects colliding in the center of velocity frame. On the left, the geometry before the collision. On the right the continued geometry after the collision. The transparent sphere indicates the points that are within the light cone of the collision.}\label{fig:limitsphere}
\end{figure}

Consequently, we see that the collision parameters put limitations on the possible choices of the junction velocities. In extreme situations there may be no possible choices at all. In what cases is it possible to find junctions that move at subluminal speeds? A sufficient condition for such a choice to be available can be obtained in the center of velocity frame. In figure \ref{fig:limitsphere} the geometry of a collision of two defects with positive deficit angle is shown. The sphere on the right hand image indicates the points that are within the lightcone of the collision point. If we can choose the junctions within this area (as is the case in the image) then they are subluminal and all defects are timelike.

However, if the deficit angle of defect $A$ is large enough, then there will be no points on the defect $B$ that lie within the lightcone of the collision. Consequently, it is impossible to choose the junction on the $B$ defect such that it is timelike. This happens when the intersection of the deficit angle of $A$ with the defect $B$ lies outside the lightcone. In terms of the collision  parameters this means
\begin{equation}
v^2 + \frac{4v^2}{1-v^2}\frac{\tan^2\psi_A/2}{\sin^2\phi}\geq 1,
\end{equation}
where $v$ is the velocity of the defects in the center of velocity frame (i.e. $v=\tanh\eta/2$). Consequently, the condition that all junctions can be chosen to be timelike is (in terms of $\eta$)
\begin{equation}\label{eq:junctionlimit}
\begin{aligned}
\tanh^2\eta &\leq \frac{\sin^2\phi \cos^2 \psi_A/2}{1-\cos^2\phi \cos^2 \psi_A/2},\quad\text{and}\\
\tanh^2\eta &\leq \frac{\sin^2\phi\cos^2 \psi_B/2}{1-\cos^2\phi \cos^2 \psi_B/2}.\\
\end{aligned}
\end{equation}
Notice that for orthogonal ($\phi=\pi/2$) collisions this reduces to the limit where the junctions in the single defect continuation become lightlike, which we found in equation \eqref{eq:1defectjunctionlimit}.

\begin{figure}[tb]
\centering\includegraphics[width=80mm]{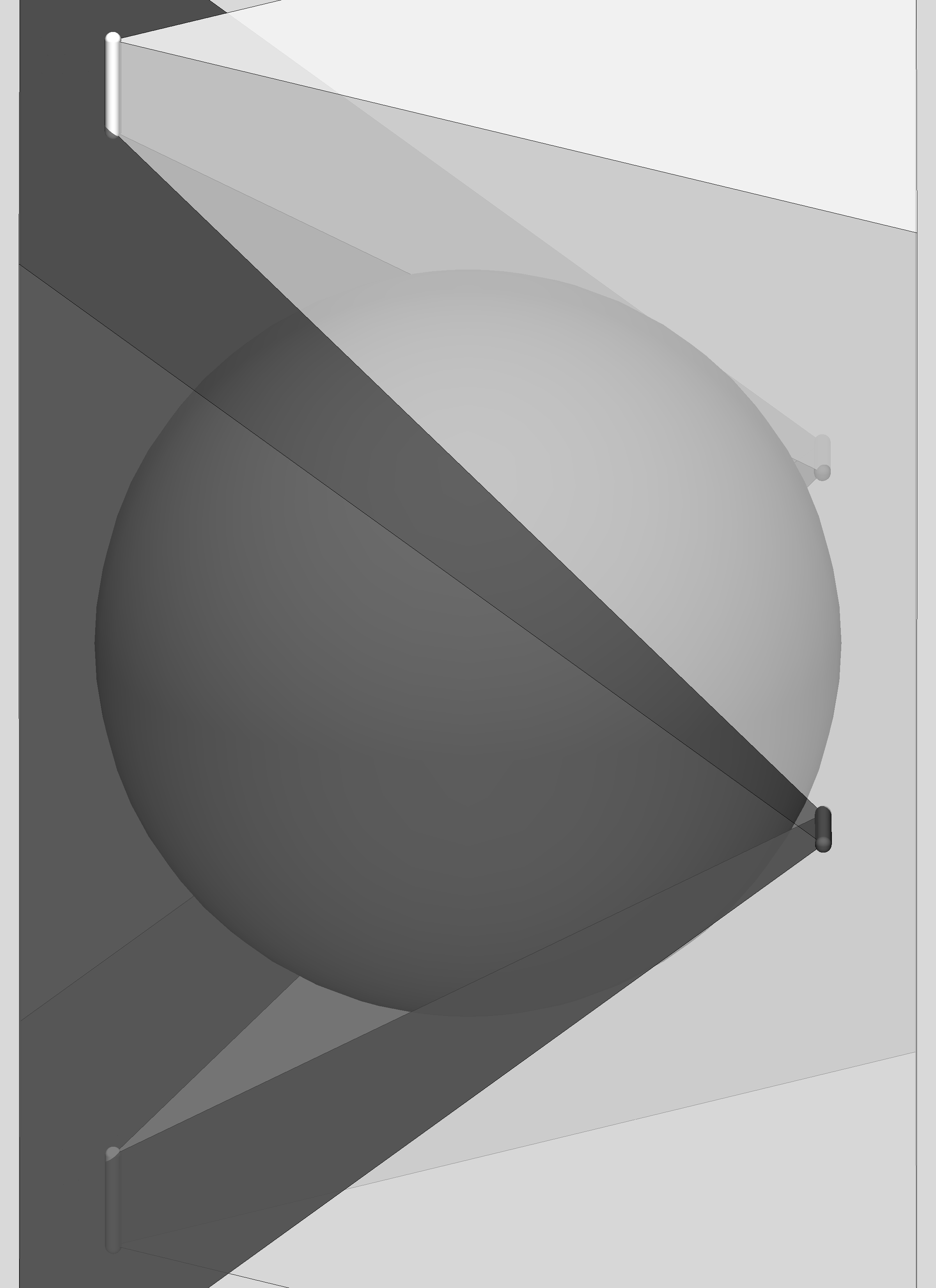}
\caption{If the deficit angles are large enough the sphere can envelope the entire future lightcone of the collision point.}\label{fig:nofuture}
\end{figure}

If the defect angles become even bigger (either because the defects have higher mass, or because the collision velocity is larger), than it can happen that the entire future lightcone of the collision point becomes contained in the wedges that are removed from the spacetime by the defects, as happens in the example shown in figure \ref{fig:nofuture}.

The limit where this happens can be obtained by finding the intersection of the edges of the two defects, and calculating the velocities of its edges. If all these velocities are greater than one, the wedges completely ``eat up'' the future lightcone of the collision. For an orthogonal collision of defects, this gives the limit
\begin{equation}
\tanh^2\eta \geq  \frac{3+\cos\psi_A +\cos\psi_B - \cos\psi_A\cos\psi_B}{4}.
\end{equation}
This is the same limit we found in equation \eqref{eq:orthotracebound} for the velocity where the single defect in the single defect continuation became spacelike. 

It is no coincidence that we find the same limits as for the single defect continuation. For an orthogonal collision, choosing the junctions to lie at the intersection of the defect with the other defects deficit angle implies that the junctions for both ends of a defect coincide. In that case the intermediate defect connecting these two junctions disappears and the other four intermediate defects coincide to form a single defect.

This result does however seem to be at odds with our earlier results for quadrangle continuations as discussed in section \ref{sec:quadsols}. The algebraic solutions that we found there seemed to produce valid results in a much larger part of the orthogonal collision parameter space than the single defect continuations, both for finding continuations with only subluminal junctions (see figure \ref{fig:junctionbounds}) and for finding continuations with only timelike defects (see figure \ref{fig:tracebounds}). At the same time, we observed that the quadrangle continuations were special cases of the tetrahedral continuation. Consequently, one would expect to find valid tetrahedral continuations for at least the same part of the orthogonal collision parameter space  as the quadrangle continuations. Something is not quite right.

The problem we see here, is the result of the limitations of the algebraic description of  a configuration of defects. Although a configuration is (almost) uniquely described by the Poincaré holonomies of its non-contractible loops, an  assignment of holonomies to the equivalence classes of loops does not always need to have a geometric realization. This happens to be the case for the supposed quadrangle continuation of orthogonal collisions in parts of the parameter space where there are no valid tetrahedral continuations. On closer inspection, these tetrahedral continuations correspond to configurations with junctions moving inward. Inward moving junctions are not, a priori, a problem, but in the case of an orthogonal collision they point to a geometric impossibility.  For an orthogonal collision the two semi-infinite ends of a defect meet after the collision (if continued far enough).  Inward moving junctions on these semi-infinite defects would imply that these defects meet before they end on their junction --- this is geometrically impossible.

The geometrical argument used here to find the bound where the tetrahedral continuation becomes incompatible with local causality can be applied to general, more complex continuations. If the deficit angles of the colliding defects completely ``eat up'' the future lightcone of the collision point, any continuation of that collision must start outside the lightcone of the collision. Since all the elements of a continuation originate at the collision point, this means that the continuation must contain at least some superluminal elements. If some of the defects in the continuation have negative curvature, these can restore parts of the lightcone of the collision, but the continuation must still start outside the lightcone.

\section{Conclusions}
In this chapter we have studied the collisions of pairs of moving conical defects. Because the defects each carry an holonomy they cannot simply pass through each other. Instead an appropriate configuration of intermediate defects must be found to continue the piecewise flat geometry after the collision. Each collision therefore creates new defects. Since in arbitrary configuration collisions are inevitable and the likelihood of the inverse process is negligible, the number of defects in a configuration steadily increases over time, ultimately approaching a fractal like structure.

We have seen that for a large part of the parameter space of collisions, we can indeed find such continuations. When these continuations exist, they involve a number of free parameters. The continuations are therefore not unique. This indicates that our piecewise flat model of gravity needs some additional physical input to select what continuation should occur for a collision. This could be taken to be an analogy of the fact that, in a continuous gravity theory, the evolution of a system is not only determined by the geometry of curved spacetime as dictated by Einstein's equation, but also by the matter lagrangian; the latter can still be chosen in many ways.

More alarmingly, we find that for certain collisions it is impossible to find a continuation that contains only timelike junctions and defects. This  occurs when the colliding defects satisfy a combination of the following conditions:
\begin{enumerate}
\item At least one of the defects is very massive.
\item The relative velocity of the defects is very high.
\item The colliding defects come close to being collinear.
\end{enumerate}
Since the local causality principle of our model forbids the appearance of spacelike defects (and to a lesser extent spacelike junctions), it appears that these collisions indicate that the model is possibly inconsistent. 

One might consider the possibility that one could restrict the model to configurations where the problematic regions of the collision parameter space are not reached. However, although this may be possible for the first two conditions, it seems extremely unlikely that one could arrange a configuration  in such a way that no collisions would come near to being collinear.

\ifx\fullTeX\undefined
\bibliographystyle{../bib/utcaps}
\bibliography{../bib/thesis}
\end{document}
\fi

\cleardoublepage 
\ifx\fullTeX\undefined
\documentclass[11pt,a4paper]{article}
\usepackage[utf8]{inputenc} 
\usepackage[english]{babel}
\usepackage{graphicx}
\usepackage{hyperref}
\usepackage{color}
\usepackage{amsmath}
\usepackage{amsthm}
\usepackage{amssymb}
\usepackage{bbm}
\usepackage{pictexwd,dcpic}

\graphicspath{{../pics/}}

\title{The linear continuum limit}
\author{Maarten van de Meent}
\date{\today} 

\begin{document}
\maketitle
\else
\chapter{Continuum limit}\label{ch:cont}
\fi
At first sight, the piecewise flat model of gravity we have proposed seems very limiting. The model effectively eliminates all local structure from general relativity. Consequently, we expect --- a priori ---  no long range gravitational fields or gravitational waves. This certainly seems to be at odds with the everyday world that we observe. Moreover, matter only seems to appear as extended objects in the form of straight cosmic strings. It is therefore a valid question, whether the model can serve as a model of real world physics at all.

However, the conical defects in this piecewise flat model of gravity represent both matter and gravitational degrees of freedom (or rather a combination of the two). Moreover, since the model allows both defects with positive and negative energy density (i.e. defects with a positive or negative deficit angle), one can easily imagine large scale configurations in which, on average, the energy--momentum (or equivalently the Ricci curvature)  vanishes, while the Weyl curvature does not. Such configurations would represent effective large scale vacuum structure in the piecewise flat model. 

An indication that one can completely reproduce general relativity in this way comes from Regge calculus.\cite{williams:1992, RW:2000} Regge calculus seeks to approximate general relativity by using arbitrary piecewise flat (more specifically simplicial) geometries. In that context, the continuum limit of piecewise flat geometries has been studied extensively.\cite{CMS:1982,CMS:1984,Feinberg1984343,Friedberg1984145} It is known that the space of Regge manifolds is dense in the space of (pseudo\nolinebreak)\nolinebreak{}Rie\hyp{}mannian manifolds (in some suitably defined way), and the Regge action may be obtained as the limit of the Einstein-Hilbert action. In particular, any (vacuum) solution of general relativity may be approximated to arbitrary precision by a suitably fine piecewise flat geometry.\footnote{Note, that although this is true on the level of geometries, the equations of motion derived from the Regge action may not converge due to persistent short wavelength oscillations.\cite{Brewin:1995fq,Brewin:2000zh} } Regge calculus, however, does not interpret the curvature defects in the piecewise flat geometries as propagating fundamental degrees of freedom, and consequently does not restrict the type of defects that may occur. In the piecewise flat model studied in this thesis we, however, allow only non-spacelike curvature defects to occur. It is not immediately clear if and how this restriction affects the conclusion that the space of piecewise flat configurations is dense in the space of solutions of general relativity. This will be investigated in this chapter.

Another issue is that the model only allows for a very particular source of energy--momentum in the form of straight cosmic strings. It would not be surprising if this in some way limits the types of matter sources that may be described by the model. This will be a second line of inquiry in this chapter.

The description of a configuration of defects simplifies dramatically in the limit where the energies of all defects are considered to be infinitesimal. This allows us to study continuous distributions of infinitesimal defects. Section \ref{sec:configs} explains how to describe such a configuration in this limit using a density function on the configuration space of an individual defect and introduces the notation used in the rest of the chapter.

Section \ref{sec:T} then constructs the energy--momentum tensor produced by an arbitrary configuration of physical defects. It finds the conditions that the energy--momentum will obey (and consequently must be obeyed by any theory that is to be approximated by this model in the limit of weak fields).

The metric perturbation produced by a general configuration of physical defects is obtained in section \ref{sec:h}. After this, in section \ref{sec:gravwave}, the results of the preceding sections are combined to find the metric perturbations that can be produced by a configuration with vanishing energy--momentum. We find that it is possible to reproduce the complete spectrum of gravitational waves found in linearized Einstein gravity.

\section[Weak field]{Defect configurations in weak field}\label{sec:configs}
In chapter \ref{ch:3+1gravity} two approaches to describing a general configuration of propagating defects were described.  Both of these share that they rapidly increase in complexity as the number of defects increases, because the description of the state of a defect involves the states of other defects as well. This complication is this model's manifestation of the non-linear nature of general relativity. 
 
However, in this chapter we want to look at geometries generated by configurations with large numbers of defects, which makes the previously employed methods prohibitively complex. Fortunately, the description of a configuration of defects drastically simplifies in the limit that all defect angles are very small. In this limit each defect can be treated as a linear perturbation to a Minkowski background. Moreover, any new intermediate defects that would be created by the collision of two defects are higher order in the defect angles of the colliding defects and can be neglected.

Consequently, in the linear limit the state of a defect line can be described while ignoring the presence of other defects. A configuration of defects can therefore be completely described by giving the number of defects in any given state. Since the effect of two defects with the same state is simply that of a single defect with the combined energy of the two defects, we can completely describe the configuration by giving the energy in each possible state. That is, if $\cs$ is the state space of a single defect with unit energy density, a configuration of defects can be described by a distribution $\rho:\cs\rightarrow\RR$ giving the energy density in each state.

To parametrize the state space $\cs$, recall that in \ref{sec:paramdefect} we showed that the state of a single defect line with energy density $\rho$ could be described by the following data: a vector $\vec{X}$ that gives the position of the defect, a vector $\vec{d}$ that gives its direction, and a vector $\vec{v}$ that gives its velocity.

The triple  $\bhh{\vec{X},\hat{d},\vec{v}}$ is enough to uniquely identify the state of the defect. However, this characterization is not unique, since the triple  
\begin{equation}\label{eq:equivtriple}
\bhh{\vec{X}+\alpha\vec{d},\beta\vec{d},\vec{v}+\gamma\vec{d}}
\end{equation}
describes exactly the same state. In fact, the triples describing the same state as $\bhh{\vec{X},\vec{d},\vec{v}}$ can be completely parameterized by the numbers $\alpha$, $\beta$, and $\gamma$. The state space of a line defect with unit density $\cs$ is thus given as the quotient of $\RR^3\times \RR^3\times\RR^3$ by the equivalence relation defined in \eqref{eq:equivtriple}.
 
A set of unique representatives for each equivalence class in $\cs$ can be formed by taking a triple $\bhh{\vec{X},\hat{d},\vec{v}}$ that satisfies the following conditions,
\begin{equation}\label{eq:conds}
\begin{aligned}
\vec{v} &\perp \hat{d}, \\
\vec{X} &\perp \hat{d}, \text{and}\\
\norm{\hat{d}}&=1.
\end{aligned}
\end{equation}
Such a representative triple is unique up to a sign of $\hat{d}$, i.e. to get a unique representative $\hat{d}$ should be viewed as an element of the real projective plane, $\RR\mathrm{P}^2$.

In this chapter it will be convenient to decompose  density functions $\rho$ on $\cs$ in a set of canonical functions that we call \emph{laminar plane waves}. A laminar plane wave is a configuration where all defects have the same direction $\hat{d}_0$ and velocity $\vec{v}_0$ and the density $\rho$ is a plane wave function with wave vector $\vec{k}_0$ with respect to the position $\vec{X}$ of the defects. That is, the density function $\rho^{\lp}[\vec{k}_0,\hat{d}_0,\vec{v}_0]$ corresponding to a laminar plane wave  with wave vector $\vec{k}_0$, direction $\hat{d}_0$, and velocity $\vec{v}_0$ can be written as a function  of $\bhh{\vec{X},\hat{d},\vec{v}}$ representing an equivalence class in $\cs$ as follows,
\begin{equation}\label{eq:lpw}
\rho^{\lp}[\vec{k}_0,\hat{d}_0,\vec{v}_0](\vec{X},\hat{d},\vec{v})=
	e^{2\pi\ii \vec{k}_0\cdot\vec{X}}\delta(\hat{d}-\hat{d}_0)\delta(\vec{v}-\vec{v}_0)
.\footnote{The density function of course needs to be real. The expansion in complex exponentials is for the sake of convenience only. To obtain physical results we should consider only the real part.}
\end{equation}
Since $\vec{X}$ en $\vec{v}$ are always perpendicular to $\hat{d}$ it is sufficient to consider only laminar plane waves with $\vec{k}_0$ and $\vec{v}_0$ perpendicular to $\hat{d}_0$. The set $\csp$ of laminar plane waves with parameters $\bhh{\vec{k}_0,\hat{d}_0,\vec{v}_0}$ is complete in the sense that any density function $\rho(\vec{X},\hat{d},\vec{v})$ on $\cs$ can be written as
\begin{align}
\rho(\vec{X},\hat{d},\vec{v})
	&=	\int_{\csp} \md\vec{k}_0\md\hat{d}_0\md\vec{v}_0 \,
		\rho^{\lp}[\vec{k}_0,\hat{d}_0,\vec{v}_0](\vec{X},\hat{d},\vec{v})\bar\rho(\vec{k}_0,\hat{d}_0,\vec{v}_0) \\
	&=	\int_{\vec{k}\perp\hat{d}}\md\vec{k}\, e^{2\pi\ii \vec{k}\cdot\vec{X}}\bar\rho(\vec{k},\hat{d},\vec{v}),
\end{align}
where $\bar\rho(\vec{X},\hat{d},\vec{v})$ is a density function on $\csp$. The function $\bar\rho(\vec{k},\hat{d},\vec{v})$ can therefore be viewed as a partial Fourier transform of $\rho(\vec{X},\hat{d},\vec{v})$.\footnote{Conversely, $\bar\rho(\vec{k},\hat{d},\vec{v})$ may be obtained from $\rho(\vec{X},\hat{d},\vec{v})$ by an inverse partial Fourier transform.}

It will also be convenient to split the parameter $\vec{v}$ in a component collinear with $\vec{k}$ and a component perpendicular to both $\vec{k}$ and $\hat{d}$,
\begin{equation}\label{eq:vsplit}
\vec{v} = \omega\frac{ \vec{k}}{\vec{k}^2} + v\frac{ \vec{k}\times\hat{d}}{\vec{k}^2},
\end{equation}
where the ambiguous direction of $\hat{d}$ is chosen such that $v$ is non-negative. The usefulness of this split becomes apparent when we do a Lorentz transform. The Lorentz transform of a laminar plane wave with parameters $\bhh{\vec{k},\hat{d},\omega,v}$ is again a laminar plane wave, but with different parameters. It can be shown that the combination $k_\mu = (\omega, \vec{k})$ transforms as a 4-vector under the Lorentz transformation. In section \ref{sec:T} we will see that this is the wave vector of the corresponding energy--momentum tensor. In the remainder of this chapter we shall denote the parameters of a laminar plane wave as $\bhh{k_\mu,\hat{d},v}$. 

The physicality condition that we impose on the defects implies that the velocity of each defect must be smaller than or equal to $c=1$. By squaring equation \eqref{eq:vsplit} we see that this implies that
\begin{equation}
\frac{\omega^2 + v^2}{\vec{k}^2} \leq 1.
\end{equation}
Consequently, we see that for a laminar plane wave of physical defects, the wave vector $k_\mu$ must be spacelike or lightlike and $v$ must be smaller than or equal to $\sqrt{k_\mu k^\mu}$.

\section{Energy--momentum}\label{sec:T}
In this section we will derive the energy--momentum tensor generated by an arbitrary configuration of physical defects in the limit that the distribution is continuous and all defect angles are small. This will tell us what conditions are imposed on the energy--momentum tensor by the requirement that the defects are physical (i.e. non-tachyonic). This puts limits on the kinds of models which can be found as the continuum limit of the piecewise linear model considered here.

In the linear weak field limit the energy--momentum tensor of a configuration of defects can be found by adding together the energy momentum tensors generated by the individual defects. Therefore, if $\hat{T}_{\mu\nu}\bb{\vec{X},\hat{d},\vec{v}}(x_\kappa)$ is the energy--momentum generated by a single defect with position $\vec{X}$, direction $\hat{d}$ and velocity $\vec{v}$ with unit energy density, then the total energy--momentum of a configuration of defects given by a density function $\rho(\vec{X},\hat{d},\vec{v})$ on $\cs$ is
\begin{equation}\label{eq:posT}
T[\rho]_{\mu\nu}(x_\kappa) = 
\int_\cs \!\!\!\!\md\vec{X}\,\md\hat{d}\,\md\vec{v}\;
\rho(\vec{X},\hat{d},\vec{v})\, \hat{T}_{\mu\nu}\bb{\vec{X},\hat{d},\vec{v}}(x_\kappa).
\end{equation}
Consequently, if we have an explicit expression for $\hat{T}_{\mu\nu}\bb{\vec{X},\hat{d},\vec{v}}(x_\kappa)$, we can compute the energy--momentum tensor for any configuration of defects. Alternatively, since the laminar plane waves form a complete basis for all configurations, the total energy--momentum of a configuration can also be obtained from  $\hat{T}^{\lp}_{\mu\nu}\bb{k_\mu,\hat{d},v}(x_\kappa)$, the energy--momentum of a laminar plane wave with wave vector $k_\mu$, direction $\hat{d}$ and perpendicular velocity $v$ through,
\begin{equation}
T[\bar\rho]_{\mu\nu}(x_\kappa) = 
\int_{\csp} \!\!\!\!\md k\,\md\hat{d}\,\md v\;
\bar\rho(k_\mu,\hat{d},v)\, \hat{T}^{\lp}_{\mu\nu}\bb{k_\mu,\hat{d},v}(x_\kappa).
\end{equation}

We will now derive the energy--momentum tensor $\hat{T}^{\lp}_{\mu\nu}\bb{k_\mu,\hat{d},v}(x_\kappa)$ generated by a single laminar plane wave. In section \ref{sec:statdefect} we derived the energy--momentum  tensor of a single stationary defect through the origin and directed along the $z$-axis. With unit energy density the result is, 
\begin{equation}
\hat{T}_{\mu\nu}\bb{\vec{0},\hat{z},\vec{0}}(x_\kappa)=\begin{pmatrix}
1 & 0 & 0 & 0 \\
0 & 0 & 0 & 0 \\
0 & 0 & 0 & 0 \\
0 & 0 & 0 & -1 \\
\end{pmatrix}\delta(x)\delta(y).
\end{equation}
The dependence on the position $\vec{X}=(X_x,X_y,0)$\footnote{Remember that $\vec{X}$ should be perpendicular to $\hat{d}$.} can be obtained by performing appropriate shifts, which yields
\begin{equation}
\hat{T}_{\mu\nu}\bb{(X_x,X_y,0),\hat{z},\vec{0}}(x_\kappa)=\begin{pmatrix}
1 & 0 & 0 & 0 \\
0 & 0 & 0 & 0 \\
0 & 0 & 0 & 0 \\
0 & 0 & 0 & -1 \\
\end{pmatrix}\delta(x-X_x)\delta(y-X_y).
\end{equation}
By combining this result with equations \eqref{eq:lpw} and \eqref{eq:posT} we obtain the energy--momentum generated by a stationary laminar plane wave with wave vector $k_\mu=(0,\vec{k})$, direction $\hat{z}$ and zero velocity,
\begin{equation}\label{eq:statwave}
\begin{aligned}
\hat{T}^{\lp}_{\mu\nu}\bb{k_\lambda,\hat{z},0}(x_\kappa) &=
\int_{\cs}\!\!\!\! \md\vec{X}\,\md\hat{d}\,\md\vec{v}\; \rho^{\lp}[k_\lambda,\hat{z},0](\vec{X},\hat{d},\vec{v})\, \hat{T}_{\mu\nu}\bb{\vec{X},\hat{d},\vec{v}}(x_\kappa)\\
&=\int_{\vec{X}\perp\hat{z}}\!\!\!\! \md\vec{X}\;\hat{T}_{\mu\nu}\bb{\vec{X},\hat{z},\vec{0}}(x_\kappa)e^{2\pi\ii \vec{k}\cdot\vec{X}}\\
&=\begin{pmatrix}
1 & 0 & 0 & 0 \\
0 & 0 & 0 & 0 \\
0 & 0 & 0 & 0 \\
0 & 0 & 0 & -1 \\
\end{pmatrix}e^{2\pi\ii \vec{k}\cdot\vec{x}}.
\end{aligned}
\end{equation}

Any other laminar plane wave can be obtained by applying appropriate Lorentz transformation $\Lambda^\mu_\nu$. To write the result we first notice that the right hand side of equation \eqref{eq:statwave} can be written as
\begin{equation}\label{eq:covar1}
\hh{u_\mu u_\nu + u^2 d_\mu d_\nu}e^{2\pi\ii k\cdot x},
\end{equation}
if we introduce the 4-vectors $u_\mu =  (1,0,0,0)$ and $d_\mu = (0,0,0,1)$.

After applying $\Lambda^\mu_\nu$ and a linear redefinition\footnote{The 4-vectors $u_\mu$ and $d_\mu$ span the plane of a single defect in the laminar plane wave. Applying a Lorentz transformation yields vectors spanning the plane of a defect in the transformed plane wave. However, these vectors will not correspond to the (non-covariant) parameters we introduced to describe the velocity and direction of the defect. To find a pair of 4-vectors $u'_\mu$  and $d'_\mu$ that span the same plane but correspond to the velocity $\vec{v}$ and direction $\hat{d}$ of the defect we need to do a linear transformation.} 
\begin{equation}
\begin{aligned}
d'_\nu &= \alpha  \Lambda^\mu_\nu d_\mu + \beta \Lambda^\mu_\nu  u_\nu,\\
u'_\mu &= \gamma \Lambda^\mu_\nu d_\mu + \delta \Lambda^\mu_\nu u_\mu
\end{aligned}
\end{equation}
such that 
\begin{equation}\label{cond:redef}
\begin{aligned}
d'_\mu d'^\mu&=1, & d'_0 &=0, \\
u'_\mu d'^\mu&=0, & u'_0 &=1,
\end{aligned}
\end{equation}
the direction $\hat{d}$ and velocity $\vec{v}$ of the new laminar plane wave can be found as
\begin{equation}
\begin{aligned}
u'_\mu &= (1,\vec{v}),\quad\text{and}\\
d'_\mu &= (0,\hat{d}).
\end{aligned}
\end{equation}
Applying the same Lorentz transformation and linear redefinition to equation \eqref{eq:covar1} yields the energy--momentum of the new laminar plane wave
\begin{equation}\label{eq:Tlp}
\hat{T}^{\lp}_{\mu\nu}[k_\lambda,\hat{d},v](x_\kappa) =
-\frac{1}{u'^2}\hh{u'_\mu u'_\nu + u'^2 d'_\mu d'_\nu}e^{2\pi\ii k\cdot x}.
\end{equation}
This gives us the energy--momentum for all laminar plane waves with $k_\mu k^\mu > 0$. Notice that a laminar plane wave of defects with wave vector $k_\mu$ only contributes to the Fourier mode of the energy--momentum tensor with wave vector $k_\mu$.  As a result the Fourier transform of the total energy--momentum of a configuration specified by the density function $\bar\rho$ on $\csp$ has the especially simple form,
\begin{equation}\label{eq:totalTk}
T_{\mu\nu}[\bar\rho](k_\lambda) = 
\int_{\hat{d}\perp\vec{k}} \!\!\!\!\md\hat{d}
	\!\!\int_0^{\sqrt{k_\lambda k^\lambda}}\s{-50}\md v \;
	 	\bar\rho(k_\lambda,\hat{d},v)\hat{T}^{pl}_{\mu\nu}[k_\lambda,\hat{d},v].
\end{equation}
The energy--momentum tensor for laminar plane wave with $k_\mu k^\mu =0$, can be found as a limit of equation \eqref{eq:Tlp}. There are two possibilities,
\begin{enumerate}
\item $k_\mu \rightarrow 0$ corresponding to a constant ``wave'' of defects with the same direction and velocity.
\item $u_\mu u^\mu \rightarrow 0$ corresponding to a laminar plane wave of lightlike defects.
\end{enumerate}
The first case is easy enough  to compute since equation \eqref{eq:Tlp} is regular in this limit. However, in the second limit the $1/u'^2$ factor blows up. This can be fixed by noting that the normalization of $\hat{T}^{\lp}_{\mu\nu}$ is arbitrary, and we are therefore free to rescale it by a factor $-u'^2$ changing equation \eqref{eq:Tlp} to
\begin{equation}\label{eq:Tlp2}
\hat{T}^{\lp}_{\mu\nu}[k_\lambda,\hat{d},v](x_\kappa) =
\hh{u'_\mu u'_\nu + u'^2 d'_\mu d'_\nu}e^{2\pi\ii k\cdot x},
\end{equation}
which is regular in the limit that $u'^2$ goes to zero. This fixes $\hat{T}^{\lp}_{00}$ to be 1, which has the additional advantage of giving $\bar\rho$ the physical interpretation of the energy density present in a particular mode of laminar plane wave.

We now have all the ingredients we need to calculate the energy--momentum tensor of a general configuration of physical defects and can explore what conditions this will satisfy. In particular, we are interested in what conditions are imposed on the energy--momentum tensor by the restriction that all defects must be physical.

One immediate condition that we observe from equation \eqref{eq:Tlp} is that $k^\mu\hat{T}^{\lp}_{\mu\nu} = 0$ because $k^\mu d_\mu = k^\mu u_\mu = 0$. Consequently, equation \eqref{eq:totalTk} implies that the total energy--momentum must satisfy
\begin{equation}
k^\mu T_{\mu\nu}[\bar\rho] = 0,
\end{equation}
for any distribution $\bar\rho$ on $\csp$. That is, energy is conserved, as one expects from any reasonable physical theory.

Other conditions can be obtained by examining $T_{\mu\nu}[\bar\rho](k_\lambda)$ mode by mode. We have already observed that laminar plane waves with wave vector $k_\mu$ only contribute to modes of the energy--momentum tensor with the same wave vector. Therefore, since laminar plane waves of physical defects must have $k_\mu k^\mu \geq 0$, it is impossible for a configuration of defects to generate an energy--momentum tensor with Fourier modes with $k_\mu k^\mu < 0$.

For modes with $k_\mu k^\mu > 0$, by Lorentz invariance, we can restrict ourselves to the special case that $k_\mu = (0,k,0,0)$ without loss of generality. The direction $\hat{d}$, which must be perpendicular to $\vec{k}$, can then be parameterized by a single angle $\phi$ with $\phi=0$ corresponding to the direction of the $\hat{z}$-axis. The Fourier transform of equation \eqref{eq:Tlp2} then becomes,
\begin{equation}\label{eq:spacelikeT}
\hat{T}^{\lp}_{\mu\nu}[k_\kappa,\hat{d},v]=
\begin{pmatrix}
1 
	& 0 
		& \frac{v}{k} \cos\phi 
			 &  \frac{v}{k} \sin\phi \\
0 
	& 0 
		& 0
			& 0 \\
 \frac{v}{k} \cos\phi
	&  0
		&  \frac{v^2}{k^2} - \sin^2\phi
			& \cos\phi\sin\phi\\
  \frac{v}{k} \sin\phi
 	& 0
 		& \cos\phi\sin\phi
 			& \frac{v^2}{k^2} - \cos^2\phi\\
\end{pmatrix}.
\end{equation}
If we expand $\bar\rho(k_\mu,\hat{d},v)$ as
\begin{equation}\label{eq:expansion}
\bar\rho(k_\mu,\hat{d},v) =
\sum_{n=0}^\infty \frac{(2n+1)}{\pi}P_n (\tfrac{2v}{\sqrt{k^\mu k_\mu}}-1)
\bhh{r_{0n} + 2 \sum_{m=1}^\infty( r_{mn} \cos m\phi + \tilde{r}_{m n}\sin m\phi)},
\end{equation}
where the $P_n$ are Legendre polynomials and the coefficients $r_{mn}$ are understood to be functions of $k_\mu$, then the total contribution of the $k_\mu$-mode of $\bar\rho$ to the energy--momentum \eqref{eq:totalTk} becomes,
\begin{equation}\label{eq:totalTexp}
T_{\mu\nu}[\bar\rho](k_\mu) =
\begin{pmatrix}
2 r_{00} 
	& 0 
		& r_{11} +r_{10}
			 & \tilde{r}_{11} +\tilde{r}_{10} \\
0 
	& 0 
		& 0
			& 0 \\
r_{11}\!+\!r_{10}
	&  0
		& \tfrac{-1}{3}r_{00}\!+\!r_{01}\!+\!\tfrac{1}{3}r_{02}\!+\!r_{20}
			& \tilde{r}_{20}\\
\tilde{r}_{11}\!+\!\tilde{r}_{10}
 	& 0
 		&\tilde{r}_{20}
 			& \tfrac{-1}{3}r_{00}\!+\!r_{01}\!+\!\tfrac{1}{3}r_{02}\!-\!r_{20}\\
\end{pmatrix}.
\end{equation}
If we compare this to the most general form of a mode energy--momentum tensor with wave vector $k_\mu=(0,k,0,0)$ that satisfies $k^\mu T_{\mu\nu}=0$,
\begin{equation}
T_{\mu\nu}(k_\kappa) =
\begin{pmatrix}
 T_{00} 
	& 0 
		&T_{01}
			 & T_{02} \\
0 
	& 0 
		& 0
			& 0 \\
T_{01}
	&  0
		&T_{22}
			& T_{23}\\
T_{02}
 	& 0
 		&T_{23}
 			& T_{33}\\
\end{pmatrix},
\end{equation}
then we find that any such energy--momentum tensor can be generated by appropriate choices of the coefficients $r_{mn}$. We therefore find the physicality condition on the defects puts no further restrictions on the modes of the energy--momentum tensor with $k_\mu k^\mu > 0$.

In the special case that $k_\mu$ is lightlike, we can assume, without loss of generality, that $k_\mu= (\omega,\omega,0,0)$. In this limit  \eqref{eq:Tlp2} becomes,
\begin{equation}
\hat{T}^{\lp}_{\mu\nu}[k_\mu,\hat{d},v]=
\begin{pmatrix}
1 
	& 1 
		& 0 
			 & 0 \\
1
	& 1 
		& 0
			& 0 \\
0
	&  0
		& 0
			&0\\
 0
 	& 0
 		& 0
 			& 0\\
\end{pmatrix},
\end{equation}
and the total contribution of a distribution of defects $\bar\rho$ to the Fourier mode of the energy--momentum with $k_\mu= (\omega,\omega,0,0)$ becomes
\begin{equation}\label{eq:totalTexp2}
T_{\mu\nu}[\bar\rho](k_\mu)=
\begin{pmatrix}
r_{00} 
	& r_{00} 
		& 0 
			 & 0 \\
r_{00}
	& r_{00} 
		& 0
			& 0 \\
0
	&  0
		& 0
			&0\\
 0
 	& 0
 		& 0
 			& 0\\
\end{pmatrix}.
\end{equation}
Apparently,  the lightlike modes of the energy--momentum tensor generated by a configuration of  physical defects are subject to additional restrictions. Not only do these modes have to be transverse, but they also cannot have any pressure or momentum perpendicular to their direction of propagation. This restriction can be formalized in the following way: For any lightlike mode of the energy--momentum tensor $T_{\mu\nu}(k_\kappa)$ and any lightlike vector $l^\mu$, the contraction $l^\mu T_{\mu\nu}(k_\kappa)$ is a non-spacelike vector. 

The other special case, $k_\mu=0$ is somewhat different since any direction $\hat{d}$ will be perpendicular to $\vec{k}=(0,0,0)$. The total contribution to the zero mode of the energy--momentum tensor will thus be found by integrating equation $\eqref{eq:Tlp2}$ over all mutually perpendicular $\vec{v}$ and $\hat{d}$. That is,
\begin{equation}
T_{\mu\nu}[\bar\rho](0) = -\int_{\vec{v}\perp\hat{d}} \md\vec{v}\md\hat{d}\; \bar\rho(0,\hat{d},\vec{v})(u_\mu u_\nu + u^2 d_\mu d_\nu).
\end{equation}
Since tensors of the form $u_\mu u_\nu + u^2 d_\mu d_\nu$ span the space of symmetric 2-tensors, we can conclude that any zero mode of the energy--momentum tensor may be produced by a configuration of defects. The condition that it contains only physical defects poses no further restrictions.

We therefore obtain the following restrictions that the energy-momentum tensor of a configuration of physical defects will satisfy in the continuum weak field limit
\begin{enumerate}
\item $k^\mu T_{\mu\nu}(k_\lambda) = 0$ for all $k_\mu$.
\item $T_{\mu\nu}(k_\lambda) = 0$ for all $k_\mu$ with $k^\mu k_\mu < 0$.
\item $l^\mu T_{\mu\nu}(k_\lambda)$ is a non-spacelike vector for all lightlike $k_\mu$ and $l_\mu$.
\end{enumerate}

The first and third condition hold for most physically reasonable theories. The first expresses conservation of energy--momentum, while the last is normally imposed as part of the null dominant energy condition which is employed in cosmology to ensure vacuum stability while allowing negative vacuum energy.\cite{CHT:2003}

The second condition is satisfied by various simple matter models used in general relativity, such as dusts. However, it is typically violated in classical wave like systems. For example, consider a standing wave solution of the Klein--Gordon equation, $\phi =  \cos(\omega t)\cos(\vec{k}\cdot\vec{x})$. Even if $\vec{k}^2 > \omega^2$, the energy--momentum tensor --- which behaves like the square of $\phi$ --- will have terms which are proportional to $\cos(2\omega t)$ and consequently will violate the second condition.

This indicates that the model cannot represent all types of matter at linear order. At this level all interactions are neglected, and we end up with a system that is very similar to a dust of non-interacting point particles. Beyond  linear order defect lines will collide in a non-trivial manner, as was discussed in chapter \ref{ch:collisions}. The energy--momentum tensor corresponding to the continuation of a collision will typically violate the second condition.

At non-linear level we need to choose a prescription, which tells us how each collision is continued, to complete the dynamics of the model. Different prescriptions will lead to different types of interacting matter. At this point it is unclear what types of limitation on the energy--momentum tensor --- if any --- will persist once all possible types of interaction are included.

\section{Metric perturbations}\label{sec:h}
We now turn to the effect that a configuration of defects $\bar\rho(k_\mu,\hat{d},v)$ has on the metric. In the limit of weak fields the metric $g_{\mu\nu}$ can be separated in a static Minkowski background $\eta_{\mu\nu}$ and a small perturbation $h_{\mu\nu}$,
\begin{equation}
g_{\mu\nu}(x_\lambda) = \eta_{\mu\nu} + h_{\mu\nu}(x_\lambda).
\end{equation}
If we consider only the linear perturbations caused by the presence of a defect, then the total perturbation caused by a continuous distribution of defects can be found as the integral of the perturbations of individual components. Consequently, if $ \hat{h}^{\lp}_{\mu\nu}[k_\lambda,\hat{d},v](x_\kappa)$ is the perturbation of the metric caused by a laminar plane wave  with wave vector $k_\mu$, direction $\hat{d}$, and perpendicular velocity $v$, then the total perturbation caused by a configuration $\bar\rho$ is given by,
\begin{equation}\label{eq:totalH}
h_{\mu\nu}[\bar\rho](x_\kappa) = 
\int_{\csp} \!\!\!\!\md k\,\md\hat{d}\,\md v\;
\bar\rho(k_\lambda,\hat{d},v)\, 
\hat{h}^{\lp}_{\mu\nu}\bb{k_\lambda,\hat{d},v}(x_\kappa).
\end{equation}
Therefore, if we know $\hat{h}^{\lp}_{\mu\nu}\bb{k_\lambda,\hat{d},v}$ for any combination of the parameters $(k_\lambda,\hat{d},v)$, than $h_{\mu\nu}[\bar\rho]$ can be computed for any configuration $\bar\rho$. 

To calculate $\hat{h}^{\lp}_{\mu\nu}\bb{k_\lambda,\hat{d},v}(x_\kappa)$ we first need to fix its linear gauge freedom, which we do by setting the gauge condition $\partial^\mu h_{\mu\nu} =0$.\footnote{There is some residual gauge freedom for the lightlike modes of the metric perturbation as we will discuss later on.} With this choice the linearized Einstein equation becomes (in its Fourier transformed form),
\begin{equation}\label{eq:lineinstein}
T_{\mu\nu}(k_\kappa)= 2\pi^2k^2\bhh{h_{\mu\nu}-h(\eta_{\mu\nu}-\frac{k_\mu k_\nu}{k^2})},
\end{equation}
where $h$ is the trace of $h_{\mu\nu}$.

When $k^\mu k_\mu \neq 0$, equation \eqref{eq:lineinstein} can be inverted to obtain the linear metric perturbation as a function of the energy--momentum tensor.  In particular, if  $\hat{T}^{\lp}_{\mu\nu}\bb{k_\lambda,\hat{d},v}$ is the Fourier mode of the energy--momentum tensor generated by a laminar plane wave, then the metric perturbation generated by that laminar plane wave is given by a single Fourier mode,
\begin{equation}
 \hat{h}^{\lp}_{\mu\nu}[k_\lambda,\hat{d},v]= \frac{1}{2\pi^2 k^2}
 \bhh{\delta_\mu^\alpha\delta_\nu^\beta -\hf (\eta_{\mu\nu}-\tfrac{k_\mu k_\nu}{k^2})\eta^{\alpha\beta}}\hat{T}^{\lp}_{\alpha\beta}[k_\lambda,\hat{d},v].
\end{equation}
We can therefore study the effects of a distribution of defects $\bar\rho(k_\mu,\phi,v)$ on a per mode basis. By plugging in the $\hat{T}^{\lp}_{\mu\nu}[k_\lambda,\hat{d},v]$ from equation \eqref{eq:Tlp2}, we find that, 
\begin{equation}\label{eq:metricT}
\hat{h}^{\lp}_{\mu\nu}[k_\lambda,\hat{d},v]= \frac{1}{2\pi^2 k^2}\bhh{u_\mu u_\nu - u^2 (\eta_{\mu\nu}-d_\mu d_\nu -\frac{k_\mu k_\nu}{k^2})},
\end{equation}
where $u_\mu = (1,\vec{v})$ and $d_\mu = (0,\hat{d})$.

When  $k^\mu k_\mu > 0$,  we can assume by Lorentz invariance that  $k_\mu = (0,k,0,0)$.  Parameterizing $\hat{d}$ as $(0,-\sin\phi,\cos\phi)$ and $\vec{v}$ as $(0,v\cos\phi,v\sin\phi)$, we obtain the explicit expression,
\begin{equation}\label{eq:spacelikeh}
\hat{h}^{\lp}_{\mu\nu}[k_\lambda,\hat{d},v]=\frac{1}{2\pi^2 k^2}\begin{pmatrix}
\frac{v^2}{k^2}
	& 0 
		& \frac{v}{k} \cos\phi 
			 & \frac{v}{k} \sin\phi \\
0 
	& 0 
		& 0
			& 0 \\
\frac{v}{k} \cos\phi
	&  0
		&  \cos^2\phi
			& \cos\phi\sin\phi\\
\frac{v}{k} \sin\phi
 	& 0
 		& \cos\phi\sin\phi
 			& \sin^2\phi\\
\end{pmatrix}.
\end{equation}

The total metric perturbation caused by a configuration of defects given by a distribution $\bar\rho$ on $\csp$ can be obtained from $\hat{h}^{\lp}_{\mu\nu}[k_\lambda,\hat{d},v]$ through the integral,
\begin{equation}\label{eq:totalHk}
h_{\mu\nu}[\bar\rho](k_\lambda) = 
\int\md\hat{d}\,\md v\;
\bar\rho(k_\lambda,\hat{d},v)\, 
\hat{h}^{\lp}_{\mu\nu}\bb{k_\lambda,\hat{d},v}.
\end{equation}
Performing this integral in the case that $k_\mu=(0,k,0,0)$ and applying the expansion of $\bar\rho$ as given in \eqref{eq:expansion} yields,
\begin{equation}\label{eq:spacelikeHdist}
h_{\mu\nu}[\bar\rho](k_\kappa)=
\frac{1}{2 \pi^2 k^2}\begin{pmatrix}
\tfrac{2}{3}r_{00}+r_{01}+\tfrac{1}{3}r_{02}
	&\s{10} 0\s{10} 
		&\s{10}  r_{10} + r_{11}\s{10} 
			 & \tilde{r}_{10} + \tilde{r}_{11}\\
0 
	& 0 
		& 0
			& 0 \\
r_{10} + r_{11}
	&  0
		& r_{00} + r_{20}
			& \tilde{r}_{20}\\
 \tilde{r}_{10} + \tilde{r}_{11}
 	& 0
 		& \tilde{r}_{20}
 			& r_{00} - r_{20}\\
\end{pmatrix}.
\end{equation}

When $k^\mu k_\mu = 0$ the linearized Einstein equation \eqref{eq:lineinstein} cannot be inverted and equation \eqref{eq:metricT} cannot be applied directly.  However, we may obtain the metric perturbation for these modes as a limiting case of the modes with $k^\mu k_\mu > 0$. 

In the case that $k_\mu$ becomes lightlike we can assume due to Lorentz invariance that it goes to $k_\mu = (\omega,\omega,0,0)$. Such a laminar plane wave is the limit of waves with momentum $k_\mu = (\omega,\kappa,0,0)$ as $\kappa \rightarrow \omega$. Since physicality requires that $0\leq v \leq \sqrt{\kappa^2 -\omega^2}$, $v$ must simultaneously  go to zero. Simply setting $v=0$ and using $k_\mu = (\omega,\kappa,0,0)$, $u_\mu = (1,\omega/\kappa,0,0)$, and $d_\mu = (0,0,-\sin\phi,\cos\phi)$ in equation \eqref{eq:metricT} yields,
\begin{equation}\label{eq:v0h}
\hat{h}^{\lp}_{\mu\nu}[k_\lambda,\phi,0]=
-\frac{1}{2\pi^2 \kappa^2}\begin{pmatrix}
0
	& 0 
		& 0
			 & 0 \\
0 
	& 0 
		& 0
			& 0 \\
0
	&  0
		&  \cos^2\phi
			& \cos\phi\sin\phi\\
 0
 	& 0
 		& \cos\phi\sin\phi
 			& \sin^2\phi\\
\end{pmatrix}.
\end{equation}
Since only the pre-factor depends on $\kappa$ the limit as $\kappa$ goes to $\omega$  is straightforward.

Letting $v$ go to zero by another route will lead to a result that (in the $\kappa \rightarrow \omega$ limit) differs from the above by a term of the following form
\begin{equation}\label{eq:vah}
\begin{pmatrix}
\xi_1
	& \xi_1 
		& \xi_2
			 & \xi_3 \\
\xi_1
	& \xi_1 
		& \xi_2
			& \xi_3 \\
\xi_2
	&  \xi_2
		&  0
			& 0\\
 \xi_3
 	& \xi_3
 		& 0
 			&0\\
\end{pmatrix},
\end{equation}
where the $\xi_i$ are arbitrary (possibly infinite) parameters. 

Such a contribution can be gauged away.  Under an infinitesimal coordinate transformation given by a vector field $\xi_\mu$, the  $h_{\mu\nu}$ transforms as,
\begin{equation}
 h_{\mu\nu} \to h_{\mu\nu} + \partial_\mu\xi_\nu +\partial_\nu\xi_\mu.
\end{equation}
The gauge condition $\partial^\mu h_{\mu\nu} =0$ implies that
\begin{equation}
k^\mu k_\mu \xi_{\nu}(k_\lambda)+ k_\nu k^\mu \xi_{\mu}(k_\lambda)=0.
\end{equation}
This completely fixes $\xi_\mu(k_\lambda)$ for $k^\mu k_\mu \neq 0$. However, when $k^\mu k_\mu= 0$ it only implies that $k^\mu \xi_{\mu}(k_\lambda) = 0$. The residual gauge transformation subject to that condition for $k_\mu = (\omega,\omega,0,0)$ takes the form of equation \eqref{eq:vah}. Contributions of that form can therefore be gauged away, and we are free to adopt \eqref{eq:v0h} as the gauge fixed form of $\hat{h}^{\lp}_{\mu\nu}[(\omega,\omega,0,0),\phi,0]$.

Inserting \eqref{eq:v0h} in the integral \eqref{eq:totalHk} and using the expansion \eqref{eq:expansion} for $\bar\rho$ yields the lightlike modes of the metric perturbation caused by a configuration of defects given by a distribution $\bar\rho$
\begin{equation}\label{eq:lightlikeHdist}
h_{\mu\nu}[\bar\rho](\omega,\omega,0,0)=
\frac{1}{2 \pi^2 \omega^2}\begin{pmatrix}
0
	& 0 
		&0
			 & 0\\
0 
	& 0 
		& 0
			& 0 \\
0
	&  0
		& r_{00} + r_{20}
			& \tilde{r}_{20}\\
0
 	& 0
 		& \tilde{r}_{20}
 			& r_{00} - r_{20}\\
\end{pmatrix}.
\end{equation}

In the case that $k_\mu \rightarrow 0$, the equation \eqref{eq:metricT} diverges. This signals a breakdown of the linear perturbation approach in the limit of constant fields. This is not unexpected, since constant non-zero energy densities will typically lead to non-trivial effects on the global level with respect to the topology or causal structure. 

\section{Gravitational waves}\label{sec:gravwave}
In the previous two sections we have obtained the energy--momentum and metric perturbation caused by a continuous distribution of defects at the linear level. At the microscopic level of individual defects the energy--momentum tensor completely fixed the  metric (up to gauge transformations) due to the ad hoc rule we imposed that the vacuum should be completely flat. We are now able to answer the question whether this property persists in the continuum limit.

Since we are considering the linear limit, it is enough to consider what additional metric structure may be present for a configuration with zero energy--momentum. In section \ref{sec:T} we obtained a complete expression for the energy--momentum of a configuration of defects described by a distribution $\bar\rho$ on $\csp$. Requiring that this expression (equations \eqref{eq:totalTexp} and \eqref{eq:totalTexp2}) vanishes implies the following conditions on the coefficients $r_{nm}(k_\mu)$ of $\bar\rho$ in the expansion \eqref{eq:expansion};
\begin{align}
\left. \begin{aligned}
 r_{00}(k_\mu) &= 0,	& r_{01}(k_\mu)+ \tfrac{1}{3}r_{02}(k_\mu)&= 0,\\
 \tilde{r}_{20}(k_\mu) &= 0,	& r_{11}(k_\mu)+ r_{10}(k_\mu)&= 0,\\
 r_{20}(k_\mu) &= 0,	& \tilde{r}_{11}(k_\mu)+ \tilde{r}_{10}(k_\mu)&= 0,
\end{aligned}\right\}&\quad\text{for $k^\mu k_\mu >0$, and}\\
\left.\begin{aligned}
r_{00}(k_\mu) & = 0,
\end{aligned}\right\}&\quad\text{for $k^\mu k_\mu =0$.}
\end{align}

The metric perturbation caused by a vacuum configuration of defects can now be found by applying these conditions to the complete expressions \eqref{eq:spacelikeHdist} and \eqref{eq:lightlikeHdist}, found in section \ref{sec:h} for the metric perturbation caused by a configuration of defects. This yields that for $k^\mu k_\mu >0$, $h_{\mu\nu}(k_\lambda)$ vanishes when $T_{\mu\nu}$ vanishes, as one would expect since the linearized Einstein equation is invertible for $k^\mu k_\mu \neq 0$.

However, for $k^\mu k_\mu =0$, the requirement that $T_{\mu\nu}(k_\lambda)$ vanishes only fixes $r_{00}(k_\mu)$  to be zero. The coefficients $r_{20}(k_\mu)$ and $\tilde{r}_{20}(k_\mu)$ are  unconstrained. Consequently, for each lightlike $k_\mu$ there exists a two parameter family of vacuum metric structures. When  $k_\mu= (\omega,\omega,0,0)$ these are described by 
\begin{equation}\label{eq:gravwaves}
h_{\mu\nu}[\bar\rho](\omega,\omega,0,0)=
\frac{1}{2 \pi^2 \omega^2}\begin{pmatrix}
0
	& 0 
		&0
			 & 0\\
0 
	& 0 
		& 0
			& 0 \\
0
	&  0
		& r_{20}
			& \tilde{r}_{20}\\
0
 	& 0
 		& \tilde{r}_{20}
 			&  - r_{20}\\
\end{pmatrix},
\end{equation}
where we immediately recognize $r_{20}$ and $\tilde{r}_{20}$ as the coefficients of the familiar $+$ and $\times$ polarizations of gravitational waves. 

This answers the question we posed in the introduction of whether any of the vacuum structure of general relativity would be recovered in the continuum limit of our model. The answer turns out to be affirmative. In fact, we find that all vacuum (i.e. Ricci flat) metrics that are available in linearized general relativity may be found as the continuum limit of a sequence of configurations of physical defects.

\section{Conclusions}
We have studied continuous distributions of conical defects. In the limit of weak fields, considering only linear contributions, the energy--momentum tensor of such a distribution turns out to satisfy certain conditions. These conditions can be expressed as follows for the Fourier transform of the energy--momentum tensor,
 \begin{enumerate}
\item $k^\mu T_{\mu\nu}(k_\lambda) = 0$ for all $k_\mu$.
\item $T_{\mu\nu}(k_\lambda) = 0$ for all $k_\mu$ with $k^\mu k_\mu < 0$.
\item $l^\mu T_{\mu\nu}(k_\lambda)$ is a non-spacelike vector for all lightlike $k_\mu$ and $l_\mu$.
\end{enumerate}
The first and third conditions are satisfied by most reasonable classical theories. The second condition, shows an inability to model wave-like phenomenon at the linear order. This property is shared by other non-interacting matter models, like dusts. Beyond linear order line defects collide in a non-trivial manner causing violations of this second condition. It would be very interesting to see if any restrictions remain once all possible interactions are included.

The metric perturbation caused by a configuration of defects turns out to be mostly fixed by its energy--momentum. That is if two continuous distributions of defects have the same energy--momentum tensor, they also produce the same linear metric perturbation, at least for most modes. 

The lightlike modes of the metric perturbation form an exception, as they are only partially fixed by the energy--momentum tensor. The metric perturbations of two configurations with the same energy--momentum may differ by a transverse traceless lightlike mode, i.e. by a gravitational plane wave. We thus see that the piecewise flat model of propagating defects --- even though it does not contain any a priori vacuum structure --- recovers the vacuum structure of general relativity in the continuum limit, at least at the linearized level.
\ifx\fullTeX\undefined
\bibliographystyle{../bib/utcaps}
\bibliography{../bib/thesis}
\end{document}
\fi

\cleardoublepage 
\ifx\fullTeX\undefined
\documentclass[11pt,a4paper]{article}
\usepackage[utf8]{inputenc} 
\usepackage[english]{babel}
\usepackage{graphicx}
\usepackage{hyperref}
\usepackage{color}
\usepackage{amsmath}
\usepackage{amsthm}
\usepackage{amssymb}
\usepackage{bbm}
\usepackage{pictexwd,dcpic}

\graphicspath{{../pics/}}

\title{Gravitational Waves}
\author{Maarten van de Meent}
\date{\today} 

\begin{document}
\maketitle
\else
\chapter{Gravitational waves}\label{ch:gravwave}
\fi

In the previous chapter we showed that our piecewise flat model of gravity contains gravitational waves as an emergent feature in the linearized continuum limit. In this chapter we extend this result to exact gravitational waves. In contrast to linearized general relativity, it is not possible to derive completely  general exact vacuum solutions of general relativity. Deriving exact solutions requires some form of additional symmetry.  For example, it is known how to construct completely general gravitational waves with cylindrical symmetry,\cite{Weber:1957} and how to construct exact gravitational plane waves.\cite{bondi:1959}

Here we will focus on plane gravitational waves and show how to construct a family of increasingly fine discrete configurations of defects that approach an exact gravitational wave in the limit that the configuration becomes continuous.

Our strategy will be to first consider gravitational plane waves in the linearized limit, and examine what linearized distributions of defects approximate these waves (section \ref{sec:linplanewaves}). Based on the linearized configuration, we will make an ansatz for an exact piecewise flat planar wavefront, which reduces to the linearized distribution in the linear limit. We show that this ansatz is indeed a bona-fide piecewise flat geometry, and that in the limit where the configuration becomes continuous it produces an impulsive gravitational wave solution of general relativity (section \ref{sec:planewavefronts}). In section \ref{sec:genplanewaves} we combine trains of these wavefronts to form general exact gravitational plane wave solutions. Finally, in section \ref{sec:otherwaves} we comment on the construction of non-planar waves and the additional subtleties that those entail.

\section{Linear plane waves}\label{sec:linplanewaves}
For simplicity we shall consider, without loss of generality, plane waves travelling in the positive $z$-direction. Expressed in lightcone coordinates $(u,v,x,y)$,\footnote{Recall our conventions that spacetime metrics have signature $(- + + +)$, lightcone coordinates are defined as $u=\tfrac{z+t}{\sqrt{2}}$ and $v=\tfrac{z-t}{\sqrt{2}}$, and we use units such that $c = \hbar = 8\pi G =1$.} such a wave takes the following form in linearized general relativity,
\begin{equation}\label{eq:gwpert}
h_{\mu\nu} = \begin{pmatrix}
0 & 0 & 0 & 0\\
0 & 0 & 0 & 0\\
0 & 0 & h_{+} & h_{\times}\\
0 & 0 & h_{\times} & -h_{+}\\
\end{pmatrix}\ee^{\ii \omega u},
\end{equation}
where $h_{\mu\nu}$ is the perturbation of the metric on the Minkowski background. In chapter \ref{ch:cont} we saw how to produce such a metric perturbation in linearized piecewise flat gravity. The basic building block was a configuration that we called a ``laminar plane wave of defects''. A laminar plane wave of defects is a space-filling configuration of parallel conical curvature defects (i.e. line defects that have the same orientation and velocity), whose energy density varies as the amplitude of a plane wave with a wave vector perpendicular to the defects. That is, given a point $x^\mu$ in our spacetime there is exactly one defect (spanned by 4-vectors $d^\mu$ and $u^\mu$) in the laminar plane wave configuration that passes through this point, because all the defects in the laminar plane wave are parallel. The linear energy density of this defect is given  by $\Omega \ee^{\ii k_\mu x^\mu}$ with some fixed wave vector $k_\mu$, which is perpendicular to $d^\mu$ and $u^\mu$.

\begin{figure}[btp]
\centering
\includegraphics[width=120mm]{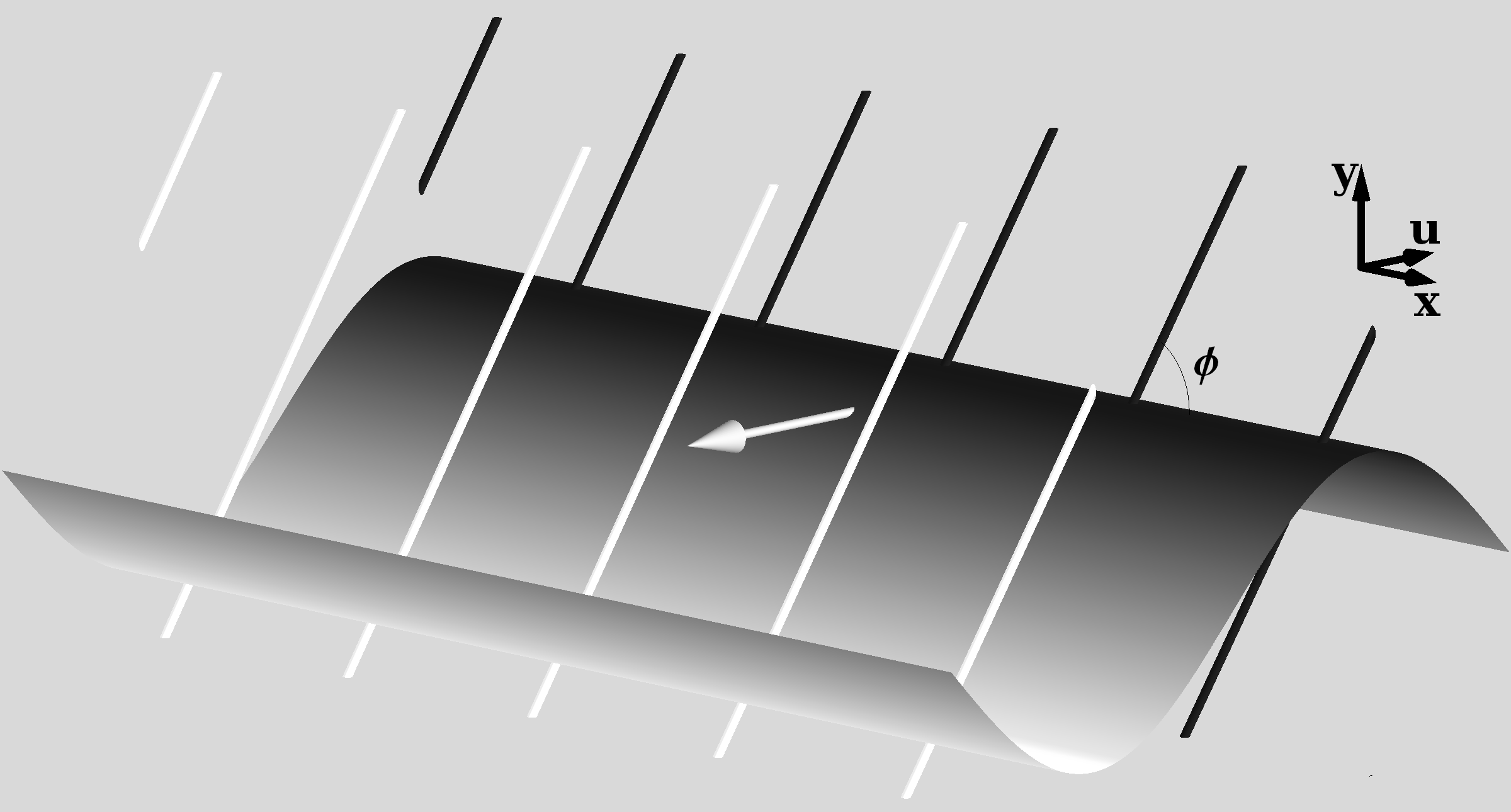}
\caption{A laminar plane wave consists of parallel line defects travelling in the same direction, whose densities vary as a plane wave in the direction of propagation.}\label{fig:laminarplanewave}
\end{figure}

In the linearized limit of our piecewise flat model an arbitrary configuration of defects may be written as a linear combination of laminar plane waves. In chapter \ref{ch:cont} we found that a gravitational wave of the form \eqref{eq:gwpert} was produced as a superposition of laminar plane waves of defects  with the velocity $u^\mu=(1,0,0,1)$ and wave vector $k^\mu=(\omega, 0,0,\omega)$. The orientation $d_\mu$ of the defects in these laminar plane waves can then be parametrized by a single angle, $\phi$, which we choose to be the angle between the defect lines and the $x$-axis.

To be a gravitational wave the total energy density amplitude of the superposition of laminar plane waves must vanish. That is, if the energy density amplitude of the laminar plane wave with orientation $\phi$ is called $\Omega(\phi)$, then this condition can be expressed as
\begin{equation}\label{eq:omegacond}
  \int_0^\pi\Omega(\phi) \md\phi = 0.
\end{equation}
It was established in section \ref{sec:gravwave} that this is the only condition on the superposition of laminar plane waves of defects required for it to describe a gravitational plane wave.  In equation \eqref{eq:gravwaves}, the polarization of the resulting gravitational wave was shown to be given by
\begin{equation}
\begin{aligned}
  h_{+} & =  r_{20} =\int_0^\pi\Omega(\phi)\cos(2\phi) \md\phi,\\
 h_{\times} & =  \tilde{r}_{20} = \int_0^\pi\Omega(\phi)\sin(2\phi) \md\phi.
\end{aligned}
\end{equation}
All higher moments of $\Omega(\phi)$ are unconstrained. That is, there is a large family of linear configurations of defects that produce the same linear gravitational wave. We therefore have quite a bit of playing room for finding a linearized configuration that can be interpreted as the limit of some exact configuration of defects.

Geometrically, it would be convenient if $\Omega(\phi)$ has as few discrete components as possible. Condition \eqref{eq:omegacond} implies that we need at least two components. This is, in fact, sufficient to create waves of any polarization. If we take
\begin{equation}
  \Omega(\phi)  =  \alpha\hh{ \delta(\phi-\phi_0) -  \delta(\phi-\phi_0-\tfrac{\pi}{2})},
\end{equation}
then the polarizations are given as $h_{+}= 2\alpha \cos2\phi_0$ and $h_{\times}= 2\alpha \sin2\phi_0$. We thus have perpendicular components of opposite energy, with the amplitude of the defects determining the amplitude of the wave, and the orientation determining the polarization.

\section{Plane wavefronts}\label{sec:planewavefronts}
\begin{figure}[btp]
\centering\includegraphics[width=120mm]{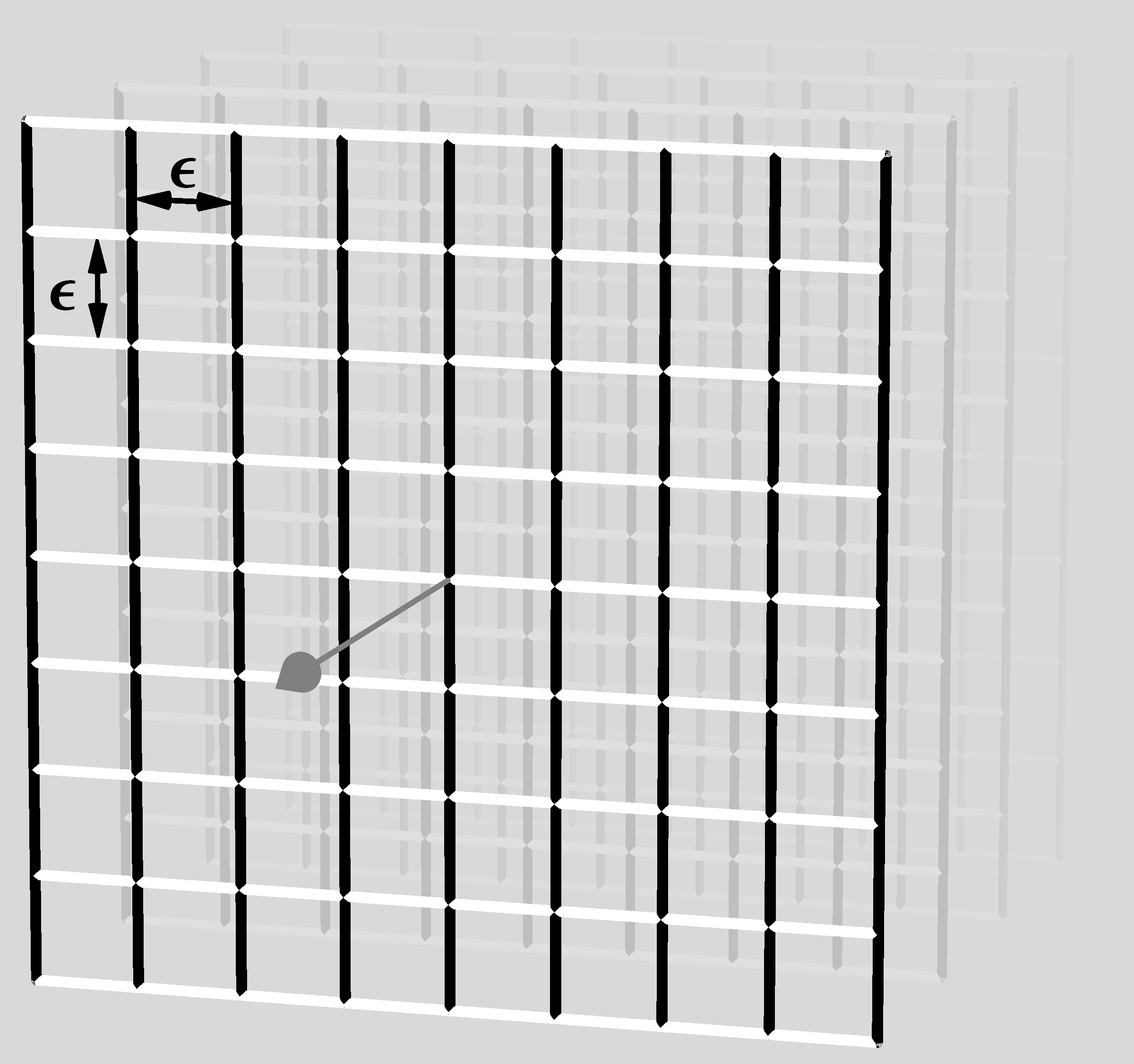}
\caption{A wavefront of perpendicular positive (white) and negative (black) energy line defects.}\label{fig:gridfront}
\end{figure}

This linear configuration prompts us to make an ansatz for an exact configuration of defects. Further discretizing the linear distribution, we guess that a configuration consisting of subsequent grids of perpendicular positive and negative energy massless defects should suffice to approximate a gravitational wave. To analyse this configuration we first consider a wavefront located at $u=\tfrac{z+t}{\sqrt{2}}=0$ consisting of a single grid (see figure \ref{fig:gridfront}).

We first need to check if this ansatz defines a proper configuration of defects. In \cite{hooft:2008} the conditions that should be met at a junction of multiple defects were discussed. These can be phrased in terms of the holonomies of loops around the defect lines. They can be summarized as saying that the product of the holonomies of a set of loops whose concatenation is contractible should be the identity. For the proposed grid configuration the junction conditions must be satisfied at each vertex of the grid.

Since all the horizontal line defects in figure \ref{fig:gridfront} have the same density, the holonomy of each simple loop around a section of a horizontal line defect should be the same, and similarly for the holonomies of simple loops around vertical line defects.   If we call the holonomy of a simple loop around a section of a horizontal grid lines $Q_a$ and the holonomies of the vertical grid lines $Q_b$, then at each vertex the junction condition is

\begin{equation}\label{eq:ewjc}
Q_a Q_b {Q_a}^{-1}{Q_b}^{-1} = \Id.
\end{equation}

The holonomy of a lightlike defect is a null rotation, i.e. a Lorentz transformation that leaves a lightlike and a perpendicular spacelike direction invariant. In the $\PSL{2,\CC}$ representation of the Lorentz group, the holonomy of a general lightlike defect moving in the $z$-direction is represented as
\begin{equation}
Q_z = \begin{pmatrix}
1 & \zeta \\
0 & 1 
\end{pmatrix},
\end{equation}
where $\zeta$ is a complex number. The argument of $\zeta$ gives the direction of the defect, while its modulus gives the energy density of the defect, i.e. the parabolic angle of the null rotation. The junction condition for $n$ lightlike defects moving in the $z$ direction with complex parameters $\zeta_1,\ldots, \zeta_n$ therefore reduces to
\begin{equation}
\zeta_1 +\cdots +\zeta_n =0.
\end{equation}
Furthermore if $Q_a$ has complex parameter $\zeta_a$ then ${Q_a}^{-1}$ has complex parameter $-\zeta_a$. Hence the junction condition for the vertices of the grid \eqref{eq:ewjc} reduces to
\begin{equation}\label{eq:gridjunction}
\zeta_a + \zeta_b - \zeta_a -\zeta_b =0,
\end{equation}
which is true for any $\zeta_a$ and $\zeta_b$. Our ansatz configuration of a grid of perpendicular lightlike defects is therefore a bona-fide configuration of defects for any value of the amplitudes of its components. In particular, it is valid for the particular case we are interested in, where the components have opposite amplitudes.

Our next step is to compare this configuration with an exact plane wave solution of the vacuum Einstein equations. For this we need to find a metric to describe the geometry of the grid of defects.

Away from the grid the spacetime is flat. Hence, if we position the grid at $u=0$, we can choose the metric for $u<0$ to have the familiar Minkowski form,
\begin{equation}\label{eq:lcMinmetric}
\md s^2 = 2\md u \md v+ \md x^2 +\md y^2.
\end{equation}
For simplicity we choose to align the grid with the $x$ and $y$-axes,\footnote{This should correspond to wavefront with ``$+$'' polarization, according to the linear analysis.} such that the positive energy defects are aligned in the $x$-direction and sit at constant values of $y = \epsilon(n  +\tfrac{1}{2})$ for integer values of $n$ and grid spacing $\epsilon$, and the negative energy defects are oriented in the $y$-direction and sit at constant values of $x = \epsilon(m+\tfrac{1}{2})$. 

\begin{figure}[tbfp]
\centering
\includegraphics[width=120mm]{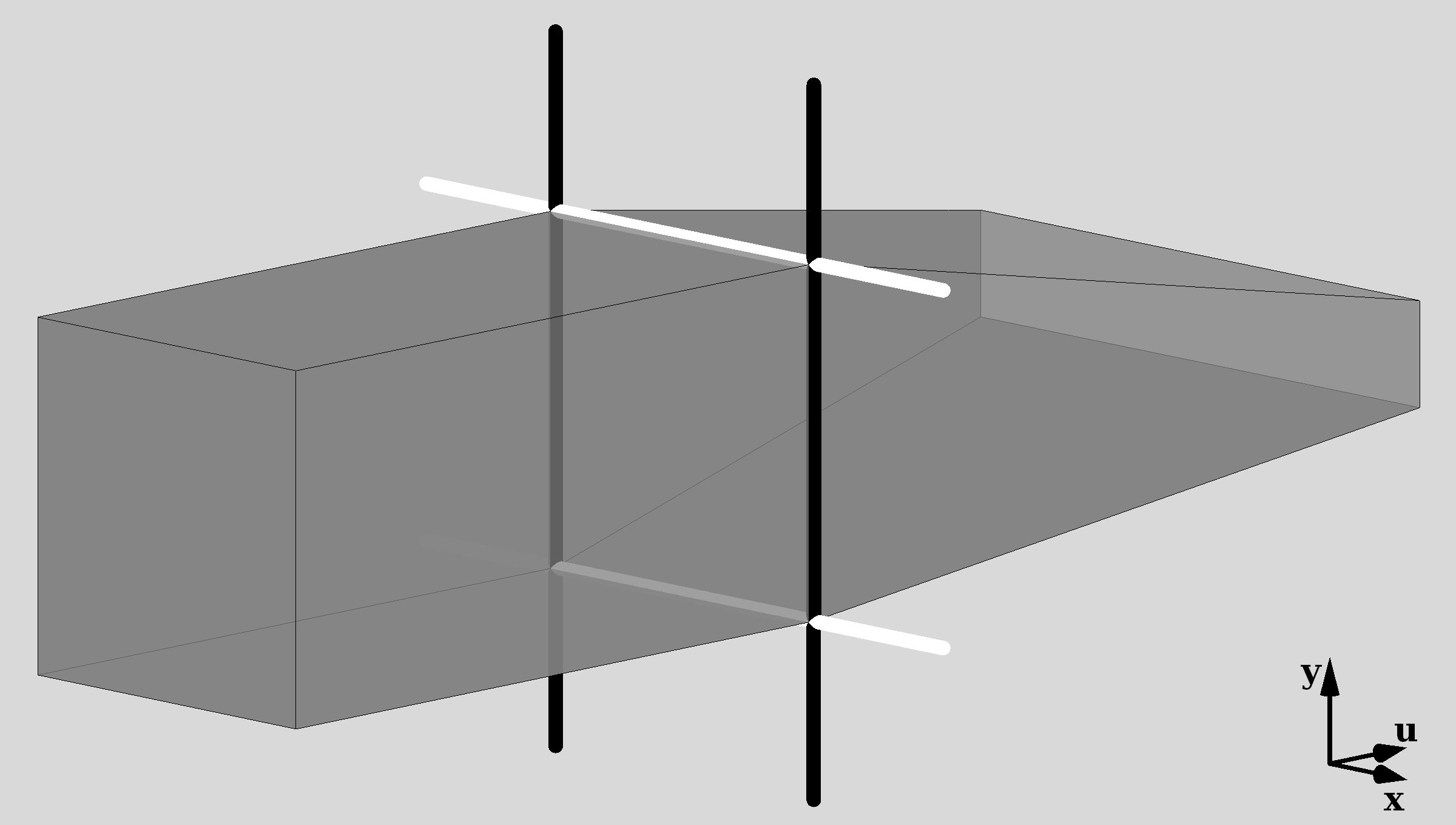}
\caption{The geometry of one ``tube'' of spacetime.The deficit angles of the positive energy defects (white), cause the geometry to be squeezed in the $y$-direction as $u$ increases. Similarly the surplus angles of the negative energy defects (black) cause the geometry to be elongated in the $x$-direction.}\label{fig:squaretube}
\end{figure}

The defect and surplus angles of the grid lines can be oriented along the direction of propagation of the grid, i.e. the positive $u$ direction. The cuts of the defect/surplus angles then divide the $u>0$ side of the grid in square ``tubes'' of spacetime (see figure \ref{fig:squaretube}). The Minkowski metric from $u<0$ side of the grid can be continued into each tube. The constant $u$ and $v$ slices in each tube start out as a square at the base ($u=0$) of the tube. As $u$ increases the square is elongated in the $x$ direction and compressed in the $y$ direction.

We now construct coordinates on each tube, which can be patched together to one coordinate system on the $u>0$ side of the spacetime. The space
\begin{equation}
T_{m,n}=\RR_{>0}\times\RR\times [\epsilon(m-\tfrac{1}{2}),\epsilon(m+\tfrac{1}{2})]\times [\epsilon(n-\tfrac{1}{2}),\epsilon(n+\tfrac{1}{2})]
\end{equation}
can be mapped to the tube behind the face of the grid centered on $x=\epsilon m$ and $y=\epsilon n$ by the mapping,
\begin{equation}
\begin{aligned}
 u &\mapsto u,\\
 v &\mapsto v,\\
 x &\mapsto \epsilon m+(1+\frac{\alpha}{\epsilon} u)\bhh{x-\epsilon m},\\
 y &\mapsto \epsilon n+(1-\frac{\alpha}{\epsilon} u)\bhh{y-\epsilon n}.
\end{aligned}
\end{equation}
The metric induced on $T_{m,n}$ by this mapping is,
\begin{equation}\label{eq:Tmnmetric}
\md s^2 = 2\md u \md v+
\hh{(1+\tfrac{\alpha}{\epsilon} u)\md x + \tfrac{\alpha}{\epsilon}(x-\epsilon m)\md u}^2+
\hh{(1-\tfrac{\alpha}{\epsilon} u)\md y -\tfrac{\alpha}{\epsilon}(y-\epsilon n)\md u}^2.
\end{equation}
Together, the patches $T_{m,n}$ cover the entire $u>0$ side of the grid. The metric \eqref{eq:Tmnmetric} may be viewed as a metric on the union of all $T_{m,n}$, $\RR_{>0}\times\RR^3$, if $m$ and $n$ are interpreted as functions of $x$ and $y$ respectively. That is,\footnote{Here $\floor{x}$ denotes the floor function, which returns the largest integer smaller than or equal to $x$.}
\begin{equation}
\begin{aligned}
m(x) &= \floor{x/\epsilon + 1/2},\\
n(y) &= \floor{y/\epsilon + 1/2}.
\end{aligned}
\end{equation}
These functions are not continuous, so we need to worry whether \eqref{eq:Tmnmetric} describes a well-defined geometry on the whole $u>0$ half-space. The appropriate condition \cite{Clarke:1987,Mars:1993mj} is that the metric induced on the common boundary of two patches is the same in both patches.

On the $x=\epsilon(m+\tfrac{1}{2})$ boundary the metric on $T_{m,n}$ induces the 3-dimensional metric,
\begin{equation}
\md s^2 = 2\md u \md v+
\hh{(1-\tfrac{\alpha}{\epsilon} u)\md y -\tfrac{\alpha}{\epsilon}(y-\epsilon n)\md u}^2.
\end{equation}
On the other side of the boundary the metric on the $T_{m+1,n}$, can easily be seen to induce the same 3-dimensional metric. Similarly, we find that the 3-dimensional metrics induced on the $y=\epsilon(n+\tfrac{1}{2})$ boundaries match, and hence that the metric \eqref{eq:Tmnmetric} is well-defined on the whole $u>0$ half-space.

Moreover, the 3-dimensional metric induced on the $u=0$ hyperplane is
\begin{equation}
\md s^2 = \md x^2+  \md y^2,
\end{equation}
which agrees with the metric induced from the Minkowski side. Consequently, the metric \eqref{eq:lcMinmetric} for $u<0$ and the metric \eqref{eq:Tmnmetric} for $u>0$ together describe a well-defined geometry on the whole spacetime, given by the metric
\begin{multline}
\md s^2 = 2\md u \md v+
\hh{(1+\tfrac{\alpha}{\epsilon} u\theta(u))\md x + \tfrac{\alpha}{\epsilon}(x-\epsilon m)\theta(u)\md u}^2\\
+\hh{(1-\tfrac{\alpha}{\epsilon} u\theta(u))\md y -\tfrac{\alpha}{\epsilon}(y-\epsilon n)\theta(u)\md u}^2,
\end{multline}
where $\theta(u)$ is the Heaviside stepfunction.

This metric should describe the geometry of our piecewise flat configuration of defects. As a check, we can calculate the Riemann tensor of this metric. Since the metric is discontinuous along the boundaries of the patches, we need to treat the curvature tensors as tensor distributions as described in \cite{Mars:1993mj}. The upshot of this is that in a case like this we can and should treat the metric components as distributions rather than normal functions. Using the fact that in a distributional sense
\begin{align}
\frac{\md}{\md x} \floor{x} &= \sum_{i\in\ZZ} \delta(x - i)\equiv \Delta(x),\quad\text{and}\\
\frac{\md^2}{\md u^2} u\theta(u) &= \delta(u),
\end{align}
we find that the Riemann tensor is given by\footnote{For notational convenience we denote the tensor product of a 2-form with itself by a square, i.e. $(\md u\wedge\md y)^2 = (\md u\wedge\md y)\otimes(\md u\wedge\md y)$.}
\begin{equation}
R = 4\alpha \delta(u)\bhh{\Delta_\epsilon(y+\tfrac{\epsilon}{2})(\md u\wedge\md y)^2 -\Delta_\epsilon(x+\tfrac{\epsilon}{2})(\md u\wedge\md x)^2},
\end{equation}
where $\Delta_\epsilon(x)$ is the Dirac comb with period $\epsilon$, i.e.
\begin{equation}
\Delta_\epsilon(x)\equiv  \sum_{i\in\ZZ} \delta(x - \epsilon i).
\end{equation}
We observe that the Riemann tensor indeed vanishes everywhere, except on the grid of defects where it has singular delta peaks as expected. We have therefore found a metric for our piecewise flat spacetime consisting of a grid of lightlike defects.

Our next step is to find the continuum limit of this spacetime. For this we will send the grid size $\epsilon$ to zero while keeping the density of defects (approximately) constant. The number of defects in a unit area of the grid scales inversely proportionately  to $\epsilon$, hence the amplitude of the defects $\alpha$ should scale linearly with $\epsilon$ to keep the total amplitude per unit area constant. Inserting $\alpha=\epsilon \alpha_0$ in the above metric and taking the limit of $\epsilon$ goes to zero, we find
\begin{equation}\label{eq:impulsivewave}
\md s^2 = 2\md u \md v+
\hh{1+\alpha_0 u\theta(u)}^2 \md x^2 + \hh{1-\alpha_0 u\theta(u)}^2 \md y^2,
\end{equation}
where $\theta(u)$ is the Heaviside step function. The Riemann tensor (distribution) is given by
\begin{equation}
R = 4\alpha_0 \delta(u)\bhh{(\md u\wedge\md y)^2 -(\md u\wedge\md x)^2},
\end{equation}
from which the Einstein tensor may be obtained, which is seen to vanish. Consequently, the metric \eqref{eq:impulsivewave} is an exact solution of the vacuum Einstein equation, a gravitational wave consisting of a single plane wavefront. Waves of this type are called impulsive  gravitational plane waves, and were first considered by Penrose.\cite{Penrose:1972}

\section[General waves]{General gravitational plane waves}\label{sec:genplanewaves}
We have succeeded in explicitly constructing a family of exact defect configurations that converges to a specific exact gravitational wave solution of general relativity in the limit that the configuration becomes continuous. In the construction we made some explicit choices for the sake of simplicity. We chose to align the grid with the coordinate axes and as a result described a plane wavefront with ``$+$'' polarization. Rotating the grid simply results in a rotation of the $\md x$ and $\md y$ terms in \eqref{eq:impulsivewave}, producing a wavefront with a different polarization. For example, rotating the grid by $\pi/4$ produces a wavefront with ``$\times$'' polarization. The result always is an impulsive plane wave with constant polarization. 

We could also take a different geometry for the grid. Equation \eqref{eq:gridjunction} shows that the components of the grid do not need to be perpendicular, nor do the components have to have matching amplitudes. As long as the curvature density of the positive curvature defects is equal to the curvature density of the negative curvature defects, the Einstein tensor will vanish in the continuum limit. Different geometries will not produce any wavefronts that could not be produced by a square grid.

Gravitational waves with longer wave packets may be constructed as a succession of wavefronts. Each wavefront contributes to the Riemann  curvature in the following way,
\begin{equation}
R = \alpha \delta(u-u_0)\bb{\bhh{\md u\wedge(\cos\phi\;\md x+\sin\phi\;\md y)}^2 -\bhh{\md u\wedge(\sin\phi\;\md x-\cos\phi\;\md y)}^2},
\end{equation}
where $\phi$ is the direction of the polarization of the wavefront.

A continuous succession of wavefronts with $\alpha$ and $\phi$ varying with $u$ therefore produces a wave with curvature,
\begin{equation}
R = \alpha(u)\bb{\bhh{\md u\wedge(\cos\phi(u)\;\md x+\sin\phi(u)\;\md y)}^2 -\bhh{\md u\wedge(\sin\phi(u)\;\md x -\cos\phi(u)\;\md y)}^2}.
\end{equation}
This is the most general curvature produced by exact gravitational plane waves of the type described by Bondi et al. in \cite{bondi:1959}.

\section{Other gravitational waves}\label{sec:otherwaves}
We have shown that it is possible to construct general gravitational plane waves in our piecewise flat model of gravity. Since a general gravitational wave will locally resemble a plane wave, this suggests that it should be possible to construct any gravitational wave. This is further supported by the observations that any gravitational wave may be obtained in the linear limit of the model, and that its is possible to find a piecewise flat approximation to any solution of general relativity.\cite{CMS:1984} Although this is all very suggestive, it is not enough to conclude that any gravitational wave may be constructed in our piecewise flat model of gravity. It is therefore instructive to explicitly examine the construction of a non-planar wave in our model and see what new issues arise.

As an example, we will consider cylindrical gravitational waves. In the case of cylindrical symmetry the solutions of the vacuum Einstein equation can found by solving a single linear differential equation.\cite{Einstein:1937qu,Weber:1957} Consequently, the complete set of possible solutions is known. In particular, it is known that it is impossible to form isolated impulsive wavefronts. Instead, impulsive wavefronts leave a wake of backscattered gravitational waves. Explicit solutions with impulsive wavefronts \cite{Alekseev:1991kz} show that the form of this backscattered component can be quite complicated. This suggests that it may be too much to hope for a simple description of the backscattered component in terms of a piecewise flat geometry.\footnote{This says nothing about whether a piecewise flat description exists. We simply observe that constructing one will be very complicated.} Instead we will focus on constructing an impulsive cylindrical gravitational wave front, while we allow the backscattered component to be any defect configuration, i.e. not necessarily one with vanishing energy--momentum.

\begin{figure}[tb]
\centering
\includegraphics[width=120mm]{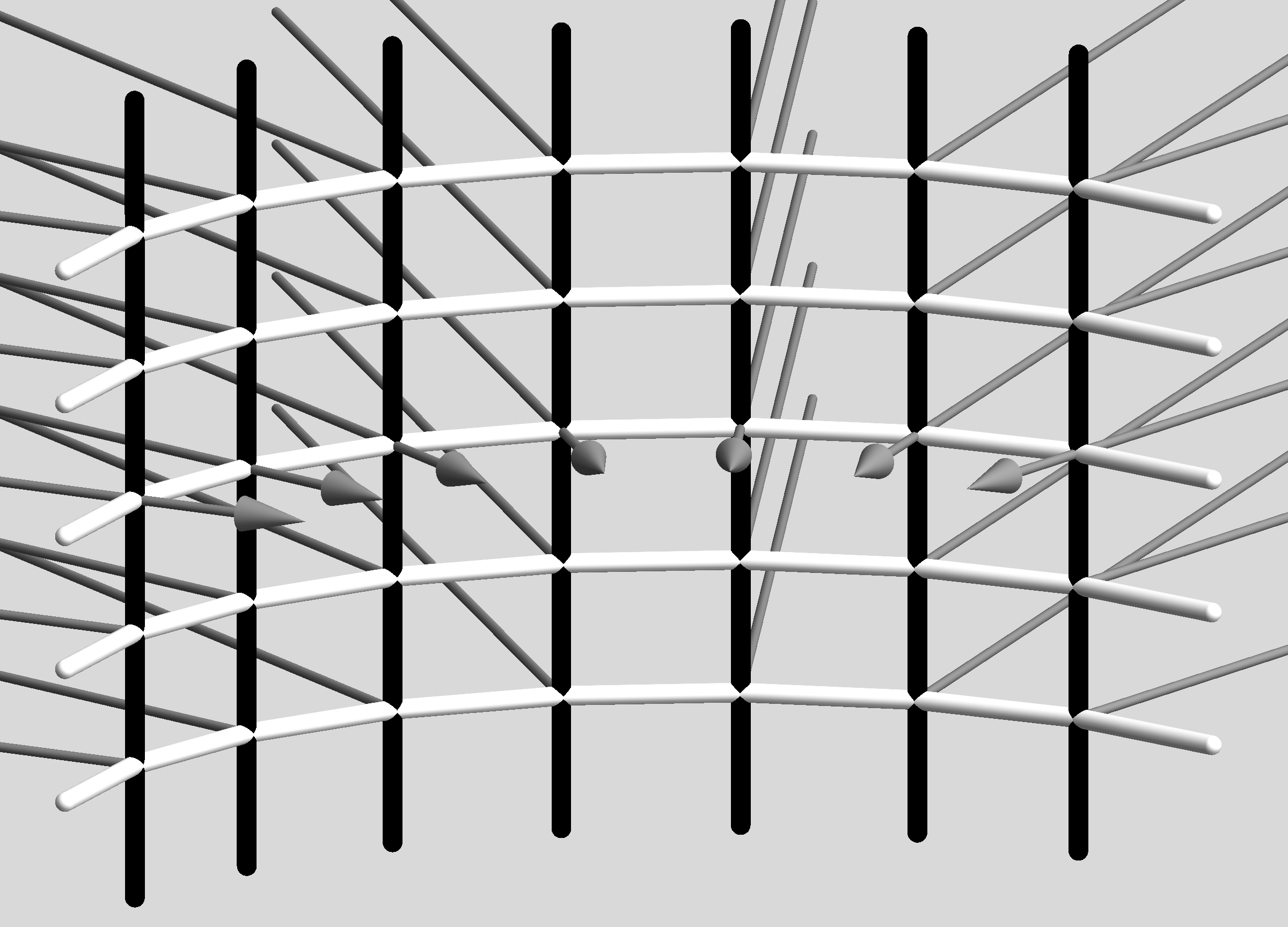}
\caption{To first order, a piecewise flat cylindrical wave front should consist of axial (black) and azimuthal (white) defects moving inwards in the radial direction. The junction conditions imply that any additional radial defect (gray) should be present at each junction.}\label{fig:cylindricalfront}
\end{figure}

Let us consider a single ingoing cylindrical impulsive wavefront located at $u=0$,\footnote{We use cylindrical lightcone coordinates defined by  $u=\tfrac{r+t}{\sqrt{2}}$ and $v=\tfrac{r-t}{\sqrt{2}}$.} which propagates into a Minkowski background (i.e. the Riemann tensor vanishes for $u<0$). In the limit of large radii ($v\rightarrow \infty$), the wavefront will resemble an impulsive plane wave. Our first order guess for the piecewise flat cylindrical wavefront therefore consists of a grid of perpendicular massless defects with positive and negative energy oriented in the axial and azimuthal directions and moving in the radial direction (see figure \ref{fig:cylindricalfront}). 

Curving a flat grid to a cylinder introduces two new complications. The first
is that due to simple geometric considerations the velocity of the azimuthal defects must be smaller than the velocity of the axial defects. The azimuthal defects, therefore, are not quite massless. However, as the grid is made finer, this effect disappears and the azimuthal defects again become massless in the continuum limit. The second effect is that the four defects of the grid meeting at each junction do not lie in one plane. A consequence of this is that the junction condition \eqref{eq:ewjc} cannot be satisfied by these four defects. To satisfy \eqref{eq:ewjc} a fifth defect pointing in the outward radial direction must be added to each junction (depicted as the gray lines in figure \ref{fig:cylindricalfront}). Just like general relativity, the piecewise flat model does not allow isolated cylindrical wavefronts. 

Repeating the steps of section \ref{sec:planewavefronts} we can find the piecewise flat metric associated to this configuration and calculate the continuum limit. The result is
\begin{equation}
\md s^2 = 2\md u \md v + \frac{(u+v+\alpha u\theta(u))^2}{2}\md\phi^2 + (1-\alpha u \theta(u))^2 \md z^2.
\end{equation}
The corresponding energy--momentum tensor is
\begin{equation}
T=\alpha \frac{v-1}{v} \delta(u) \md u^2 - 2\alpha \frac{1}{(1-\alpha u)(v+(1+\alpha)u)}\theta(u)\md u\md v.
\end{equation}
As expected the energy--momentum vanishes in the $u<0$ region, but does not in the `backscattered' $u>0$ region due to the presence of the radial defects. However, the energy--momentum also has a non-vanishing delta peak at $u=0$ for $v\neq 1$, i.e. it fails to model a cylindrical impulsive gravitational wavefront. This failure is the result of the two singular components of the Riemann tensor,
\begin{equation}
R = 4 \alpha \delta(u) (\md u\wedge\md z)^2  -4\frac{\alpha}{v}\delta(u)(\md u\wedge\frac{v}{\sqrt{2}}\md\phi)^2 +\theta(u)\bhh{\cdots},
\end{equation}
scaling differently with $v$. The expected behaviour --- as can be deduced from linear cylindrical waves, or by expanding the exact cylindrical solution in \cite{Alekseev:1991kz} for small values of $u$ --- is for both curvature components to scale inversely with $\sqrt{v}$.

On the level of the piecewise flat geometry, this happens because the number of axial defects stays constant while the radius changes, causing their density to scale inversely with the radius. The density of the azimuthal defects, on the other hand, stays constant. To change the scaling of the impulsive front, the energy densities of the defects must change as the defects move inward. The only way this can happen is if the defects periodically emit new defects into the backscattered region.

\begin{figure}[tbp]
\centering
\includegraphics[width=120mm]{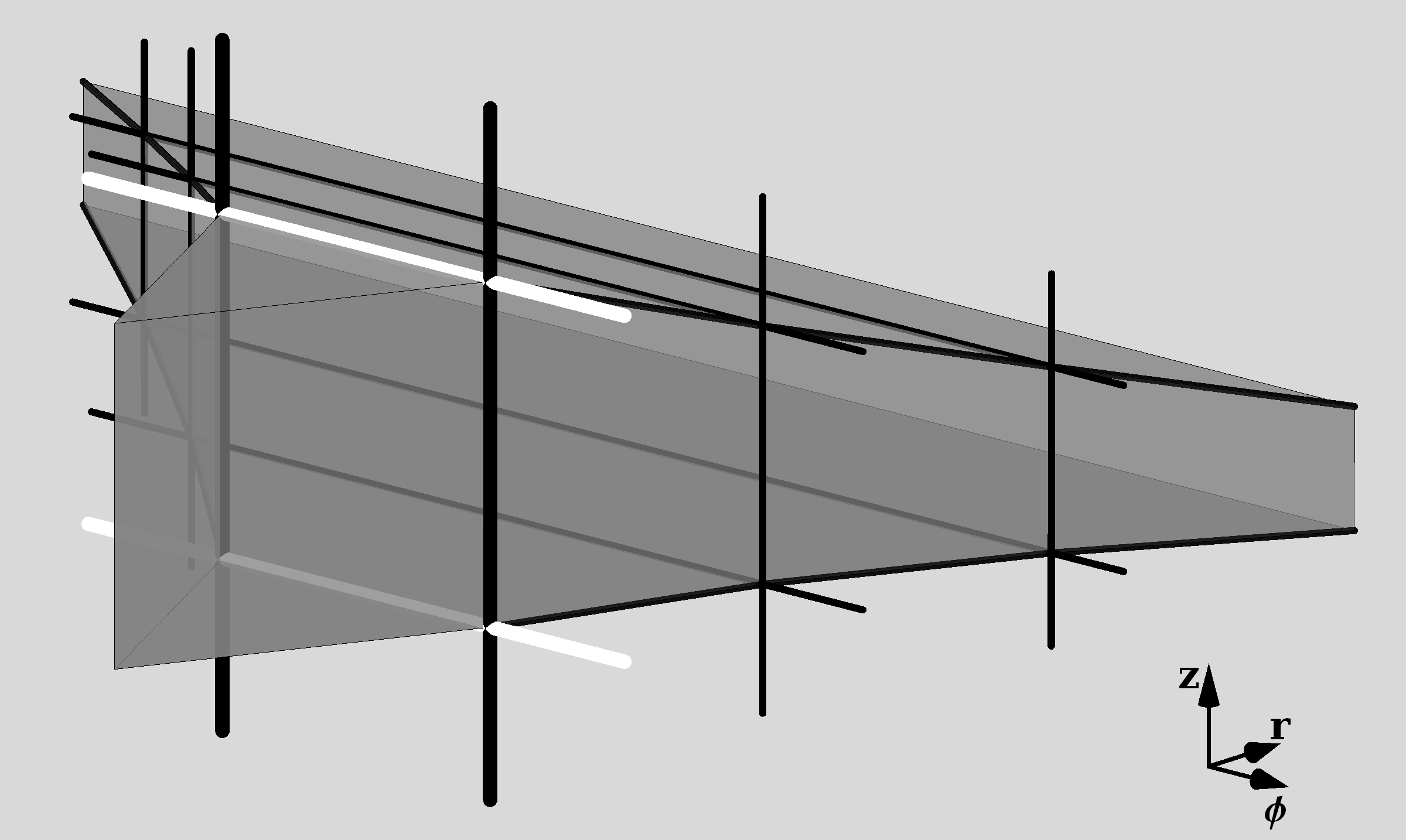}
\caption{The geometry of a single cell of the piecewise flat geometry approximating an impulsive cylindrical gravitational wave front.}\label{fig:cylindricaltube}
\end{figure}

We implement this by having the ingoing impulsive front emit an outward moving lightlike front of perpendicular defects. A single cell of the piecewise flat geometry then takes shape shown in figure \ref{fig:cylindricaltube}. By adjusting the frequency and the energies of the emitted perpendicular components we can achieve any scaling with $v$ that we want. To achieve the proper scaling for the curvature of a cylindrical wave front, we choose the emitted components such that the metric in the continuum limit becomes,
\begin{equation}
\md s^2 = 2\md u \md v + \frac{(u+v+\alpha\sqrt{v} u\theta(u))^2}{2}\md\phi^2 + (1-\alpha \frac{u}{\sqrt{v}} \theta(u))^2 \md z^2.
\end{equation}
The singular part of the corresponding energy--momentum tensor is
\begin{equation}
T = 0 +\theta(u)\bhh{\cdots}.
\end{equation}
The impulsive wavefront has thus vanished from the energy--momentum tensor. Yet, the singular part of the Riemann curvature,
\begin{equation}
R = 4 \frac{\alpha}{\sqrt{v}} \delta(u) (\md u\wedge\md z)^2  -4\frac{\alpha}{\sqrt{v}}\delta(u)(\md u\wedge\frac{v}{\sqrt{2}}\md\phi)^2 +\theta(u)\bhh{\cdots},
\end{equation}
does not vanish, and has the expected scaling for a cylindrical impulsive wavefront. The described configuration indeed contains a cylindrical impulsive wavefront.

The described piecewise flat geometry is just one way an impulsive cylindrical gravitational wavefront could evolve in our piecewise flat model of gravity. Recall that the model fundamentally describes gravity coupled to matter. As discussed in chapter \ref{ch:collisions}, the dynamics of the model is incomplete.\footnote{This is related to the different ways in which a piecewise flat geometry can be continued after the collision of two defects.} It needs to be complemented by additional rules which relate hoe the matter content interacts, much like general relativity needs to be complemented by equations of motion for the matter content. Different dynamical completions of the model lead to different evolutions of cylindrical wavefronts. The specific evolution described here corresponds a matter interaction where interacting gravitational waves can produce other types of matter.  

One might ask if there exists a dynamical completion of the model that corresponds to pure gravity with no matter present at all. Of course, this is of academic interest only since the physical world also contains matter. In such a completion the backscattered component of the evolving cylindrical wavefront must also satisfy the vacuum Einstein equation. This requires the addition of additional defects to the configuration. These will break the remaining residual symmetry of the piecewise flat geometry. As a consequence it will no longer be possible to describe the geometry as an infinite collection of identical piecewise flat cells, substantially adding to the complexity of the description. This breaking of symmetry indicates that cylindrical gravitational waves may not be the best setting to further study the dynamical completion of the piecewise flat model.

Nonetheless, the study of the interaction of gravitational waves in the piecewise flat model may provide essential guiding insights for dynamically completing the model. A more promising setting for this study is the interaction of two plane wavefronts. This problem is well documented in the case of general relativity coupled to various types of matter.\cite{Khan:1971vh,Griffiths:1991,Barrabes:2002yg,Stephani:2003} Moreover, the intrinsic plane symmetry is much more suited to a piecewise flat geometry. Furthermore, it is well known that the future of plane wave collisions can contain regions which are locally diffeomorphic to black hole solutions.\cite{Khan:1971vh, Stephani:2003} Studying plane wave collisions may therefore offer crucial insight in one of the major open problems of the piecewise flat model; the construction of non-trivial stationary vacuum solutions such as black holes. However, the study of dynamical completions of the model is a major project on its own and will not be attempted here.

\section{Conclusions}
In this chapter we have extended our previous linear result from chapter \ref{ch:cont} for obtaining gravitational waves in our piecewise flat model for gravity in 3+1 dimensions to an explicit construction of a family of exact piecewise flat configurations that approaches exact plane wave solutions of general relativity. This suggests that the construction of general gravitational waves may be possible. As an example, we have constructed the evolution of a cylindrical gravitational wavefront allowing the backscattering to produce other types of matter. 

Understanding the construction of gravitational waves is an important step towards showing that the proposed model can indeed serve as description of gravity in 3+1 dimensions. Further study of the interactions of gravitational waves may provide essential guiding insights for completing the incomplete dynamics of the model.
\ifx\fullTeX\undefined
\bibliographystyle{../bib/utcaps}
\bibliography{../bib/thesis}
\end{document}
\fi

\cleardoublepage 
\ifx\fullTeX\undefined
\documentclass[11pt,a4paper]{article}
\usepackage[utf8]{inputenc} 
\usepackage[english]{babel}
\usepackage{graphicx}
\usepackage{hyperref}
\usepackage{color}
\usepackage{amsmath}
\usepackage{amsthm}
\usepackage{amssymb}
\usepackage{bbm}
\usepackage{pictexwd,dcpic}

\graphicspath{{../pics/}}

\title{Conclusions and Outlook}
\author{Maarten van de Meent}
\date{\today} 

\begin{document}
\maketitle
\else
\chapter{Conclusions and Outlook}\label{ch:openproblems}
\fi
In this thesis we have studied the piecewise flat gravity model in 3+1 dimensions first suggested by 't Hooft in \cite{hooft:2008}. The primary idea behind the model is to reproduce the local finiteness of gravity in 2+1 dimensions by requiring that the Riemann curvature tensor vanishes in empty regions of spacetime (i.e. areas where the energy--momentum is zero).

This is achieved by studying a system of straight cosmic strings propagating in a 3+1 dimensional spacetime. Geometrically these strings correspond to conical defects of the spacetime with codimension 2. Unlike other piecewise flat approaches to gravity like Regge calculus, these conical defects are interpreted as physical degrees of freedom. This implies that they must obey the rules of causality, which means that we can allow only defects that are non-spacelike.

In chapter \ref{ch:3+1gravity} we have described two distinct ways of describing an arbitrary configuration of conical defects. Each has its own advantages and disadvantages. 

In the first approach we described a configuration of defects using the Poincaré holonomies of loops around the defects. A general configuration of defects is then given as a group homomorphism of the fundamental group of the spacetime minus the defects to the Poincaré group. The requirement that all defects are non-spacelike translates to the condition that there must exist a complete set of generators of the fundamental group whose assigned holonomy is either rotationlike or a null rotation.

One of the advantages of this approach is that the holonomies are gauge invariant quantities.\footnote{Strictly speaking, there is still a global $\ISO{3,1}$ invariance connected to the choice of frame of the base point of the loops.} Moreover, the consistency conditions at junctions in the network of defects take a clear algebraic form in terms of the holonomies. A major downside is that the holonomies entangle the geometric data of the different defects in an essentially non-local way. This makes this method of description very complicated for configurations consisting of more than a handful of defects.

A second more important drawback is that the holonomies do not properly distinguish between defects with positive and negative (Ricci scalar) curvature. This means that there may be geometrically distinct configurations of defects that have algebraically indistinguishable holonomies. Possibly even more disturbingly, we found in chapter \ref{ch:collisions} that there exist assignments of the holonomies that appear to be consistent algebraically, but which do not allow a consistent geometric realization.

The second approach emphasized the piecewise flat structure of the spacetime rather than the defects. It described a spacetime filled with defects as a collection of (3+1)-dimensional polytopes which were glued together along their faces. This gives a much more local description of the geometry. Moreover, it fully distinguishes between different configurations.

The downside of this geometric  approach is that there are many piecewise flat structures that describe the same configuration. This means that even though this description has only a finite number of degrees of freedom per unit volume, it still contains many unphysical (gauge) degrees of freedom.

\section{Collisions}
In chapter \ref{ch:collisions} we studied the collision of a pair of defects. Since the defects carry non-commuting holonomies they cannot simply pass through each other. Instead an interpolating configuration of defects must be found that continues the piecewise flat geometry after the collision. Various topological configurations of intermediate defects were examined --- ranging from a simple single defect continuation for orthogonal collisions to more complicated `tetrahedral' continuations involving six new intermediate defects.

A common feature of these continuations is that they introduce new degrees of freedom. The current physical requirements of the model do not uniquely determine values of all the new parameters. The model as it currently is, is therefore incomplete. Additional physical input is required to uniquely determine the evolution of a configuration through a collision.

Although this is in some sense alarming, this indeterminism may be a blessing in disguise. The extra physical data needed to pick a continuation of a collision may allow for the inclusion of different types of interaction in the model. For example, if the defects carried additional conserved charges, this would put new constraints on the possible continuations. If no such freedom were present, and the evolution of the model were completely determined, than the model would allow for only one type of matter content with no further room for model building.

A more pressing matter, however, is the observation that certain collisions demand continuations which seem to be inconsistent with the principle of local causality ---  a fundamental tenet of the model. More specifically, we found that if the parameters\footnote{The mass densities of the colliding defects and their relative angle and velocity.} of a collision exceed the bounds set by equation \eqref{eq:junctionlimit}, then some of the junctions in the considered continuations move faster than the speed of light. This situation occurs when the mass densities of the defects are large, their relative velocity is near relativistic, and/or the defects are close to parallel.

If a junction moves faster that the speed of light, then there exists a Lorentz frame in which the junction is instantaneous in time. In that frame the junction corresponds to a defect spontaneously splitting into multiple defects. Since this happens at arbitrary distances from the collision, the split will appear to have no local cause. Superluminal junctions therefore seem to be in conflict with the local causality principle of our model.

The generic argument of section \ref{sec:limits} shows that the appearance of superluminal junctions is universal for all possible continuations. It therefore appears that our proposed piecewise flat model cannot consistently find continuations for all types of collisions. Can such collisions be avoided by a fortunate  choice of initial conditions? If the bound \eqref{eq:junctionlimit} only involved the masses and the velocities of the defects this might seem a possibility. However, the bound set by \eqref{eq:junctionlimit} is also violated by low energy collisions of almost collinear defects. The author sees no plausible physical argument that would disallow the creation of nearly collinear pairs of defects. It therefore seems unlikely that the collisions leading to continuations with superluminal junctions can be avoided altogether.

Is there then a way that the superluminal  junctions that appear can be reconciled with local causality? As we argued above, this appears to be impossible. It is, however, not completely beyond the realm of possibility. Even though the causal past of a superluminal junction does not contain the actual collision event, it will contain parts of both colliding defects before the collision. Consequently, the junction could have received information of the impending collision. This does require some sort of signalling mechanism. One (simplistic) example of such a mechanism would be for the surface where the wedge of spacetime was removed  to form a conical geometry, to carry physical information.\footnote{The natural thing to do is to treat this as a mere constructional artefact without geometrical or physical significance.} One could then always choose the junctions to occur exactly at the intersection of the wedge with the colliding defect. The junctions would still be superluminal, but it would just signify the meeting of two physical entities (the wedge and the defect). As such there would be no problem with causality, just like there is no problem with the tip of a laser beam flashing across the moon moving faster than the speed of light. Such a solution does however seem very artificial. 

Besides superluminal junctions, we also observed that for the particular continuations studied some of the intermediate defects became superluminal for extreme values of the collisions parameters. This obviously is inconsistent with the base requirement of our model that no superluminal defects appear. However, the general argument of section  \ref{sec:limits} does not imply that this must be the case for any continuation. In fact, there appears little geometric obstruction to constructing more complicated continuations to remove the superluminal defects.

In all it seems that the appearance of superluminal junctions and defects signals that the proposed  piecewise flat model of gravity is --- in some fundamental way --- flawed. Fixing this may require some fundamental modifications to the model.

Another issue, which we have paid little attention to in this thesis, is related to the fact that each collision introduces a number of new defects.  Because the inverse process (some configuration of defects cleanly splitting in two defects) is very unlikely, this will cause the spacetime structure to become finer over time\footnote{Unless compensated by a comparable spacetime expansion, but this would require extreme fine tuning.} ultimately approaching some fractal-like structure. This contrasts significantly with the situation in 2+1 dimensions.  In 2+1 dimensions, defects almost never collide head-on, so that no such fragmentation takes place.

In his original paper \cite{hooft:2008} 't Hooft suggested to counteract this by restricting the holonomies of the defects to some discrete subset of the Poincaré group. This would significantly increase the probability of processes where defects disappear, and the hope was that at some point an equilibrium in the structure would be reached. As an added bonus, restricting to a discrete subset of the Poincaré group would introduce a minimal angle at which defects can collide, getting rid of the problematic nearly collinear collision --- although it would do little to prevent high energy collisions. It is, however, not clear that this can be implemented in a consistent way, while finding continuations for all collisions.

\section{Matter content}
In the linearized limit of the model collisions do not pose problems, because at linear order defects can simply move through each other. In chapter \ref{ch:cont} 
we utilized this limit to examine the continuum limit of our piecewise flat model of gravity.

In particular, we examined the type of energy--momentum distribution that could be generated by an arbitrary configuration of defects. The result was that the model could reproduce the same  energy--momentum distribution as a gas of non-interacting point particles. That is, even though the fundamental excitations of our model are straight strings with an infinite extent, it is still possible to approximate the matter configurations produced by point masses. For this conclusion it was essential that the model include defects with both positive and negative curvature.

Not surprisingly, we cannot reproduce interacting forms of matter in the linearized limit. This is exactly, because collisions are the only form of interaction between the defects, and these became trivial in the linear limit. Once we include higher order terms collisions come back into the game and will produce some form of interaction.

At this point, however, it is unclear what types of interaction may be reproduced. It seems that the extra input needed to find a continuation for a collision may in fact give us some freedom in describing different types of interaction. 

\section{Gravitational waves}

One thing that we observed in our study of the linearized model, was that it was possible to find non-trivial configurations with vanishing energy--momentum. By calculating the linear perturbation to the metric caused by an arbitrary linearized configuration of defects, we were able to show that these configurations with vanishing energy--momentum correspond to linear gravitational waves.

So, even though the model removed gravitational waves as fundamental local excitations, these waves appear as an emergent feature in the continuum limit of the model.

In chapter \ref{ch:gravwave} we have explicitly extended this conclusion to exact gravitational wave solutions by constructing a family of defect configurations that can approximate an arbitrary (exact) gravitational plane wave. Obtaining general exact wave solutions is much more difficult, but we sketched the outlines of the construction of a cylindrical gravitational wavefront from defects. This construction highlighted the fact that a cylindrical wave front in general relativity cannot appear without backscattering. The backscattered component of such a front appears to be very complicated, and will likely contain the collisions of lightlike defects. 

Collisions of lightlike defect occur exactly on the boundary of the collision parameter space where superluminal effects start to occur. This boundary case was not included in our general study of chapter \ref{ch:collisions}, this needs further study.

Moreover, the chaos of the backscattering of a cylindrical wave front is not the  best environment to study the collisions of piecewise flat wavefronts. A more controlled setting is the collision of impulsive wave fronts. There is extensive literature on these in general relativity.\cite{Khan:1971vh, Griffiths:1991, Stephani:2003, Chandrasekhar:1984at, Szekeres:1970vg, Szekeres:1972uu} One thing that is known about these is that the resulting spacetime often locally corresponds to a black hole metric.\cite{Ferrari:1988nu} Studying such collisions in piecewise flat gravity may therefore also give crucial insight in the construction of static gravitational fields using configurations of conical defects.

\section{Quantization}
In this thesis we have focused on the classical aspects of the proposed piecewise flat model of gravity in 3+1 dimensions. One of the motivations behind the model is that it might be easier to quantize due to the similarity with gravity in 2+1 dimensions. At the same this similarity suggests that the quantization of this model will be subject to a similar set of subtleties. In this section we will comment on how some of our conclusions affect the possible quantization of the model.

First notice that indeed the model has only a finite number of degrees of freedom per unit volume. This means that its quantization is a form of ordinary quantum mechanics rather than quantum field theory, bypassing any problems with renormalization. 

We have explored different approaches to describing a configuration of defects. As with the 2+1 dimensional model of gravitating point particles, each of these will have their own implications for quantization. 

The first of these was the description of a configuration of defects by the assignments of Poincaré holonomies to equivalence classes of loops around the defects. In particular, we found that a single defect can be described by its Poincaré holonomy.

This description is strongly reminiscent of the description of a gravitating point particle in 2+1 dimensions using its Poincaré holonomy. This suggests that like in 2+1 dimensions,\cite{Matschull:1997du} the Lorentz holonomy may act as the covariant momentum of the defect. Consequently, quantization using this type of description may lead to similar features as observed in \cite{Matschull:1997du}, namely discretization  of time and space and non-commutativity of the position variables.

This  would obviously need to be checked more carefully using a dynamical formulation of the model starting from an action. However, there also are some reasons to doubt the above conjecture. For one we know that the Lorentz holonomy of a defect encodes not only the energy and velocity of a defect, but also its orientation. The latter, normally would be considered a position rather than a momentum variable. 

Moreover, we know that the description of a configuration of defects in terms of its holonomies is not entirely unambiguous once we allow for defects with negative curvature. Furthermore, we also observed that there exist consistent assignments of holonomies that do not correspond to a geometrically realizable configuration of defects. Both these observations suggest that the holonomies may not be the right variables to use for quantization.

The second approach to describing configurations of defects that we explored was in terms of piecewise flat manifolds. This description certainly is unambiguous, and due to its geometric nature there is no doubt about the geometric realization of any configuration. The downside of this approach is that it has a lot of residual gauge freedom; there are many different descriptions of the same configuration of defects. 

Any quantization using this description therefore needs to take into account these gauge degrees of freedom. This could be done by choosing a gauge fixing prior to the quantization. A partial approach to this could be to try to find a foliation of the piecewise flat cell complex by Cauchy surfaces in such a way that each slice is again a piecewise flat manifold and that the time in each cell runs at the same rate. This (if possible) should yield something that is very similar to 't Hooft's polygon model in 2+1 dimensions --- which one could refer to this as a polytope model. 

Regardless of the approach to describing a configuration of defects, any quantization of the model will require an action. In this thesis we have attacked most problems in the terms of kinematics, rather than formulating it in terms of dynamics as the result of some action. Nonetheless the common wisdom is that an action is required to properly formulate the dynamics of the model and its quantization. As an alternative 't Hooft has suggested \cite{hooft:2008} that this model could be quantized using his ``pre-quantization'' formalism.\cite{hooft:2007}

\section{Applications outside of quantum gravity}

The problems with consistency of the model when considering collisions may signal that it is not suitable as a fundamental model of gravity in 3+1 dimensions. Nonetheless, the studies of chapters \ref{ch:cont} and \ref{ch:gravwave} have shown that the model can at least approximate some systems in general relativity.

As such, the fact that this model describes these systems in terms of discrete excitations evolving with time, may prove useful in (numerical) analysis of such system. For example, the ability to decompose gravitational wavefronts in terms of lightlike defects opens up the possibility of studying non-uniform wavefronts.

As another example, it is relatively easy to find defect configurations that approximate FRW solutions of general relativity with equation of state parameter $w=-1/3$. Since these configurations consist of discrete defects, they are inhomogeneous at small scales. Such configurations may therefore serve as interesting test cases  for approaches to cosmological averaging. The use here may however be limited by the fact that these models may belong to a somewhat pathological subset of configurations due to the absence of gravitational fields between the defects.

\ifx\fullTeX\undefined
\bibliographystyle{../bib/utcaps}
\bibliography{../bib/thesis}
\end{document}
\fi

\renewcommand{\publ}{}

\appendix
\cleardoublepage
\ifx\fullTeX\undefined
\documentclass[11pt,a4paper]{article}
\usepackage[utf8]{inputenc} 
\usepackage[english]{babel}
\usepackage{graphicx}
\usepackage{hyperref}
\usepackage{color}
\usepackage{amsmath}
\usepackage{amsthm}
\usepackage{amssymb}
\usepackage{bbm}
\usepackage{pictexwd,dcpic}

\graphicspath{{../pics/}}

\title{Appendix: The $\PSL{2,\CC}$ representation of the Lorentz group}
\author{Maarten van de Meent}
\date{\today} 

\begin{document}
\maketitle
\else
\phantomsection
\addcontentsline{toc}{chapter}{Appendix}
\chapter{Some facts about \texorpdfstring{$\PSL{2,\CC}$}{PSL2C}}\label{app:PSL2C}
\fi
The group of proper orthochronous Lorentz transformations in 3+1 spacetime dimensions, $\SO{3,1}^+$ is isomorphic to the group of projective special linear transformations of the complex plane, $\PSL{2,\CC}$, also known as the group of Möbius transformations. Each Möbius transformation,
\begin{equation}
z\mapsto \frac{a z+ b}{c z + d},
\end{equation}
can be represented as a complex $2\times 2$ matrix,
\begin{equation}
\begin{pmatrix}
a & b \\
c & d \\
\end{pmatrix}.
\end{equation}
This identification is unique up to an overall scalar factor. We can use this freedom to ensure that the determinant of the matrix, $a d-b c$, is equal to one. This condition fixes the overall scalar factor up to a sign. We can therefore represent Lorentz transformation by special linear  $2\times 2$ matrices modulo an overall sign, which are known as projective linear transformations. In terms of the groups,
\begin{equation}
\SO{3,1}^+ \cong \SL{2,\CC}/\ZZ_2 \equiv \PSL{2,\CC}.
\end{equation}

This representation of the Lorentz group is often very useful when dealing with the algebraic conditions on holonomies of defects. This is mostly because of the practical reason that $\PSL{2,\CC}$ matrices have only four (complex) components compared to the sixteen (real) components of a  $\SO{3,1}$ matrix. In this appendix we review some basic facts about this representation and clarify some of the conventions used. See chapter 1 of \cite{Penrose:1984} for an extensive discussion of this representation albeit with slightly different terminology.

\section{Lorentz transformations}
Minkowski space can be identified with the space of complex Hermitian  $2\times 2$ matrices as follows
\begin{equation}\label{eq:txyztoX}
(t,x,y,z)\mapsto X=\begin{pmatrix}
t+z & x - \ii y \\
x+\ii y & t-z
\end{pmatrix}.
\end{equation}
Under this  identification  the determinant  of $X$ is equal to (minus) the Minkowski norm
\begin{equation}
\det X = t^2 - x^2 - y^2 - z^2.
\end{equation}
Matrices $Q$ in $\SL{2,\CC}$  act on the space of Hermitian $2\times 2$ matrices through the adjoint action
\begin{equation}\label{eq:adjoinaction}
\Ad_Q X = Q X Q^\dag.
\end{equation}
Since the determinant of $Q$ is one, this action preserves the determinant of $X$. Consequently, using the identification \eqref{eq:txyztoX}, this action can be interpreted as a linear norm preserving action of $\SL{2,\CC}$ on Minkowski space. The action \eqref{eq:adjoinaction} and the identification \eqref{eq:txyztoX} therefore define a group homomorphism from $\SL{2,\CC}$ to the Lorentz group. This group homomorphism is in fact a 2--1 surjection onto the identity component of the Lorentz group with kernel $\set{\Id, -\Id}$. Consequently, we find that
\begin{equation}
\PSL{2,\CC} \equiv \SL{2,\CC}/\set{\Id, -\Id} \cong \SO{3,1}^+.
\end{equation}

\section{Generators}
The Lie group $\PSL{2,\CC}$ is generated by the three Pauli matrices
\begin{align}
\sigma_x = \frac{1}{2}\begin{pmatrix} 0 & 1\\ 1&0\end{pmatrix}, &&
\sigma_y = \frac{1}{2}\begin{pmatrix} 0 & -\ii\\ \ii&0\end{pmatrix}, &&
\sigma_z = \frac{1}{2}\begin{pmatrix} 1 & 0\\ 0&-1\end{pmatrix}.
\end{align}
If we denote $\vec{\sigma}\equiv (\sigma_x,\sigma_y,\sigma_z)$, then any element $Q$ of $\PSL{2,\CC}$ may be written as\footnote{Note that this a non-trivial statement since the exponential map from $\Sl{2,\CC}$ to $\SL{2,\CC}$ is not onto.}
\begin{equation}
Q = \exp\bhh{\vec\sigma\cdot(\vec{b}+\ii\vec{r})}.
\end{equation}
This parametrization has some useful characteristics. If $\vec{b}=0$, then
\begin{equation}
Q^\dag 
 = \hh{\exp\bhh{\ii \vec\sigma\cdot\vec{r}}}^\dag
 = \exp\bhh{-\ii \vec\sigma\cdot\vec{r}}
 = Q^{\m 1}.
\end{equation}
So, $Q$ is unitary. Conversely, if $Q$ is unitary, then $\vec{b}=0$. Consequently, the elements with $\vec{b}=0$ form a subgroup. This subgroup corresponds to the subgroup of rotations of the Lorentz group. An element with $\vec{b}=0$ corresponds to a rotation about the $\vec{r}$ axis with angle $\norm{\vec{r}}$.

   Similarly, $Q$ is hermitian if and only if $\vec{r}=0$. Hermitian elements of $\PSL{2,\CC}$ correspond to Lorentz boosts in the  $\vec{b}$ direction with rapidity $\norm{\vec{b}}$.

Consequently, a general rotationlike  Lorentz transformation will correspond to a $\PSL{2,\CC}$ matrix
\begin{equation}
\begin{aligned}
R & = \exp(\vec\sigma\cdot\vec{b})\exp(\ii\vec\sigma\cdot\vec{r})
			\exp(-\vec\sigma\cdot\vec{b})\\
&=\exp\bhh{\ii\exp(\vec\sigma\cdot\vec{b})(\vec\sigma\cdot\vec{r})			
			\exp(-\vec\sigma\cdot\vec{b})}\\
&=\exp\bhh{\sinh\norm{\vec{b}}\frac{\vec{b}\times\vec{r}}{\norm{\vec{b}}}
	+\ii\cosh\norm{\vec{b}}\,\vec{r}
},
\end{aligned}
\end{equation}
with $\vec{b}\cdot\vec{r}=0$. Conversely, we can conclude that any
\begin{equation}
Q = \exp\bhh{\vec\sigma\cdot(\vec{b}+\ii\vec{r})},
\end{equation}
with $\vec{b}\cdot\vec{r}=0$ and $\norm{\vec{b}} < \norm{\vec{r}}$ corresponds to a rotationlike Lorentz transformation. Similarly, one finds that elements with $\vec{b}\cdot\vec{r}=0$ and $\norm{\vec{b}} > \norm{\vec{r}}$ correspond to boostlike Lorentz transformations.

In the boundary case that $\vec{b}\cdot\vec{r}=0$ and $\norm{\vec{b}} = \norm{\vec{r}}$ one finds that
\begin{equation}
\begin{aligned}
\bhh{\vec\sigma\cdot(\vec{b}+\ii\vec{r})}^2 
 &= (\vec{b}+\ii\vec{r})\cdot(\vec{b}+\ii\vec{r})\Id_{\CC^2} \\
 &= (\norm{\vec{b}}^2 - \norm{\vec{r}}^2 +2\ii \vec{b}\cdot\vec{r})\Id_{\CC^2}\\
 &= 0.
\end{aligned}
\end{equation}
That is, $\vec\sigma\cdot(\vec{b}+\ii\vec{r})$ is nilpotent. Consequently, the corresponding $\PSL{2,\CC}$ element is unipotent, and by extension so is the corresponding Lorentz transformation. The eigenvalues of  the corresponding Lorentz transformation thus are all equal to one; It is a null rotation.

\section{Conjugacy classes}

In the piecewise flat model of gravity studied in this thesis, it is often necessary to determine the conjugacy class of a Lorentz transformation. In the $\PSL{2,\CC}$ representation of the Lorentz group, the conjugacy class of an element can be determined on basis of the trace.

Two matrices in $\SL{2,\CC}$ belong to the same conjugacy class if and only if they have the same Jordan normal form. The Jordan normal form of a general  $2 \times 2$ matrix, $M$, is 
\begin{equation}
\begin{pmatrix}
\lambda_1 & 0 \\ 0 & \lambda_2
\end{pmatrix},\quad\text{or}\quad
\begin{pmatrix}
\lambda & 1 \\ 0 & \lambda
\end{pmatrix}.
\end{equation}
Lets first consider the first case. If $M$ is in $\SL{2,\CC}$, then $\det M = \lambda_1 \lambda_2 = 1$, and consequently $\lambda_2 = 1/\lambda_1$. The trace is thus given by
\begin{equation}
\tr M = \lambda_1 + \frac{1}{\lambda_1},
\end{equation}
and we obtain
\begin{equation}\label{eq:trlambda}
\lambda_{1,2} = \frac{\tr M \pm \sqrt{(\tr M)^2-4}}{2}.
\end{equation}
Furthermore, if $M$ represents an element of $\PSL{2,\CC}$ we can use the overall sign freedom to fix $\im \tr M >0$.

In the second case, $\det M =\lambda^2=1$, implies that $\lambda =\pm 1$, and we can use the overall sign in $\PSL{2,\CC}$ to fix  $\lambda = 1$. As a result, the trace of $M$ is 2.  Plugging $\tr M =2$ in equation \eqref{eq:trlambda} we see that in the diagonalizable case this implies that $M$ is the identity.

We therefore obtain the following lemma:
\begin{lemma}
If $M_1$ and $M_2$ are non-identity elements of $\PSL{2,\CC}$ represented as $\SL{2,\CC}$ matrices with $\im \tr M_{1,2} >0$ or $\im \tr M_{1,2} = 0$ and $\re \tr M_{1,2} >0$, then $M_1$ and $M_2$ lie in the same conjugacy class if and only if  $\tr M_1 =\tr M_2$.
\end{lemma}

If we write an element $Q$ of $\PSL{2,\CC}$ as
\begin{equation}
Q = \exp\bhh{\vec\sigma\cdot(\vec{b}+\ii\vec{r})},
\end{equation}
then the trace is given by
\begin{equation}
\tr Q = 2\cos\sqrt{
\norm{\vec{r}}-\norm{\vec{b}}-2\ii\vec{r}\cdot\vec{b}
}.
\end{equation}
Consequently, a $\PSL{2,\CC}$ transformation $Q$ not equal to the identity is ...
\begin{itemize}
\item ... \emph{rotationlike} ($\norm{\vec{r}}>\norm{\vec{b}}$ and $\vec{r}\cdot\vec{b}=0$), if and only if 
\begin{equation}
\im\tr Q = 0\quad\text{and}\quad\abs{\tr Q} < 2.
\end{equation}
\item ... \emph{boostlike} ($\norm{\vec{r}}<\norm{\vec{b}}$ and $\vec{r}\cdot\vec{b}=0$), if and only if
\begin{equation}
\im\tr Q = 0\quad\text{and}\quad\abs{\tr Q} > 2.
\end{equation}
\item ... a \emph{null rotation} ($\norm{\vec{r}}=\norm{\vec{b}}$ and $\vec{r}\cdot\vec{b}=0$), if and only if 
\begin{equation}
\im\tr Q = 0\quad\text{and}\quad\abs{\tr Q} = 2.
\end{equation}
\end{itemize}
This characterization of elements of $\PSL{2,\CC}$ is very useful when we need to test if a defect described by a certain holonomy is timelike, spacelike, or lightlike.

\ifx\fullTeX\undefined
\bibliographystyle{../bib/utcaps}
\bibliography{../bib/thesis}
\end{document}
\fi


\selectlanguage{dutch}
\cleardoublepage
\hyphenation{di-mens-ion-aal di-mens-ion-ale}
\chapter*{Nederlandse samenvatting}\addcontentsline{toc}{chapter}{Nederlandse samenvatting}\markboth{Nederlandse samenvatting}{}
De moderne natuurkunde rust op twee standaardtheorieën. Aan de ene kant geeft de kwantumveldentheorie aan hoe materie zich gedraagt op (sub)atomaire schaal. Aan de andere kant is er de algemene relativiteitstheorie van Albert Einstein, die  de gravitationele interactie tussen planeten, sterren en hele melkwegstelsels beschrijft. Beide theorieën zijn in de afgelopen eeuw tot de kleinste details van hun voorspellingen nagemeten en hebben iedere experimentele toets waaraan ze zijn onderworpen glansrijk doorstaan.

Echter, als beide theorieën gelijktijdig op één systeem worden toegepast, dan lijken ze elkaar tegen te spreken. In experimenten leidt dit tot nu toe niet tot problemen omdat de inconsistenties pas zichtbaar worden in systemen die tegelijkertijd heel erg klein en heel erg zwaar zijn. De dichtheid waarbij beide theorieën belangrijk worden, heet de Planckdichtheid, ongeveer $10^{96}$ kilogram per kubieke meter. Dit is de dichtheid die wordt bereikt als je de massa van het hele zichtbare heelal samendrukt tot de grootte van één atoomkern.

Systemen met een dergelijke dichtheid kunnen natuurlijk niet in een lab gerealiseerd worden, en het is daarom niet zo gek dat de tegenstrijdigheden tussen deze twee theorieën niet naar voren komen in experimenten. Toch zijn er in de natuurkunde waarschijnlijk situaties waarin zulke  extreme omstandigheden voorkomen. De algemene relativiteitstheorie voorspelt bijvoorbeeld dat de materie in het centrum van een zwart gat tot een punt wordt samengedrukt. Het omgekeerde doet zich voor in het vroege heelal,dat volgens dezelfde theorie vanuit een toestand met oneindige dichtheid is geëxpandeerd tot het heelal van nu.

Om de wisselwerking tussen deeltjes in zulke extreme omstandigheden te beschrijven is een theorie nodig die zowel de kwantumveldentheorie als de algemene relativiteitstheorie omvat. Een dergelijke overkoepelende theorie wordt ook wel een theorie van kwantumgravitatie genoemd. De moderne theoretische natuurkunde is al meer dan een halve eeuw op zoek naar een dergelijke theorie. Tot nog toe zonder een definitief succes.

Een centraal probleem bij het opstellen van een overkoepelende theorie is dat als de conventionele technieken om een kwantumtheorie te formuleren worden toegepast op de algemene relativiteitstheorie, deze leiden tot onacceptabele oneindigheden in de voorspellingen. Deze oneindigheden vinden hun oorsprong in  de geometrische aard van de algemene relativiteitstheorie. In dit proefschrift wordt  een alternatief model voor zwaartekracht ontwikkeld dat dit probleem probeert te vermijden.

\section*{Zwaartekracht is kromming van de ruimtetijd}
De algemene relativiteitstheorie kan in een notendop worden samengevat  als: ``Zwaartekracht is kromming van de ruimtetijd.'' Wat wordt daarmee bedoeld? 

We beschouwen eerst het begrip kromming. Stel een wereld met maar twee ruimtelijke dimensies, zoals \emph{Flatland} (Platland) beschreven door Edwin A. Abbott.\cite{Abbott:1885} En stel, dat die platlanders leven op het oppervlak van een bol (in plaats van op een plat vlak zoals in het werk van Abbott). Hoe zouden de twee-dimensionale inwoners dat kunnen zien en meten?  Één mogelijkheid is  het meten van de hoeken van geometrische figuren, een favoriete bezigheid van de inwoners van Platland. In tegenstelling tot de gebruikelijk constatering dat de som van de hoeken van een driehoek 180 graden is, zullen de platlanders constateren dat deze som afhangt van de oppervlakte van de driehoek; des te groter het oppervlak van de driehoek, des te groter de som van de hoeken. Dergelijke afwijkingen van de gebruikelijke Euclidische meetkunde worden door wiskundigen samengevat onder de algemene noemer `kromming'. 

Een ander effect dat de platlanders zouden kunnen waarnemen, is dat op een bol parallelle lijnen elkaar in twee punten snijden. Nu liggen deze punten op tegenovergestelde punten van de bol. Dus als de bol erg  groot is, zullen ze dit niet snel merken. Maar, een oplettende platlander zal zien dat twee passanten, die hem in een rechte lijn parallel aan elkaar voorbij lopen, na een tijdje dichter naar elkaar toe zijn bewogen. Mocht de platlander in de veronderstelling verkeren in een plat vlak te leven, dan zou hij concluderen dat er kennelijk een soort kracht op de twee passanten werkt die ze naar elkaar toe trekt; een soort zwaartekracht.

Volgens de algemene relativiteitstheorie is de zwaartekracht in onze drie-dimens\-ionale wereld een vergelijkbaar effect. Alleen gaat het niet om kromming van alleen de ruimte, maar om kromming van de ruimte en de tijd samengenomen tot een vier-dimensionale ruimte, de ruimtetijd. De banen van planeten om de zon zijn volgens de relativiteitstheorie in werkelijkheid rechte lijnen in deze ruimtetijd. Maar, doordat wij bij onze waarnemingen (impliciet) veronderstellen dat de ruimtetijd `vlak' is, lijken deze banen voor ons gekromd en concluderen wij dat er kennelijk een kracht op de planeten werkt: de zwaartekracht.

De algemene relativiteitstheorie stelt verder dat er een wisselwerking bestaat tussen de kromming van de ruimtetijd en massa. Aan de ene kant bepaalt de kromming hoe massa's door de ruimtetijd bewegen, aan de andere kant bepaalt de aanwezigheid van massa hoe de ruimtetijd zich kromt. Dit laatste wordt vastgelegd door de zogenaamde Einsteinvergelijking.

De kromming wordt niet volledig vastgelegd door de Einsteinvergelijking. Een  deel van de kromming kan zich vrijelijk bewegen, als een soort golf door de ruimte. Deze golven worden zwaartekrachtsgolven genoemd. Het zijn juist deze zwaartekrachtsgolven die zorgen voor (niet renormaliseerbare) oneindigheden bij het opstellen van een kwantummechanische tegenhanger van de algemene relativiteitstheorie. 

\section*{Zwaartekracht in twee dimensies}
Als we een theorie niet goed begrijpen, kan het leerzaam zijn om te bekijken hoe de theorie zich gedraagt als we het aantal ruimtelijke dimensies veranderen. Dit geldt ook voor de algemene relativiteitstheorie.

In een wereld met maar twee ruimtelijke dimensies zoals Platland --- en dus met een drie(=2+1)-dimensionale ruimtetijd --- is de algemene relativiteitstheorie veel eenvoudiger. In tegenstelling tot de theorie in 3+1 dimensies, legt de Einsteinvergelijking in 2+1 dimensies de kromming van de ruimtetijd geheel vast in termen van de massaverdeling: Platland kent geen zwaartekrachtsgolven.

De kromming bevindt zich alleen op die punten in de ruimtetijd waar massa is. Lege stukken ruimtetijd zijn geheel vlak. De kromming op de plaats van een puntdeeltje is een zogenaamd `conisch defect'. Dat wil zeggen: op de plek van de puntdeeltje ontbreekt er een taartpunt uit de ruimtetijd zoals bij de punt van een kegel (zie figuur \ref{fig:conicaldefect} op bladzijde \pageref{fig:conicaldefect}). De grootte van de ontbrekende hoek is evenredig met de massa van het puntdeeltje.

De geometrie van een ruimtetijd van een verzameling puntdeeltjes kan gezien worden als allemaal blokjes vlakke ruimtetijd die langs hun vlakken aan elkaar geplakt zijn. De puntdeeltjes bewegen zich dan langs de ribben van deze blokjes. Een dergelijke constructie wordt een \emph{stuksgewijs vlakke} geometrie genoemd. 

Omdat ruimtetijd waar zich geen massa bevindt helemaal vlak is, is er in 2+1 dimensies geen vrijheid voor zwaartekrachtsgolven. Dit doet vermoeden dat het formuleren van een kwantummechanische versie van de algemene relativiteitstheorie makkelijker is in 2+1 dimensies. Dit werd voor het eerst bevestigd door Edward Witten in 1988.\cite{Witten:1988hc} Sindsdien wordt gravitatie in 2+1 dimensies veel gebruikt als voorbeeldtheorie om conceptuele problemen in kwantumgravitatie te onderzoeken.

\section*{Stuksgewijs vlakke zwaartekracht in drie dimensies}
Het idee achter het model dat is beschreven in dit proefschrift is het construeren van een model voor zwaartekracht in drie ruimtelijke dimensies (en dus 3+1 ruimtetijddimensies) dat de stuksgewijs vlakke structuur van algemene relativiteitstheorie in 2+1 dimensies bezit. Omdat zo'n model geen zwaartekrachtsgolven zal bezitten, is het mogelijk makkelijker om een kwantummechanische tegenhanger te formuleren. In dit proefschrift gaan we echter niet verder op de mogelijkheid van het formuleren van een kwantummechanische versie van het model, maar richten we ons op de klassieke eigenschappen en interne consistentie van het model.

Het model wordt bepaald door drie basis regels:
\begin{enumerate}
\item De ruimtetijd is vlak op plekken waar geen materie is.
\item De gehele ruimtetijd (zowel de regio's met als zonder materie) voldoet aan de Einstein vergelijking.
\item Alle materie voldoet aan lokale causaliteit. (Elke gebeurtenis kan voorspeld worden door de gebeurtenissen in zijn verleden). Materie kan niet sneller bewegen dan het licht.
\end{enumerate}
De combinatie van regels 1 en 2 legt een strikte beperking op aan de soort materie die we toe kunnen staan. De enige vorm van materie die is toegestaan, zijn rechte staven met een constante dichtheid en snelheid. De derde regel stelt vervolgens dat deze staven maximaal met de lichtsnelheid mogen reizen.

De geometrie van zo'n rechte staaf is wederom een conisch defect. De gehele geometrie van een collectie staven bestaat uit vlakke (3+1)-dimensionale brokken ruimtetijd die langs hun ((2+1)-dimensionale) randen aan elkaar geplakt zijn. De rechte staven bevinden zich vervolgens op de (1+1)-dimensionale grensvlakken. Dit is wederom een stuksgewijs vlakke geometrie. 

\subsection*{Botsingen}
Een nieuw fenomeen ten opzichte van gravitatie in 2+1 dimensies is, dat in dit model de staven kunnen botsen. Hoofdstuk \ref{ch:collisions} richt zich op de vraag wat er gebeurt als twee staven met elkaar botsen. Het is geometrisch niet toegestaan dat de staven door elkaar heen bewegen. In plaats daarvan moet uit de botsing een nieuwe stuksgewijs vlakke geometrie ontstaan met nieuwe staven. Dit blijkt altijd mogelijk te zijn, echter met een aantal op- en aanmerkingen.

Ten eerste, is de stuksgewijs vlakke geometrie die ontstaat na een botsing niet uniek: er bestaat een enorm scala aan mogelijkheden. De regels die we aan het model hebben opgelegd, leggen de dynamica van het model niet volledig vast. Dit is vergelijkbaar met de situatie in de algemene relativiteitstheorie waar de Einsteinvergelijking aangevuld moet worden met een bewegingsvergelijking voor de aanwezige materie om de dynamica volledig te bepalen. Om een volledig model te krijgen zal ons stuksgewijs vlakke model voor zwaartekracht aangevuld moet worden met regels die zeggen welke mogelijke vervolgtoestand gekozen moet worden bij elke botsing.

Ten tweede blijken er extreme botsingen te bestaan, waarvoor alle mogelijk vervolgtoestanden knooppunten van staven bevatten die sneller bewegen dan het licht. Dit lijkt in strijd met onze lokale causaliteitsregel en stelt daarmee de consistentie van dit model ter discussie. Deze botsingen doen zich voor als de botsende staven extreem hoge energie hebben en/of bijna parallel zijn. De extreemheid van deze omstandigheden doet vermoeden dat er mogelijk een verband is met het ontstaan van singulariteiten (zoals zwarte gaten) in de algemene relativiteitstheorie.

\subsection*{Zwaartekrachtsgolven}
Een directe consequentie van regel 1 van het stuksgewijs vlakke model voor zwaartekracht is dat er geen zwaartekrachtsgolven zijn toegestaan. Dit was onze opzet. Het is echter ook een probleem als we willen dat het model de werkelijkheid beschrijft. Er bestaat namelijk sterk (indirect) bewijs dat er zwaartekrachtsgolven bestaan in de natuur.

Het model biedt echter een uitweg. In principe staat het model zowel staven met positieve energie als staven met negatieve energie\footnote{De staven representeren een combinatie van materie- en gravitatievrijheidsgraden en kunnen daardoor effectief negatieve energie hebben.} toe. Dit geeft de mogelijkheid tot het creëren van configuraties waarin (gemiddeld genomen) de energie van de staven met positieve energie opgeheven wordt door de energie van staven met negatieve energie. De fysische interpretatie van een dergelijke configuratie is een puur zwaartekrachtsveld. Deze mogelijkheid is onderzocht in hoofdstukken \ref{ch:cont} en \ref{ch:gravwave}.

In hoofdstuk \ref{ch:cont} is de limietsituatie van het model onderzocht waarin alle staven heel erg licht zijn. In deze limiet vereenvoudigt de beschrijving van het model sterk doordat alle interactie tussen de staven verdwijnt. Het is daardoor mogelijk  een beschrijving te geven van een algemene configuratie van staven.

Met deze beschrijving kunnen we uitrekenen wat de energie-impulsverdeling van een algemene configuratie is. Deze komt overeen met  de energie-impulsverdeling van een gas van puntdeeltjes zonder interactie. Dit is opmerkelijk aangezien onze fundamentele bouwstenen, de staven, oneindig lang waren. Tevens blijkt uit de berekening dat er inderdaad een scala aan configuraties bestaat, die netto geen energie en impuls bezitten. Door uit te rekenen wat het effect (tot op eerste orde) op de geometrie van een algemene configuratie is, hebben we kunnen laten zien dat deze configuraties zonder energie en impuls precies overeenkomen met de vacuümoplossingen van de gelineariseerde Einsteinvergelijking. Dat wil zeggen, ze beschrijven (lineaire) zwaartekrachtsgolven.

In hoofdstuk \ref{ch:gravwave} is vervolgens aangetoond dat het in het volledige model mogelijk is configuraties te construeren die een vlakke zwaartekrachtsgolf willekeurig dicht benaderen. Tevens wordt een opzet gegeven voor de constructie van cilindrische zwaartekrachtsgolven. Dit is een sterke aanwijzing dat de constructie van willekeurige zwaartekrachtsgolven mogelijk is binnen het model.

Dus hoewel het model op de fundamenteel schaal van individuele vlakke stukjes ruimte geen plaats biedt voor zwaartekrachtsgolven, ontstaan deze effectief toch als men kijkt naar grote schalen waar configuraties bestaan uit heel veel kleine vlakke stukjes.

\section*{Conclusies}
De formulering van een stuksgewijs vlak model voor zwaartekracht in 3+1 dimensies is een gedeeltelijk succes. Het model dat we hebben beschreven heeft de eigenschappen die we wilden op korte afstand, maar lijkt op grote schalen toch zwaartekracht te beschrijven zoals we die kennen uit de algemene relativiteitstheorie.

Het model is echter niet compleet en heeft extra regels nodig om de dynamica van een systeem volledig te beschrijven. Verder is het model mogelijk niet consistent bij hoge energieën. Dit laatste is mogelijk een indicatie dat dit model niet zonder verdere modificaties geschikt is als basis voor een fundamentele theorie voor kwantumgravitatie.
\selectlanguage{english}

\cleardoublepage

\chapter*{Acknowledgements}\addcontentsline{toc}{chapter}{Acknowledgements}\markboth{Acknowledgements}{}
The work in this thesis would not have been possible without the support of all the amazing people around me. I would like to take this opportunity to explicitly thank a few of them. I apologize in advance to the people I forget to mention.

Firstly, I would like to thank my promotor Gerard 't Hooft. Despite his insane travel schedule, he always made time to discuss ideas with me and provide me with feedback on my progress. For this I am also deeply indebted to Gerard's secretaries, Wilma and Els. Without them keeping some resemblance of order in his schedule, it would have been next to impossible to find a time slot for our next meeting.

I would further like to thank the rest of the staff at the Institute for Theoretical Physics in Utrecht for always showing interest in my projects. In particular, I owe thanks to Renate Loll and her group for including me in their group meetings. These were always interesting. So long, and thanks for all the pizza.

Then there are my fellow occupants of room 309 at the institute (Clem, Dima, Jan, and Simone). The work climate in our office was always excellent, and not just because our schedules had a limited overlap.  I would also like to thank the other PhD students of my `generation': Jeroen, Jurjen, Maaike, and Stefanos. In Dutch we have a saying: ``Gedeelde smart is halve smart.'' I guess this also holds for the pain of writing a PhD thesis.

Thanks also goes out to all the other PhD students at the institute. There is nothing like somebody else's problems to take your mind of your own.\footnote{In some sense this is the antipode of the Somebody Else's Problem effect popularized by Douglas Adams.} And if nobody had problems (or wanted to talk about them), there always was the New York Times crossword puzzle (thank you Hedwig, for introducing that as a break room activitity). I would also like to express my appreciation for the rest of the PhD students in the DRSTP. The national meetings and PhD schools always were fun and interesting.

Not unimportantly, I wish to thank my family. My parents, for always having faith in me and supporting me. My brother, for having more faith in my abilities than anyone alive, including me. My in-laws, for always showing an interest in my research, despite me never quite being able to explain it to you. In particular, I thank Harm for proof reading this document.

Finally and most importantly, Anneleid. Putting up with me this past year cannot have been easy. I have been tired and grumpy a lot of the time, and still you not only always took care of me, but also proof read most of my drafts despite having no background in physics or mathematics to speak of. I love you.

\cleardoublepage

\chapter*{Curriculum vitae}\addcontentsline{toc}{chapter}{Curriculum vitae}\markboth{Curriculum vitae}{}
Maarten van de Meent was born on June 29, 1982 in Delft, the Netherlands. At the age of four his family moved to Bilthoven, where he grew up and attended the Werkplaats Kindergemeenshap both for primary and secondary school (VWO). In 2001 he enrolled in the mathematics and physics programs at the University of Amsterdam. After obtaining bachelor degrees in both programs, he continued with the master's program in mathematical physics. He wrote his graduate thesis \emph{``Topological Strings and Quantum Crystals''} under supervision of Robbert Dijkgraaf, and obtained his MSc degree with distinction (cum laude) in 2006.

In the academic year 2006--2007 he attended the pure mathematics program of the University of Oregon, where he obtained a second master's degree. During this year he was supported by generous scholarships from the VSBfonds, Fulbright, and Huygens programs.

After his return to the Netherlands in 2007, he started work on his PhD research at the Institute of Theoretical Physics in Utrecht, under supervision of Gerard 't Hooft. The results of this research can be found in this thesis.

\cleardoublepage
\addcontentsline{toc}{chapter}{Bibliography}
\markboth{Bibliography}{}\raggedright
\printbibliography
\cleardoublepage
\end{document}